\documentclass[superscriptaddress,aps,prx,nofootinbib,notitlepage,10pt,longbibliography]{revtex4-2}
\pdfoutput=1
\usepackage{graphicx}
\usepackage{amsmath}
\usepackage{amsfonts}
\usepackage{amsthm}
\usepackage{amssymb}
\usepackage{amsthm}
\usepackage{mathtools}
\usepackage{array}
\usepackage{comment}
\usepackage{cprotect}
\usepackage{placeins}
\usepackage{multirow}
\usepackage{bbm}
\usepackage[caption=false]{subfig}
\usepackage[colorlinks]{hyperref}
\hypersetup{
	pdfstartview={FitH},
	pdfnewwindow=true,
	colorlinks=true,
	linkcolor=blue,
	citecolor=blue,
	filecolor=blue,
	urlcolor=blue}
\usepackage[all]{hypcap}
\usepackage{tikz}
\usepackage{verbatim}
\usepackage{subfig}
\usetikzlibrary{arrows}
\usepackage{soul}
\usepackage{textcomp}
\usepackage{float}
\usepackage{qcircuit}
\usepackage{enumerate}
\usepackage{dcolumn}
\newcolumntype{d}[1]{D{.}{.}{#1}}
\usepackage{longtable}
\usepackage{tabularx}
\newcolumntype{Y}{>{\centering\arraybackslash}X}

\usepackage{listings}
\usepackage{color}
\usepackage[utf8]{inputenc}
\usepackage{comment}
\usepackage{siunitx}
\definecolor{codegreen}{rgb}{0,0.6,0}
\definecolor{codegray}{rgb}{0.5,0.5,0.5}
\definecolor{codepurple}{rgb}{0.58,0,0.82}
\definecolor{backcolour}{rgb}{0.95,0.95,0.92}

\usepackage{xcolor,colortbl}

\DeclareMathOperator{\sos}{SOS}

\DeclareMathOperator*{\argmin}{argmin}

\lstdefinestyle{mystyle}{
  backgroundcolor=\color{backcolour},   commentstyle=\color{codegreen},
  keywordstyle=\color{magenta},
  numberstyle=\tiny\color{codegray},
  stringstyle=\color{codepurple},
  basicstyle=\footnotesize,
  breakatwhitespace=false,
  breaklines=true,
  captionpos=b,
  keepspaces=true,
  numbers=left,
  numbersep=5pt,
  showspaces=false,
  showstringspaces=false,
  showtabs=false,
  tabsize=2
}

\usepackage[capitalise]{cleveref}

\newcommand{\eq}[1]{Eq.~\hyperref[eq:#1]{(\ref*{eq:#1})}}
\renewcommand{\sec}[1]{\hyperref[sec:#1]{Section~\ref*{sec:#1}}}
\DeclareRobustCommand{\app}[1]{\hyperref[app:#1]{Appendix~\ref*{app:#1}}}
\newcommand{\tab}[1]{\hyperref[tab:#1]{Table~\ref*{tab:#1}}}
\newcommand{\fig}[1]{\hyperref[fig:#1]{Figure~\ref*{fig:#1}}}
\newcommand{\figa}[2]{\hyperref[fig:#1]{Figure~\ref*{fig:#1}#2}}
\newcommand{\figx}[2]{\hyperref[fig:#1]{Figure~\ref*{fig:#1}(#2)}}
\newcommand{\thm}[1]{\hyperref[thm:#1]{Theorem~\ref*{thm:#1}}}
\newcommand{\lem}[1]{\hyperref[lem:#1]{Lemma~\ref*{lem:#1}}}
\newcommand{\cor}[1]{\hyperref[cor:#1]{Corollary~\ref*{cor:#1}}}
\newcommand{\defn}[1]{\hyperref[def:#1]{Definition~\ref*{def:#1}}}
\newcommand{\alg}[1]{\hyperref[alg:#1]{Algorithm~\ref*{alg:#1}}}

\def\bra#1{\mathinner{\langle{#1}|}}
\def\ket#1{\mathinner{|{#1}\rangle}}

\newcommand{\proj}[1]{\ket{#1}\!\bra{#1}}
\newcommand{\ketbra}[1]{\ket{#1}\!\bra{#1}}

\def\Be{\textsc{{Be}}}

\newlength{\oldwidth}
\newlength{\oldheight}
\begin{document}

\makeatletter
\def\l@subsubsection#1#2{}
\def\l@f@section{}%
\def\toc@@font{\footnotesize \sffamily}%
\makeatother

\title{Fast quantum simulation of electronic structure by spectrum amplification}

\author{Guang Hao Low}
\email{guanghaolow@google.com}
\thanks{Equal contribution}
\affiliation{Google Quantum AI, Venice, CA 90291, United States}

\author{Robbie King}
\affiliation{Google Quantum AI, Venice, CA 90291, United States}
\affiliation{Department of Computing and Mathematical Sciences, Caltech, CA 91125, United States}

\author{Dominic W. Berry}
\affiliation{School of Mathematical and Physical Sciences,
Macquarie University, Sydney, NSW 2109, Australia}

\author{Qiushi Han}
\affiliation{Department of Computer Science, University of Illinios at Urbana-Champaign, Urbana, IL 61801, United States}

\author{A. Eugene DePrince III}
\affiliation{Google Quantum AI, Venice, CA 90291, United States}
\affiliation{Department of Chemistry and Biochemistry, Florida State University, Tallahassee, FL, United States}

\author{Alec White}
\affiliation{Quantum Simulation Technologies Inc., Boston, MA, USA}

\author{Ryan Babbush}
\affiliation{Google Quantum AI, Venice, CA 90291, United States}

\author{Rolando D. Somma}
\affiliation{Google Quantum AI, Venice, CA 90291, United States}

\author{Nicholas C. Rubin}
\email{nickrubin@google.com}
\thanks{Equal contribution}
\affiliation{Google Quantum AI, Venice, CA 90291, United States}

\begin{abstract}
The most advanced techniques using fault-tolerant quantum computers to estimate the ground-state energy of a chemical Hamiltonian involve compression of the Coulomb operator through tensor factorizations, enabling efficient block-encodings of the Hamiltonian. A natural challenge of these methods is the degree to which block-encoding costs can be reduced. We address this challenge through the technique of spectrum amplification, which magnifies the spectrum of the low-energy states of Hamiltonians that can be expressed as sums of squares. Spectrum amplification enables estimating ground-state energies with significantly improved cost scaling in the block encoding normalization factor $\Lambda$ to just $\sqrt{2\Lambda E_{\text{gap}}}$, where $E_{\text{gap}} \ll \Lambda$ is the lowest energy of the sum-of-squares Hamiltonian. To achieve this, we show that sum-of-squares representations of the electronic structure Hamiltonian are efficiently computable by a family of classical simulation techniques that approximate the ground-state energy from below. In order to further optimize, we also develop a novel factorization that provides a trade-off between the two leading Coulomb integral factorization schemes-- namely, double factorization and tensor hypercontraction-- that when combined with spectrum amplification yields a factor of 4 to 195 speedup over the state of the art in ground-state energy estimation for models of Iron-Sulfur complexes and a CO$_{2}$-fixation catalyst. 
\end{abstract}

\maketitle

\section{Introduction}
High-accuracy simulation of the electronic structure of molecules and materials is one of the primary motivations for building a quantum computer. This motivation originates from the difficulty of robustly simulating electrons with classical techniques at low enough cost, to accommodate large systems necessary for quantitative prediction of chemical phenomena such as reaction chemistry, molecular spectra, and ground-state bonding information. The phase estimation algorithm (PEA)~\cite{kitaev1995quantum}, implemented on a fault-tolerant quantum computer, provides one such route towards quantitative electronic structure simulation at polynomial cost in the system size~\cite{PhysRevLett.79.2586, aspuru2005simulated}. The asymptotic algorithmic advantage of the PEA led to a large body of work examining the constant factors associated with ground-state energy estimation. While initial quantum resource estimates were exceedingly high~\cite{PhysRevA.90.022305}, the focus on optimizing quantum algorithms has resulted in drastic reductions. As a representative example, over the past six years, the quantum resources (gate complexity) required for PEA of the Iron-Molybdenum co-factor FeMoco have steadily improved by factors of 2\textendash4 with each algorithmic innovation~\cite{PhysRevLett.123.070503, kivlichan2019phase, PRXQuantum.2.030305, caesura2025faster, PhysRevResearch.3.033055}, resulting in a seven orders of magnitude reduction in resources. These reductions enable efficient ground-state energy estimation of classically intractable systems that are predicted to run on future, fault-tolerant quantum computers faster than lower-accuracy classical methods~\cite{berry2024rapid, goings2022reliably}. Computational prediction of chemistry usually requires many calculations and thus even further reductions will be required to enable routine large-scale simulation~\cite{PRXQuantum.4.040303}. Herein, we introduce a new algorithmic strategy and take the first steps towards an optimized implementation, resulting in a substantial reduction over prior art.

At the core of recent algorithmic improvements in second quantized electronic structure simulation is the exploitation of tensor decompositions of the Coulomb operator yielding an efficient linear combination of unitaries (LCU) representation. Using the LCU representation, a unitary (a quantum walk operator) implementing a function of the Hamiltonian can be efficiently constructed and used by the PEA~\cite{PhysRevLett.121.010501, BabbushPRX18, berry2018improved}. Ultimately, the quantum cost of performing phase estimation on a walk operator constructed using the LCU representation scales linearly in the normalization constant associated with a `block-encoding' of the Hamiltonian~\cite{low2019hamiltonian, berry2019qubitization}.  For electronic structure simulation, the normalization constant is intimately related to the degree of compression of the two-electron integral operator when constructing the LCU. The two primary strategies for compressing the large two-electron integrals using tensor factorizations are Double-Factorization (DF)~\cite{PhysRevResearch.3.033055} and Tensor-HyperContraction (THC)~\cite{PRXQuantum.2.030305}, each providing normalization constants that scale as low-order polynomials of system size. The freedom of the LCU representation and symmetries of chemical Hamiltonians allow for modification of the Hamiltonian and the LCU normalization constant, by adding operators that annihilate any state with the desired symmetries. This strategy, called the block-invariant symmetry shift (BLISS), was recently demonstrated to reduce LCU normalizations by a factor of 2 when using particle number symmetry shifts~\cite{loaiza2023block, patel2024guaranteed}. Bounds identifying the limits of symmetry shifting also suggest that current integral factorizations combined with symmetry shifting are within a factor of 4 of the optimal LCU normalization~\cite{2024CortesLowerBound, caesura2025faster}. Therefore, finding the algorithmic improvements necessary to make quantum chemistry calculations routine will not be possible solely through more sophisticated tensor factorizations.

To address the stagnating returns of symmetry-shifting and compression, we show that further quantum speedups are achievable by the quantum algorithmic idea of `spectrum amplification' (SA). Fundamentally, SA allows one to relax the precision requirements in certain estimates, potentially resulting in significantly improved resource counts. The concept was initially introduced to achieve a quadratic quantum speedup within the context of adiabatic quantum computation~\cite{somma2013spectral}, and was more recently used to obtain improved quantum algorithms for simulating dynamics of low-energy states~\cite{low2017hamiltoniansimulationuniformspectral,Zlokapa2024hamiltonian}. SA first produces a `square root’ of the Hamiltonian, so that the low-energy spectrum is amplified (i.e., $\sqrt{x} \gg x$ for $x \ll 1$). A precise estimate of a small eigenvalue of the original Hamiltonian can then be obtained by squaring a coarser estimate of an eigenvalue of its square root. The latter problem is expected to be solved more efficiently using the PEA~\cite{gu2021fast}; however, to obtain a net gain, we must ensure that simulating the Hamiltonian square root is not considerably more difficult than simulating the original Hamiltonian. Otherwise, the improvements due to requiring less precision in the estimate could be negated by the overhead due to simulating the square root.  

To this end, we note that SA is directly applicable to Hamiltonians represented as 
a sum-of-squares (SOS):
\begin{align}\label{eq:ham_sos_equality}
H=\sum_{\alpha}O^\dagger_{\alpha} O^{}_\alpha + E_{\sos} =H_\text{sqrt}^\dagger H^{}_\text{sqrt}+E_{\sos} \;,
\end{align}
where $E_{\sos}$ is an energy shift ensuring that $(H-E_{\sos})$ is positive semidefinite. The matrix representations of operators $O_\alpha$ are not necessarily square and we can define $H_{\rm sqrt}\coloneqq \sum_\alpha \ket {\chi_\alpha} \otimes O_{\alpha}$ for any orthogonal set of states $\ket {\chi_\alpha}$\footnote{Equation~\eqref{eq:ham_sos_equality} is also known as 
a `gap-amplifiable' Hamiltonian in Ref.~\cite{Zlokapa2024hamiltonian}.}. SA focuses then on simulating $H_{\rm sqrt}$ rather than $H$. 
To quantify the potential improvement due to SA, 
we can consider a dimensionless parameter $\Delta_\text{gap} \coloneqq (E_{\rm gs}-E_{\sos})/\lambda^2_{\rm sqrt}$, where $E_{\rm gs} \ge E_{\sos}$ is the ground-state energy of $H$, and $\lambda_{\rm sqrt}$ is a normalization factor that will appear in the process 
that block-encodes $H_\text{sqrt}$~\cite{low2017hamiltoniansimulationuniformspectral}. Ideally, we would like to satisfy $\Delta_\text{gap}\ll 1$, in which case $\sqrt{\Delta_\text{gap}} \gg \Delta_\text{gap}$, allowing us to achieve a large amplification of the low-energy spectrum and exploit the power of SA. 
However, this is not always easy to obtain; SOS representations for Hamiltonians of practical relevance with this property might not be efficiently computed~\cite{somma2013spectral,Zlokapa2024hamiltonian}.
And, even in cases where $H_{\rm sqrt}$ {\em can be} efficiently computed, 
we need to ensure that it {\em can be} simulated efficiently, which is not obvious using known block-encoding techniques.  In this work we are able to address these difficulties with three results.

First, we introduce practical quantum algorithms for spectrum amplification of the SOS representation.
As $H_\text{sqrt}$ has singular values $\sqrt{\Delta_{\text{gap}}}\lambda_\text{sqrt}=\sqrt{E_{\rm gs}-E_{\sos}}$, an asymptotic improvement in ground-state energy estimation can already be obtained by singular value estimation of $H_{\text{sqrt}}$~\cite{GSLW2019QSVT,gu2021fast}.
However, current methods for block-encoding matrix square-roots~\cite{2017ChowdhuryGibbs,Zlokapa2024hamiltonian} and for singular value estimation have much larger constant factors than standard block-encoding and phase estimation techniques~\cite{BabbushPRX18}.
We resolve this by introducing rectangular block-encodings of $H_\text{sqrt}$ that are conceptually simpler and cheaper than previous approaches based on Hermitian dilations~\cite{2009HHL}.
A quantum walk unitary can then be defined on rectangular block-encodings with eigenphases $\arccos(\sqrt{\Delta_\text{gap}})$~\cite{GSLW2019QSVT}.
A subtlety arises from this kind of quantum walk as it maps right singular values to left singular values and vice-versa, which complicates a naive application of standard phase estimation techniques~\cite{BabbushPRX18}.
Nevertheless, we prove that two quantum walk steps restores the use of standard phase estimation techniques whilst maintaining the SA advantage.

Second, we elucidate how to express a second quantized representation of $H$ as an SOS Hamiltonian and compute an optimal $\Delta_{\text{gap}}$. The protocol for achieving the equality of Eq.~\eqref{eq:ham_sos_equality} is related to SOS relaxations of non-commutative polynomial optimization problems which have been extensively used in the quantum chemistry, condensed matter, and quantum information communities~\cite{coleman1977convex, mazziotti2006quantum, pironio2010convergent}. The optimization relaxations rely on an SOS representation of non-negative model Hamiltonians which are used to construct an optimization protocol based on semidefinite programming. The primal variables of the semidefinite program (SDP) are the pseudoexpectation values defining approximations to the quantum marginals of a state~\cite{pironio2010convergent, kummer1967n, erdahl1978representability,eisert2023note, PhysRevA.75.032102}. With minor modification, the dual of the lower bound program provides the solution to Eq.~\eqref{eq:ham_sos_equality} with an optimal $\Delta_{\text{gap}}$ given a particular form for the SOS algebra defining the functional form of $O_{\alpha}$. It is well understood that expanding the algebra leads to tighter lower bound estimates $E_{\sos}$ for bosonic and fermionic Hamiltonians~\cite{pironio2010convergent, PhysRevLett.108.263002} and that $E_{\sos}$ can be made arbitrarily close to $E_{\text{gs}}$, and thus $\Delta_{\text{gap}} \ll 1$, with exponential classical resources~\cite{liu2006consistency}. In the context of SA, minimizing the $\Delta_\text{gap}$ by using a large SOS algebra is in tension with a low-cost block-encoding of the SOS Hamiltonian. We find for interesting quantum systems represented by the spin-free Hamilonian
\begin{align}\label{eq:spinfree_Hamiltonian}
H=\sum_{pq}h^{(1)}_{pq}E_{pq}+\frac{1}{2}\sum_{pqrs}h_{pqrs}^{(2)}E_{pq}E_{rs},
\quad
E_{pq}\coloneqq \sum_{\sigma\in\{0,1\}}a^\dagger_{p\sigma}a_{q\sigma},
\end{align}
where $h_{pq}^{(1)}$ and $h_{pqrs}^{(2)}$ are coefficients for the one- and two-electron operators respectively,
that a spin-free variant of the level-2 sum-of-squares algebra (where $O_{\alpha}$ are of quadratic many-body order) already gives small gaps and leads to improvements in ground-state energy estimation. The spin-free variant of the level-2 SOS algebra is the simplest algebra in the SOS hierarchy for chemistry Hamiltonians and more bespoke constructions of the SOS algebra provides a route to investigate the trade-off between gap minimization and block-encoding costs.

\begin{table}
\label{tab:Comparison_highlighs}
\begin{tabularx}{\textwidth}{|c|c|Y|Y|Y|Y|Y|Y|}
\hline
\multirow{2}{*}{Year} & \multirow{2}{*}{Innovation} & \multicolumn{3}{c|}{FeMoco-54~\cite{Reiher2017Elucidating}} & \multicolumn{3}{c|}{FeMoco-76~\cite{li2019electronic}} \\
\cline{3-8}
 & & Qubits & Toffolis& Reference &  Qubits & Toffolis & Reference\\
 \hline\hline
2017 & First resource estimate by Trotterization~\cite{Reiher2017Elucidating} & 111 & $5.0\times 10^{13}$ & \cite{Reiher2017Elucidating} & - & - & - \\
2019 & Qubitization of Single-Factorization~\cite{berry2019qubitization} & 3320 & $9.5\times 10^{10}$ & \cite{PRXQuantum.2.030305} & 3628 & $1.2\times 10^{11}$ & \cite{PRXQuantum.2.030305} \\
2020 & Qubitization of Double-Factorization (DF)~\cite{PhysRevResearch.3.033055} & 3600 & $2.3\times 10^{10}$ & \cite{PhysRevResearch.3.033055} & 6404 & $5.3\times 10^{10}$ & \cite{PRXQuantum.2.030305} \\
2020 & Tensor-Hyper-Contraction (THC)~\cite{PRXQuantum.2.030305} & 2142 & $5.3\times 10^{9}$ & \cite{PRXQuantum.2.030305} & 2196 & $3.2\times 10^{10}$ & \cite{PRXQuantum.2.030305} \\
2024 & Symmetry compression of DF~\cite{Rocca2024SymmetryCompressedDF} &1994 & $2.6\times 10^9$& \cite{Rocca2024SymmetryCompressedDF}& -& - & - \\
2025 & Symmetry compression of THC~\cite{caesura2025faster} &-& -& - & 1512& $4.3\times 10^9$ & \cite{caesura2025faster}\\ \hline
This work & Spectrum amplification \&  DFTHC & 1137 &$3.41\times 10^{8}$ & &  1459 & $9.99\times 10^{8}$ &   \\
\hline\hline
\multicolumn{2}{|c|}{Improvement of this work over ~\cite{Rocca2024SymmetryCompressedDF} and ~\cite{caesura2025faster}\footnote{Our approach is distinct and independent of recent results by Caesura et.\ al~\cite{caesura2025faster}.} } & 1.8$\times$ & 7.0$\times$& & 1.0 $\times$ & 4.3 $\times$ & \\
 \hline
\end{tabularx}
\caption{Improvements in Toffoli and qubit costs on benchmark molecules for performing phase estimation targeting ground-state energies with a standard deviation of at most chemical accuracy $\epsilon_\text{chem}=1.6$mHa. Total Toffoli counts are based on a phase estimation uncertainty of $\sigma_{\text{PEA}}=1$mHa.}
\end{table}

Third, we introduce a new tensor factorization, DFTHC, that realizes an SOS Hamiltonian representation that further optimizes the block-encoding cost. 
The representation found by semidefinite programming minimizes $\Delta_{\text{gap}}$, but does not optimize other quantum-algorithm-relevant parameters.
DFTHC is a variational ansatz that allows for a direct minimization of the block-encoding normalization factor $\lambda_\text{sqrt}$, almost achieves the optimal $E_{\sos}$, and finds extremely compact approximations to the two-body interaction tensor.
DFTHC is also readily combined with BLISS. For benchmarks on molecules with up to $N=150$ orbitals, DFTHC finds chemically accurate representations with block-encoding Toffoli gate costs scaling like $\sim N^{0.96}$, $\lambda^2_\text{sqrt}~\sim N^{1.46}$, and $E_\text{gap}\sim N^{0.88}$.

Combining the three aforementioned results, we report the quantum resources needed for ground-state energy estimation of FeMoco in~\cref{tab:Comparison_highlighs} compared with prior works. The efficiency of DFTHC+BLISS+SA enables a factor of 4 reduction in costs for the larger FeMoco active space~\cite{li2019electronic} using the simplest SOS algebra of the hierarchy. To verify the robustness of the method we also compute resource estimates for small Iron-Sulfur clusters~\cite{li2017spin}, a CO$_{2}$ Ruthenium catalyst~\cite{PhysRevResearch.3.033055} series, and the CPD1 active space of P450~\cite{goings2022reliably}; all of which have been targets of recent resource estimation work and demonstrate the scaling of the effective LCU normalization. For these systems, improvement factors of up to 200 are obtained for un-symmetry-shifted Hamiltonians. Like prior work~\cite{berry2019qubitization,PhysRevResearch.3.033055,PRXQuantum.2.030305} the reported Toffoli and qubit complexities correspond with the task of obtaining an estimate of $E_\text{gs}$ with a standard deviation $\sigma_E$ of at most chemical accuracy $\epsilon_\text{chem}=1.6$mHa. This is accomplished using the PEA to accuracy $\sigma_\text{PEA}$, which projects an input trial state $|\psi_\text{trial}\rangle$ onto the true ground state with probability almost $p=|\langle\psi_\text{gs} |\psi_\text{trial}\rangle|^2$ and then returns an estimate of $E_\text{gs}$ with standard deviation $\sigma_\text{PEA}$. For many interesting electronic structure problems, we can obtain $p \approx 1$ by preparing matrix-product-states (MPS) trial states based on modest bond dimension DMRG calculations which are still estimated to need further energy refinement~\cite{morchen2024classification, berry2024rapid}. For consistency with previous work, we assume $p=1$ in our resource estimates and the cost of PEA is then $\frac{\pi}{2\sigma_\text{PEA}}\lambda_\text{eff}\mathrm{Cost}[\textsc{Be}]$,
where $\mathrm{Cost}[\textsc{Be}]$ is the cost of block encoding a function of the $H_{\text{sqrt}}$.

Attaining chemical accuracy requires bounding error contributions from $\sigma_\text{PEA}$ and the DFTHC tensor approximation error. Tightly and efficiently bounding the error associated with an approximate Hamiltonian in DFTHC form presents a numerical and theoretical challenge that can ultimately impact the overall cost of phase estimation. In order to fully quantify how to deploy the DFTHC factorization we challenge the commonly used truncation threshold metric based on the high-spin CCSD(T) correlation energy $\epsilon_\text{corr}$ of the factorized representation.
If errors are systematic, chemical accuracy is attained if these error satisfy
\begin{align}\label{eq:intro_linear_error_bound}
\sigma_E\le\sigma_\text{PEA}+|\epsilon_\text{corr}| \le \epsilon_\text{chem}.
\end{align}
enabling us to select $\sigma_\text{PEA}=1.0$mHa, and $|\epsilon_\text{corr}|\le 0.6$mHa consistent with prior resource estimates~\cite{PhysRevResearch.3.033055,PRXQuantum.2.030305}.
However, we demonstrate that it is possible to choose much larger $\sigma_\text{PEA}=1.4$mHa by randomizing bits-of-precision truncation and initial conditions in the DFTHC optimization, so that $\epsilon_\text{corr}$ is a random variable with standard deviation $\sigma_\text{corr}\le 0.774$mHa and zero mean.
The randomization protocol has the added benefit that the variances of random errors satisfy
\begin{align}
\sigma_E\le\sqrt{\sigma_\text{PEA}^2+\sigma_\text{corr}^2}\le  \epsilon_\text{chem}.
\end{align}
We present resource estimates for this approach in a later section as it is not directly comparable to previous work but constitutes an improved strategy for accounting for errors in ground-state energy estimation.

The remainder of the paper is structured as follows:~\cref{sec:phase_esitimate_sos} reviews the general SA protocol and describes circuit improvements framed by modifications to singular value estimation,~\cref{sec:SOS_representations} describes the SDP approach to minimizing the SOS gap for electronic structure problems and demonstrates that $\Delta_{\text{gap}}\ll 1$ are achievable at polynomial classical cost,~\cref{sec:DFTHC} introduces DFTHC, the quantum circuits for its block-encoding, and numerical result. In~\cref{sec:results} we present resource estimates for a variety of benchmark systems and report improvement factors over prior work. We also highlight potential challenges in using high-spin correlation energy from CCSD(T) and improved strategies for optimizing errors for each algorithmic component to achieve chemical precision. 

\section{Quantum circuits for spectrum amplification}\label{sec:phase_esitimate_sos}

We describe SA within the context of SOS representations for Hamiltonians, discuss block-encodings and PEA,
and present practical quantum circuits for their implementations.  
Let
\begin{align}\label{eq:gap_amplifiable_Hamiltonian}
H_{\text{SA}} \coloneqq \sum_{\alpha=0}^{L-1}O_{\alpha}^{\dagger}O^{}_{\alpha},
\quad
H=H_{\text{SA}}+E_\text{SOS},
\end{align}
where $O_{\alpha}$ corresponds to a linear combination of operators. 
Assuming an SOS representation like Eq.~\eqref{eq:gap_amplifiable_Hamiltonian} for an arbitrary Hamiltonian acting on a finite-dimensional space is without loss of generality; in~\cref{sec:SOS_representations}, we show how this representation can be computed efficiently.  
As $O_{\alpha}^{\dagger}O^{}_{\alpha} \succeq 0$ is positive semidefinite, the eigenvalues of $H_{\rm SA}$ are non-negative. %
The special case where the lowest eigenvalue of $H_{\text{SA}}$ is zero corresponds to $H_{\text{SA}}$ being `frustration free',
in which case the smallest eigenvalues of all $O_{\alpha}^{\dagger}O^{}_{\alpha}$'s are also zero. For a given $H$, it is always possible
to shift it and find a frustration-free $H_{\rm SA}$, but this might require exponential-cost preprocessing and thus is not feasible. 
Despite this, it might be possible to efficiently find an $H_{\rm SA}$ that can be written as an SOS and that is `near' frustration free, with lowest eigenvalue $E_\text{gap}=E_\text{gs}-E_\text{SOS}$ close to zero, which is the approach considered in this work. 
Note that the square root $H_\text{sqrt}$ of $H_{\rm SA}=H_\text{sqrt}^\dagger H_\text{sqrt}$ has a singular value $\sqrt{E_\text{gap}}$, having an amplified sensitivity to variations in $E_\text{gs}$.

The quantum circuits we present for SA are based on the block-encoding framework~\cite{low2019hamiltonian}. We say a finite-dimensional operator $O$ is block-encoded with normalization factor $\lambda >0$ such that $\|O\| \le \lambda$, if there is a unitary quantum circuit 
\begin{align}
\Be\left[{O}/{\lambda}\right]=
\left(
\begin{array}{cc}
O/\lambda&\cdots\\
\vdots&\ddots
\end{array}
\right) \in \mathbb C^{n \times n}
\quad\implies\quad
\Pi_l \Be[O/\lambda]\Pi_m=O/\lambda,
\end{align}
where $O/\lambda \in \mathbb C^{l \times m}$ appears as a block of  $\Be\left[{O}/{\lambda}\right]$, with $l \le n$ and $m\le n$.
The operators $\Pi_l$ and $\Pi_m$ are orthogonal projectors onto subspaces of dimension $l$ and $m$, specified by the (all-zero) strings $\ket{0}_l$, and $\ket{0}_m$, respectively. (Note that $l \ne m$ if $O$ is not square.)
Highly-efficient block-encoding of $H_\text{sqrt}$ are possible to construct. Assume we are given access to controlled-block-encodings $\Be[O_\alpha/\lambda_\alpha]$, for some $\lambda_\alpha \ge \|O_\alpha\|$. Then, one may apply unitary iteration and coherent alias sampling state preparation~\cite{BabbushPRX18} to synthesize the `SELECT' and `PREPARE' circuits
\begin{align}\label{eq:rectangular_BE_primitives}
\textsc{Sel} \coloneqq \sum_{\alpha=0}^{L-1}\ket{\alpha}\! \bra{\alpha}\otimes\Be\left[\frac{O_\alpha}{\lambda_\alpha}\right],
\quad
\textsc{Prep}\ket{0}_\mathrm{a}=\frac 1 {\lambda_\text{sqrt}} \sum_{\alpha=0}^{L-1}{\lambda_\alpha}\ket{\alpha}_\mathrm{a}\ket{\text{garb}_\alpha}_\mathrm{a}
=\frac 1 {\lambda_\text{sqrt}} 
\sum_{\alpha=0}^{L-1}{\lambda_\alpha}\ket{\chi_\alpha}_\mathrm{a} \;,
\end{align}
where $
\lambda_\text{sqrt}: = \sqrt{\sum_{\alpha} \lambda_\alpha^2}$, `${\rm a}$' refers to ancillary qubits, and $\ket{\text{garb}_\alpha}_\mathrm{a}$ represents a `garbage' (unit) state that is irrelevant.
Then, we can let $H_\text{sqrt}= {\sum_{\alpha=0}^{L-1}|\chi_\alpha\rangle \otimes
 O_{\alpha}}$ and 
use these operations to construct the desired block-encoding of $H_\text{sqrt}$; that is,
\begin{align}\label{eq:rect_BE_primitives}
\Pi_l \textsc{Sel} \Pi_m \cdot \textsc{Prep}\ket{0}_\mathrm{a}&=\frac 1 {\lambda_\text{sqrt}} {\sum_{\alpha=0}^{L-1} |\chi_\alpha\rangle \otimes O_{\alpha}} \implies 
\textsc{Sel} \cdot \textsc{Prep}=\Be\left[\frac{H_\text{sqrt}}{\lambda_\text{sqrt}}\right] \coloneqq U \;.
\quad\end{align}
Note that $U$ acts on an enlarged space and the (column) dimension of the block containing $H_{\text{sqrt}}/\lambda_{\text{sqrt}}$ is larger than that of the blocks containing the individual $O_{\alpha}/\lambda_{\alpha}$'s.
Also, observe that $H_\text{sqrt}^\dagger H_\text{sqrt} = H_\text{SA}$ as desired and, 
if $E_k \ge 0$ is an eigenvalue of $H_{\rm SA}$, then $\sqrt{E_k}$ is a singular value of $H_\text{sqrt}$. 
Our definition and block-encoding of $H_\text{sqrt}$ using rectangular matrices is more efficient than previous works based on a Hermitian dilation~\cite{somma2013spectral,Zlokapa2024hamiltonian}, which use more qubits and are at least a factor of two more expensive to block-encode as shown in~\cref{sec:equivalent_SA_walks}.

\begin{figure}[H]
    \centering
    \includegraphics[width=0.85\linewidth]{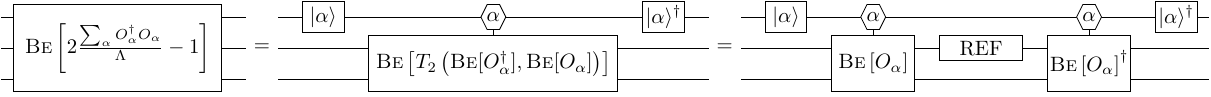}
    \caption{Controlled block encoding of the second Chebyshev polynomial of $O_{\alpha}$. The control register contains the control state, $\vert \alpha\rangle$, of Eq.~\eqref{eq:rect_BE_primitives} which provides the appropriate weighting to each $T_{2}$ in the LCU. For simplicity, $\lambda_\alpha=1$.}
    \label{fig:single_step_qubitization}
\end{figure}

As its core, the standard
PEA estimates the eigenphases of a unitary operator.
For Hermitian matrices $H$ with eigenvalues $E_1,E_2,\ldots$, it is well known that qubitization~\cite{low2019hamiltonian} converts a block-encoding $\Be[H/\lambda]$ to a quantum walk with spectrum $e^{\pm i\arccos{(E_k/\lambda)}}$ \cite{2012ChildsQuantumWalk}.
This reduction was recently employed to construct  circuits for quantum eigenvalue estimation of a Hamiltonian represented as an LCU~\cite{PRXQuantum.2.030305, BabbushPRX18, PhysRevResearch.3.033055}.
The eigenphases of the quantum walk have positive and negative values, but this has no effect on the estimated Hamiltonian eigenvalues, and will be ignored in the following.
In~\cref{sec:quantum_walk_rectangular}, we describe a quantum walk $W:=\textsc{Ref}_{\mathrm{a}\Be}U^\dagger\textsc{Ref}_{\Be}U$ on block-encodings for rectangular matrices like $H_\text{sqrt}$~\cite{GSLW2019QSVT}, where $\textsc{Ref}_{\Be}$ and and $\textsc{Ref}_{\mathrm{a}\Be}$ are reflections over the subspaces spanned by the the columns and rows of $U$ that block-encode $H_\text{sqrt}/\lambda_{\rm{sqrt}}$, respectively.
There, we show that $W$ has  spectrum
\begin{align}\label{eq:two_walk_phases}
e^{\pm i\arccos{(E_k/\Lambda-1)}},\quad \Lambda \coloneqq \frac{1}{2}\lambda^2_\text{sqrt}.
\end{align}
In fact, one may show that the first three operations of the quantum walk provide a block-encoding of $H_\text{SA}/\Lambda-1$:
\begin{align}
\Be\left[\frac{H_\text{SA}}{\Lambda}-1\right]\equiv U^\dagger\textsc{Ref}_{\Be}U=\textsc{Prep}\cdot\textsc{Sel}^\dagger\cdot\textsc{Ref}_{\Be}\cdot\textsc{Sel}\cdot\textsc{Prep}^\dagger.
\end{align}
This is equivalent to constructing a controlled-block encoding $\sum_\alpha\proj{\alpha}\otimes\Be\left[2 O^\dagger_\alpha O^{}_\alpha/\lambda^2_\alpha-1\right] = \sum_\alpha\proj{\alpha}\otimes\Be\left[T_{2}\left(\Be[O^{\dagger}_{\alpha}],\Be[ O^{}_\alpha]\right)\right]$ using a single step of qubitization as shown in~\cref{fig:single_step_qubitization},
followed by taking a linear combination block-encodings of $T_{2}(\Be[O_{\alpha}], \Be[O_{\alpha}^{\dagger}])$, using preparation $\textsc{Prep}$ and `unpreparation' $\textsc{Prep}^\dagger$ of the control state $\textsc{Prep}\ket{0}_a$ in~\cref{eq:rect_BE_primitives}. 
This generalizes similar constructions in prior work: Oblivious amplitude amplification~\cite{BerryFOCS15} when $H_\text{sqrt}$ is unitary, or for obtaining a $\frac{1}{2}$ improvement in the block-encoding normalization for Hermitian $H_\text{sqrt}$~\cite{PhysRevResearch.3.033055}.

We now perform the PEA on $W$, which provides an estimate of the eigenphase with standard deviation $\sigma_{\text{PEA}}$, using $\frac{\pi}{2\sigma_{\rm PEA}}$ total queries to controlled $U$ and $U^\dagger$~\cite{BabbushPRX18}.
The Hamiltonian $H_{\rm SA}$ has lowest eigenvalue $E_\text{gap}$, and this results in eigenphases $|\phi_\text{gap}|=\arccos(E_\text{gap}/\Lambda-1)$ for the walk operator $W$.
We use a first-order error propagation formula to obtain the  query cost for estimating $E_\text{gap}$ with variance $\sigma^2_{E_\text{gap}}$, being proportional to 
\begin{align} \label{eq:lambda_eff}
\text{PEA queries to $W$}=
\frac{\pi\lambda_\text{eff}}{2\sigma_{E_\text{gap}}}
,\quad
\lambda_\text{eff}=\sqrt{E_\text{gap}(2\Lambda - E_\text{gap})} \;.
\end{align}
Thus, the `effective lambda' $\lambda_{\text{eff}}$ for performing PEA on a single step of qubitization depends on the block-encoding scaling factor for $H_{\text{SA}}$ and $\sqrt{E_\text{gap}}$.
There can be some inaccuracy in the error propagation formula near a divergence of the derivative when $E_{\rm gap}\rightarrow 0$.
However, we show in~\cref{sec:spectrum_amplification_divergent_derivative} that the fractional corrections to $\sigma_{E_{\text{gap}}}$ are negligible for the parameters we consider. 

In summary, SA  has a smaller effective scaling constant $\lambda_{\text{eff}}$ when considering the uncertainty scaling of the $\arccos$ function of the Hamiltonian.  
The main idea in the next section is to construct an SOS representation $H_{\text{SA}}$ that has a low value of $E_\text{gap}$.
We will see that methods for computing good lower bounds to $E_\text{gs}$ translate directly to maximizing $E_\text{SOS}$, which in turn minimizes $E_\text{gap}$.

\section{Sum-of-squares representations of the chemistry Hamiltonian}\label{sec:SOS_representations}

Given a Hamiltonian $H$ that is a $2k$-degree polynomial in a set of operators $\{\mathfrak{o}_{j}\}$, which do not necessarily commute, the Hamiltonian plus a constant shift 
can be represented as an SOS:
\begin{align}\label{eq:ham_in_sos_form}
H - E_{\text{SOS}} = \sum_{\alpha=1}^{L-1}O_{\alpha}^{\dagger}O^{}_{\alpha},
\end{align}
where each $O_{\alpha}$ is itself a polynomial in $\{\mathfrak{o}_{j}\}$ of degree at least $k$, and $E_{\text{SOS}}$ is a constant. 
While this is not generally possible for continuous commutative variables (the Motzkin polynomial being one example~\cite{motzkin1967arithmetic}), operators that are constructed from polynomials of non-commutative variables can be represented as SOS of Hermitian operators~\cite{helton2002positive}. More generally, it is possible to represent the Hamiltonian with $O_{\alpha}$ involving polynomials of degree $\geq k$ which can lead to $E_{\text{SOS}}$ being closer to the lowest eigenvalue of $H$. Such a representation of the Hamiltonian constitutes a proof for the lowest eigenvalue of the Hamiltonian being at least $E_{\text{SOS}}$. In fact, the problem of expressing the Hamiltonian in this form is the dual problem of identifying a lower bound to $E_{\text{gs}}$ through the pseudomoment method~\cite{kummer1967n,erdahl1978representability, coleman1977convex}. The variational two-electron reduced-density-matrix method is one such example relevant to electronic structure, where a hierarchy of operator witnesses constrains the 2-particle reduced-density-matrix (2-RDM) to have a fermionic $n$-electron wavefunction preimage while directly optimizing with respect to the 2-RDM degrees of freedom. We now describe the efficient protocol for determining an SOS representation of a Hamiltonian through semidefinite programming.

We first note that by fixing a degree $k$, we define $\sos^k$ to be the subset of operators which can be written as a sum of squares of degree $k$ polynomials in $\{\mathfrak{o}_{j}\}$. Given the generator set involving operators in $\sos^{k}$
\begin{align}\label{eq:sos_generator}
O_{\alpha} = \sum_{d=0}^{k/2}\sum_{j}c_{j,\alpha, d}\prod_{\ell=0}^{d}\mathfrak{o}_{j,d,\alpha},
\end{align}
we can express the determination of the optimal shifted Hamiltonian as a semidefinite program
\begin{align}\label{eq:sos_sdp_program}
\max\;& E_{\text{SOS}} \\
\mathrm{s.t.}\; &H - E_{\text{SOS}} \mathbb{I} = \left(\Vec{\mathfrak{o}}^{\dagger}\right)^{T}G\Vec{\mathfrak{o}}, \nonumber \\
&G \succeq 0, \nonumber
\end{align}
where $\Vec{\mathfrak{o}}$ is the vector of operators in Eq.~\eqref{eq:sos_generator}. The SOS generators, $\mathfrak{o}_{j,d,\alpha}$, can be an arbitrary algebra as long as it is possible to represent $H$ in that algebra. Considering fermionic Hamiltonians, the natural SOS generating algebra is the set of fermionic ladder operators, and thus a specific example of $\Vec{\mathfrak{o}}$ is
\begin{align}
\Vec{\mathfrak{o}} = \left(\mathbb{I}, a_{i\sigma},a_{j\tau}^{\dagger},a_{i\sigma}a_{j\tau},a_{i\sigma}a_{j\tau}^{\dagger},...\right)^{T}.
\end{align}
where $a_{i\sigma}$ represents an annihilator for fermionic mode $i$ with spin projection $\sigma$.
The equality constraint on the Gram matrix, $G$, in Eq.~\eqref{eq:sos_sdp_program} is short hand for relating coefficients of normal ordered operators. We include the identity operator in this set as it can arise from the anticommutation relations. As a simple example, consider a spinless quadratic Hamiltonian
\begin{align}\label{eq:quad_sos_ham}
H = \sum_{j}v_{j}a_{j}^{\dagger} a^{}_{j} ,
\end{align}
and the SOS algebra $\{a^{}_{i},a_{i}^{\dagger}\}$. One can determine an SOS representation of Eq.~\eqref{eq:quad_sos_ham} by solving the program in Eq.~\eqref{eq:sos_sdp_program} to find the optimal lower bound. Factorizations of the Gram matrix $G$ provide the SOS representation of the Hamiltonian via Eq.~\eqref{eq:sos_generator}. In this example, it is also possible to analytically solve for the optimal $E_{\text{SOS}}$ and $O_{\alpha}$. First, note that operators associated with positive $v_{j}$ are already in SOS form. For negative $v_{j}$ the anticommutation relations can be applied to return an SOS operator with a positive coefficient plus an identity term. More explicitly,
\begin{align}\label{eq:H1normalorder}
H&= \sum_{j: v_j \ge 0} v_j a^\dagger_j a^{}_j + \sum_{j:v_j<0} (-v_j) a^{}_j a^\dagger_j + \sum_{j:v_j<0}v_j = \sum_\alpha O^\dagger_\alpha O^{}_\alpha +E_{\sos} \;,
\end{align}
 where the $O_\alpha$'s are the $a^\dagger_j$'s or $a^{}_j$'s depending on the sign of $v_j$, and the shift is $E_{\sos}=\sum_{j:v_j<0}v_j$.
This proves that for diagonal quadratic Hamiltonians, and using the SOS algebra $\{a^{}_i,a^\dagger_i\}$, the optimal $E_{\text{SOS}}$ can be determined. The result can be extended to non-diagonal quadratic Hamiltonians involving terms like $a^\dagger_i a^{}_j$, as they can be diagonalized efficiently. For Hamiltonians beyond quadratic many-body order more sophisticated SOS generating algebras are required to represent the Hamiltonian.  Returning to the electronic structure Hamiltonian of Eq.~\eqref{eq:spinfree_Hamiltonian},
we can consider the entire level-$2$ SOS algebra spanned by generators
\begin{align}\label{EQN:BASIS} 
\left\{a^{}_{j\sigma},a^\dagger_{j\sigma}\right\}\cup
\left\{a^{}_{i\tau}a^{}_{j\sigma},a^{\dagger}_{i\tau}a^{}_{j\sigma},a^{}_{i\tau}a^{\dagger}_{j\sigma},a^{\dagger}_{i\tau}a^{}_{j\sigma}\right\}
\end{align}
with an element of the algebra defined by a particular linear combination of the generators
\begin{align}\label{EQN:BASIS}
O_\alpha = \sum_{j,\sigma}c_{j\sigma}^{\alpha}a_{j\sigma} + \sum_{j\sigma}d_{j\sigma}^{\alpha}a_{j\sigma}^{\dagger} + \sum_{i,j,\sigma,\tau}b_{i\sigma j\tau}^{\alpha}a_{j\tau}a_{i\sigma}+\sum_{i,j,\sigma,\tau}d_{i\sigma j\tau}^{\alpha}a_{j\tau}^\dagger a_{i\sigma}^\dagger+\sum_{i,j,\sigma,\tau}e_{i\sigma j\tau}^{\alpha}a_{j\tau}^\dagger a_{i\sigma}+\sum_{i,j,\sigma,\tau}f_{i\sigma j\tau}^{\alpha} a_{j\tau} a^\dagger_{i\sigma} 
\end{align}

where the coefficients $\{c, d, b, d, e, f\}$ are complex values and thus
\begin{align}\label{eq:sos_hamiltonian_non_normal_ordered}
H_\text{SOS} \coloneqq \sum_\alpha O_\alpha^\dagger O_\alpha = \sum_{i,j,k,l,\sigma,\tau,\kappa,\lambda} {}^{D_2}G^{i\sigma j\tau}_{k\kappa l\lambda} a^\dagger_{i\sigma} a^\dagger_{j\tau}  a_{l\lambda}a_{k\kappa} + {}^{Q_2}G_{i\sigma j\tau}^{k\kappa l\lambda}  a_{k\kappa} a_{l\lambda} a^\dagger_{j\tau} a^\dagger_{i\sigma}  + ...,
\end{align}
as is done commonly in the chemistry application of outer approximations--2-RDM constraints that are known as DQG~\cite{PhysRevA.65.062511}. While optimization over this set of SOS-generators gives a tight $E_{\text{gap}}$ mirroring the high quality solutions from the variational 2-RDM method~\cite{PhysRevA.65.062511, rubin2014comparison}, the block-encoding would require a full spinful representation of $H_{\text{sqrt}}$ as the different spin sectors of the SOS Hamiltonian would contain different coefficients.  Furthermore, the largest DQG-lower bound calculations that have been performed are on the order of $60$ orbitals~\cite{mullinax2019heterogeneous} and thus it is potentially challenging to scale the classical preprocessing needed to employ SA. 
Therefore, we consider the simplified algebra spanned by
\begin{align}
\left\{\mathbb{I}, a_{j\sigma}, a_{j\sigma}^{\dagger}, \sum_{\sigma}a_{i\sigma}^{\dagger}a_{j\sigma}\right\}
\end{align}
with elements defined by
\begin{align}\label{eq:spin_free_algebra_sos_generator}
O_{\alpha} \coloneqq \sum_{j,\sigma}c_{j\sigma}^{\alpha}a_{j\sigma} + \sum_{j,\sigma}d_{j\sigma}^{\alpha}a_{j\sigma}^{\dagger} + \left(e^{\alpha} \mathbb{I} + \sum_{i,j}g_{ij}^{\alpha}\sum_{\sigma}a_{i\sigma}^{\dagger}a_{j\sigma}\right).
\end{align}
The number of generators that span this algebra is slightly larger than the minimum SOS algebra to represent the Hamiltonian. The addition of $\mathbb{I}$ makes this algebra equivalent to the spin-free algebra augmented with hole-rotation generators $\sum_{\sigma}a^{}_{i\sigma}a_{j\sigma}^{\dagger}$.  

The equality constraint of the SDP, Eq~\eqref{eq:sos_sdp_program}, mapping operators in the algebra of Eq.~\eqref{eq:spin_free_algebra_sos_generator} to the chemistry Hamiltonian Eq.~\eqref{eq:spinfree_Hamiltonian} can be obtained by re-ordering the fermionic ladder operators according to the anticommutation relations. We provide a complete derivation of the linear constraints in Appendix~\ref{app:spin-free-sdp-def}. In order to numerically verify that $E_{\text{gap}}$ is small for challenging electronic structure simulations we use a combination of variational wavefunctions, primarily low bond dimension MPS, to upper bound the size of the gap. In Table~\ref{tab:var_energies} we report known or computed variational upper bounds for common benchmarked systems.  The lower bounds are computed from the solution of the semidefinite program defined in Eq.~\eqref{eq:sos_sdp_program} using the spin-free algebra defined in Eq.~\eqref{eq:spin_free_algebra_sos_generator} modified to the isospectral algebra containing the hole rotation generators. To solve the semidefinite program we use the cuLoRADS solver~\cite{han2024accelerating} which implements a GPU-accelerated low-rank factorization-based solver~\cite{han2024low}. In all SDPs we use a rank on the primal solution scaling as $\mathcal{O}(\log(m))$ where $m$ is the number of linear constraints.  This scaling is inspired by the existence of approximate solutions with that order of rank~\cite{so2008unified}. Rank scaling as the logarithm of the number of constraints is substantially smaller than the rank necessary to certify that the local minimum corresponds to the global minium~\cite{waldspurger2020rank}. Despite this, we find stable solutions with a tight gap. In~\cref{tab:var_energies} we report variational upper bounds, spin-free algebra lower bounds, and energy gaps $(E_{\text{gs}} - E_{\sos})$ that we use in the next section to inform the search for symmetry-shifted Hamiltonians.  In the final column of Table~\ref{tab:var_energies} we report a lower bound on the SA effective LCU normalization. The lower bound is computed using the reported gaps and the lower bound to the LCU normalization computed by single determinant wavefunctions for the maximum and minimum energy state in a given particle and spin sector~\cite{2024CortesLowerBound}. 
\begin{table}[htb]
    \centering
    \scalebox{1}{
    \begin{tabular}{|c|c|D{.}{.}{6.5}|D{.}{.}{6.4}|c|D{.}{.}{2.3}|}
    \hline \hline
    \multirow{2}{*}{System} &  \multirow{2}{*}{Method} & \multicolumn{3}{c|}{Variational energy [Ha]}      & \multicolumn{1}{c|}{$\lambda_{\text{eff}}$  lower}
    \\
    \cline{3-5}
    && \multicolumn{1}{c|}{above } & \multicolumn{1}{c|}{ below}  & $E_\text{gap}$&  \multicolumn{1}{c|}{bound [Ha]}
    \\
    \hline
      Fe$_2$S$_2$ [30e, 20o]         & DMRG M=1500                                       &      -116.60551           & -117.582         & 0.976       &  3.632 \\
      Fe$_4$S$_4$ [54e, 36o]        & DMRG M=600                                        &       -327.21257          & -329.079         & 1.866       &  7.472 \\ 
      FeMoco [54e, 54o]             & DMRG M=500                                        &       -269.06020    & -272.524  & 3.464       &  16.025 \\   
      FeMoco [113e, 76o]            & DMRG M=6000 ~\cite{berry2024rapid}                &       -1119.05933    & -1123.300             & 4.240       &  21.691 \\ 
      CO$_2$[XVIII] [64e, 56o]       & DMRG M=500                                        &       -295.24441        & -296.941           & 1.695        &  11.065 \\ 
      CO$_2$[XVIII] [100e, 100o]     & DMRG M=500                                        &       -632.32045           & -635.134    & 2.813$^*$ & 21.478 \\ 
      CO$_2$[XVIII] [150e, 150o]     & CISD                                             &       -1350.65168            & -1355.571       & 4.919$^*$ & 37.365 \\ 
      CPD1-P450G [47e, 43o]         & DMRG M=1500~\cite{goings2022reliably}             &       -232.38295    & -235.136           & 2.753       & 11.720 \\ 
      CPD1-P450X [63e, 58o]         & DMRG M=500                                        &       -419.84903    & -424.550             & 4.700          & 21.141 \\ 
      \hline\hline
     \end{tabular}
    }
    \caption{Variational energies from above and below computed with different methods. The lower bounding energies are obtained by solving the SDP program~\eqref{eq:sos_sdp_program} defined over the spin-free algebra described in Eq.~\eqref{eq:spin_free_algebra_sos_generator}. The SDP finds a lower bound to the lowest energy in Fock space versus the variational energies are for a fixed particle sector. The gaps in the penultimate column are the difference between the upper and lower bound. All variational energies are for the active space Hamiltonians without a core shift $H_\text{ecore} = 0$. The asterisk ($^{*}$) indicates the systems where, due to a runtime constraints we did not converge the SDP for the $\ge100$ orbital CO$_{2}$ systems and instead computed the gap using DFTHC as described in~\cref{sec:E_gap_estimation}. The last column reports a lower bound on the SA $\lambda_{\text{eff}} \geq \sqrt{E_{\text{gap}}(\Delta_{u} - E_{\text{gap}})}$ using the LCU normalization lower bound $\Delta_{\mu}/2$ computed using the protocol in Ref.~\cite{2024CortesLowerBound}. The bound for FeMoco-[113e,76o] was computed using using the $S=3/2$ sector, for CPD1-P450 both the X and G Hamiltonian bounds were computed using $S=5/2$. For all other systems the bound was computed using the singlet $S=0$ sector.  }
    \label{tab:var_energies}
\end{table}

The lower bounds in Table~\ref{tab:var_energies} are far from accuracies required for thermochemistry, but this is not the most important metric for SA. The parameter scaling walk operator queries is the square root of the ground-state energy of $H_{\text{SOS}}$ and thus extremely tight gaps--close to frustration free representations of $H$--are not necessary to achieve a large speedup. The bounds reported in Table~\ref{tab:var_energies} demonstrate that SOS representations with small enough ground-state energies can be found with polynomial classical computing effort. Tighter lower bounds could be derived by the addition of penalty parameters or BLISS~\cite{loaiza2023block} like operators to the SOS representation. An alternative to penalty parameters is directly converting solutions from the pseudomoment problem (like variational 2-RDM) to a $H_{\text{SOS}}$ Hamiltonian. In Appendix~\ref{sec:sos-dual-moment-conversion} we describe how the primal non-commutative polynomial optimization problem can be rewritten in a standard form such that all linear constraints are necessarily zero when a primal feasible solution is found. Using the optimal dual variables of the primal problem, the gram matrix $G$ can be constructed with BLISS-like operators appearing as the equality constraints mapped to SOS operators.  The SOS mapping of equality constraints can be non-trivial or require significant modification of the standard primal program meaning it is not always a computational benefit to solve the primal problem and then convert to the dual SOS.

\section{Double factorized-tensor-hypercontraction sum-of-squares: DFTHC-SOS}
\label{sec:DFTHC}

We introduce the Double-Factorized-Tensor-Hypercontraction (DFTHC) of two-body electronic structure tensor, which is designed to span the spin-free generating set~\cref{eq:spin_free_algebra_sos_generator}.
DFTHC generalizes Double-Factorization~\cite{Motta2021LowRank, PhysRevResearch.3.033055}, its variations~\cite{Oumarou2024acceleratingquantum, Rocca2024SymmetryCompressedDF}, and also the factorized form of Tensor-Hypercontraction~\cite{PRXQuantum.2.030305}.
For spin-free Hamiltonians, the DFTHC representation is
\begin{align}\label{eq:DFTHC_representation}
H_{\text{DFTHC}}&=\sum_{pq}h^{(1)\prime}E_{pq} + \frac{1}{2}\sum_{r\in[R]}\sum_{c\in[C]}\left(W^{(rc)}\mathbb{I}+\sum_{b=0}^{B-1}w^{(rc)}_{b}\sum_{pq}u^{(r)}_{b,p}u^{(r)}_{b,q}E_{pq}\right)^2-\frac{1}{2}\sum_{r\in[R]}\sum_{c\in[C]}\left|W^{(rc)}\right|^2,\\
h^{(2)\prime}_{pqmn}&=\sum_{r\in[R],c\in[C]}\left(\sum_{b\in[B]}w^{(rc)}_{b}u^{(r)}_{b,p}u^{(r)}_{b,q}\right)\left(\sum_{b\in[B]}w^{(rc)}_{b}u^{(r)}_{b,m}u^{(r)}_{b,n}\right),\label{eq:DFTHC_h2_tensor}
\end{align}
where $h^{(2)\prime}$ is the approximated two-body tensor, Rank, Bases, and Copies $(R,B,C)$ parameters may be varied freely, and all parameters $h^{(1)\prime}, W,w,u$ are real with $\|\vec{u}^{(r)}_{b}\|_2=1$.
Note the inclusion of the identity component $W^{(rc)}_{B}\mathbb{I}$ as a free parameter, which is necessary according to~\cref{eq:spin_free_algebra_sos_generator}.
We also find it convenient to define $w^{(rc)}_{B}=W^{(rc)}+\sum_{b=0}^{B-1}w^{(rc)}_b$.
Thus, double-factorization the is special case of varying $R$ while fixing $B=N$, $w^{(rc)}_{B}=0$, and $C=1$~\cite{Motta2021LowRank,PhysRevResearch.3.033055} or $C=2$~\cite{Rocca2024SymmetryCompressedDF} or $C=N$~\cite{Oumarou2024acceleratingquantum}, in addition to imposing a mutual orthogonality constraint on $\vec{u}^{(r)}_{b}\cdot \vec{u}^{(r)}_{b^\prime}=\delta_{b,b'}$ for each rank component.
Furthermore, $(R,B,C)=(1,R_{\text{THC}},R_{\text{THC}})$ captures the factorized THC representation of two-body terms
$ \sum_{c\in[R_{\text{THC}}]}\left(\sum_{b\in{[R_{\text{THC}}]}}w^{(c)}_{b}\sum_{pq}u_{b,p}u_{b,q}E_{pq}\right)^2$.

Let us now express $H_\text{DFTHC}$ in the SOS form and evaluate its normalization factor $\Lambda$. 
We represent regular fermion operators with Majorana operators
\begin{align}
a_{p\sigma}=\frac{\gamma_{p\sigma0}+i\gamma_{p\sigma1}}{2},
\quad 
a^\dagger_{p\sigma}=\frac{\gamma_{p\sigma0}-i\gamma_{p\sigma1}}{2},
\quad\{\gamma_{p\sigma x},\gamma_{q\tau y}\}=2\delta_{pq}\delta_{\sigma\tau}\delta_{xy}.
\end{align}
In the following, we also find it convenient to define the rotated Majorana operator $\gamma_{\vec{u}\sigma x}\coloneqq \sum_{j}u_j\gamma_{j\sigma x}$ for any real unit vector $\vec{u}$.
This representation is convenient for compiling to quantum circuits as each Majorana operator maps to a Pauli operator.
Let the eigendecomposition of $h^{(1)\prime}_{pq}=\sum_{s\in\{-,+\}}\sum_{r}s w^{(r)}_su^{(r)}_{s,p} u^{(r)}_{s,q}$ be split into semipositive and negative parts where all $w^{(r)}_\pm\ge0$.
Then $h^{(1)\prime}_{pq}E_{pq}$ is diagonalized and, following~\cref{eq:H1normalorder}, we find its optimal $E_{\sos}$ contribution by 
\begin{align}
h^{(1)\prime}_{pq}E_{pq}&=\sum_{\sigma\in\{0,1\}}\left(\sum_{j}w^{(j)}_+a^\dagger_{\vec{u}^{(j)}_+\sigma}a_{\vec{u}^{(j)}_+\sigma}+\sum_{j}w^{(j)}_-a_{\vec{u}^{(j)}_-\sigma}a^\dagger_{\vec{u}^{(j)}_-\sigma}
\right)-2\sum_{j}w^{(j)}_- \;.
\end{align}
It is possible to verify the SOS representation: 
\begin{align}
H\approx H_\text{DFTHC}&=\sum_{G\in\{\mathrm{D}_1,\mathrm{Q}_1\}}\sum_{r}\sum_{\sigma\in\{0,1\}}{O_{G^\sigma,r}}^\dagger{O_{G^\sigma,r}}+\sum_{r\in[R],c\in[C]}O^\dagger_{\text{SF},rc}O_{\text{SF},rc}+E_\text{SOS},
\\
{O_{\mathrm{D}_1^\sigma,r}}&=\sqrt{w^{(r)}_{+}}\frac{\gamma_{\vec{u}^{(r)}_+\sigma 0}+i\gamma_{\vec{u}^{(r)}_+\sigma 1}}{2},
\qquad
{O_{\mathrm{Q}_1^\sigma,r}}={\sqrt{w^{(r)}_{-}}\frac{\gamma_{\vec{u}^{(r)}_-\sigma 0}-i\gamma_{\vec{u}^{(r)}_-\sigma 1}}{2}},\label{eq:SOS_generator_D1Q1}
\\
O_{\text{SF},rc}&=\frac{1}{\sqrt{2}}\left(w^{(rc)}_B+\frac{i}{2}\sum_{b\in[B]}\sum_{\sigma\in\{0,1\}}w^{(rc)}_b\gamma_{\vec{u}^{(r)}_b\sigma 0}\gamma_{\vec{u}^{(r)}_b\sigma 1}\right),\\
E_\text{SOS}&=-2\sum_{r}w^{(r)}_--\frac{1}{2}\sum_{r\in[R]}\sum_{c\in[C]}\left|W^{(rc)}\right|^2.\label{eq:SOS_generator_SF}
\end{align}

As the operators $\gamma_{\vec{u}^{(j)}_s\sigma x},\gamma_{\vec{u}^{(r)}_b\sigma 0}\gamma_{\vec{u}^{(r)}_b\sigma 1}$ are also unitary, the generators~\cref{eq:SOS_generator_D1Q1,eq:SOS_generator_SF} may be block-encoded using the LCU technique to obtain 
\begin{align}
\textsc{Be}\left[\frac{O_{G^\sigma_1,r}}{\lambda_{G^\sigma_1,r}}\right],\quad
\lambda_{G^\sigma_1,r}&=\sqrt{w_s^{(j)}},\quad \forall (s,G)\in\{(+,\mathrm{D}),(-,\mathrm{Q})\},\quad
\\
\textsc{Be}\left[\frac{O_{\text{SF},rc}}{\lambda_{\text{SF},rc}}\right],\quad\lambda_{\text{SF},rc}&=\frac{1}{\sqrt{2}}\sum_{b\in[B+1]}|w^{(rc)}_b|.
\end{align}
Following~\cref{eq:rectangular_BE_primitives} and~\cref{eq:two_walk_phases}, we obtain the block-encodings
\begin{align}\label{eq:DFTHC_block_encoding}
\textsc{Be}\left[\frac{H_\text{sqrt}^\dagger H_\text{sqrt}}{\Lambda}-1\right],
\quad
\Lambda
=
\frac{1}{2}\left(\sum_{r\sigma}\lambda_{\mathrm{D}^\sigma_1 }^2+\lambda_{\mathrm{Q}^\sigma_1 }^2+\sum_{rc}\lambda_{\text{SF},rc}^2\right)
=
\|h^{(1)\prime}\|_1+\frac{1}{4}\sum_{r\in[R]}\sum_{c\in[C]}\left|\sum_{b\in[B+1]}|w^{(rc)}_b|\right|^2.
\end{align}

At a high-level, the quantum circuit that block-encodes the DFTHC representation~\cref{eq:DFTHC_block_encoding} has some components similar to that of double-factorization, with the main difference arising from block-encoding the one-body terms in an SOS format.
We provide all details of the block-encoding circuit in~\cref{fig:fullcircuit}
 with a step-by-step description of the compilation in Appendix~\ref{sec:Block-encoding-costs}.
From the summary of block-encoding costs in~\cref{tab:femoco54_cost_breakdown}, we see that cost is dominated by the following table-lookup oracles~\cite{Low2024tradingtgatesdirty} and $\textsc{Sel}$: 
\begin{itemize}
    \item $\textsc{Rot}$ controlled by $N+RB$ indices to outputs $(N-1)$ $b_\text{rot}$ rotation angles, where each of the $N-1$ is represented by $b_{\text{rot}}$ bits, that rotate $\gamma_{0\sigma x}$ to $\gamma_{\vec{u}^{r}_b\sigma x}$.
    \item $\textsc{Qroam}$ controlled by $N+R(B+1)C$ indices to output $b_\text{Q}=\lceil\log_2 B\rceil+b_\text{coeff}$ coefficients $w^{(r)}_s,w^{(rc)}_b$.
    \item $\textsc{Sel}$ which takes the table-lookup outputs to block-encode the generators $O_{\mathrm{D}_1^\sigma,r},O_{\mathrm{Q}_1^\sigma,r},O_{\mathrm{SF},rc}$.
\end{itemize}
The Toffoli counts of these components then each scale very differently with $N,R,B,C$.

\begin{figure}
    \centering
    \includegraphics[width=\linewidth]{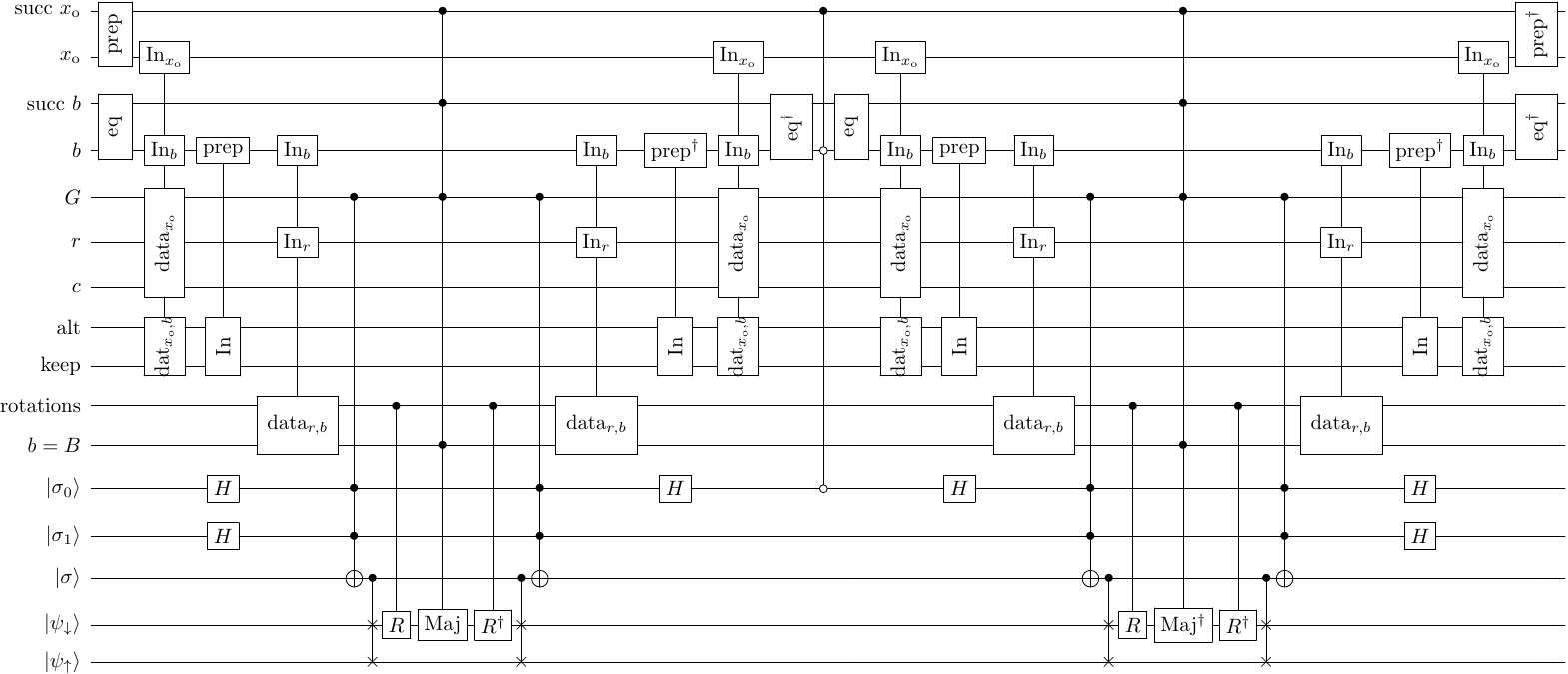}
    \caption{The complete quantum circuit for the block-encoding.
    First, `prep' is the state preparation on $x_{\text{o}}$.
    The operation `eq' creates an equal superposition on the $b$ register with succ $b$ flagging success.
    Then the QROM on $x_{\text{o}}$ and $b$ is used to output alt and keep data for the coherent alias sampling, as well as $G,r,c$ (which is independent of $b$).
    The operation with `In' on alt and keep, with `prep' on $b$ indicates the inequality test and controlled swap used in coherent alias sampling.
    Then $b,r$ are used to output the data for the rotations as well as flipping a flag qubit to indicate if $b=B$.
    There are two spin qubits $\ket{\sigma_0}$ and $\ket{\sigma_1}$, with one used for SF (the inner sum) and the other for $\text{D}_1$ and $\text{Q}_1$ (the outer sum).
    The controlled operation copies it out into $\ket{\sigma}$ depending on $G$, and that is used to control the swap between the spin up and spin down system registers.
    The operation `Maj' indicates the circuit for performing the Majorana operator as shown in Fig.~\ref{fig:Majorana}.
    The operations are all inverted, except for the Hadamard on $\ket{\sigma_1}$ and prep on $x_{\text{o}}$, which are part of the outer sum.
    There is a reflection on the control qubits used in the inner block-encoding, then the entire procedure is performed again.
    Note that we apply the Hermitian conjugate of this operator using Maj$^\dagger$.
}
    \label{fig:fullcircuit}
\end{figure}

Hence, DFTHC is beneficial following the intuition that the expressive power of the factorization can be greatly improved by making the costs of ${\textsc{Rot}}$ and ${\textsc{Qroam}}$ comparable. 
By taking advantage of spacetime trade-offs~\cite{Low2024tradingtgatesdirty} and measurement-based uncomputation~\cite{berry2019qubitization} to reduce the Toffoli gate cost of $\textsc{Qroam}$ using the idle qubits that later occupied by the output of $\textsc{Rot}$, the Toffoli counts of these components are
\begin{itemize}
    \item Computing and uncomputing $\textsc{Rot}$ twice: Each computation costs $\approx N+RB$.
    \item Computing and uncomputing $\textsc{Qroam}$ twice: Each computation costs $\approx \frac{RBC}{\lambda+1}+\lambda b_\text{Q}$ for any choice of $\lambda\le\frac{(N-1)b_\text{Rot}}{b_\text{Q}}$.
    \item Applying $\textsc{Sel}$ twice: Each application costs $\approx 4(N-1)b_\text{Rot}$.
\end{itemize}
For instance, the cost ${\textsc{Rot}}$ strongly dominates in DF, whereas ${\textsc{Qroam}}$ strongly dominates in THC.

DFTHC achieves then chemical accuracy with a most compact representation for molecules such as FeMoco [113e,76o] when $(R,B,C)$ is roughly in-between the equivalent DF and THC parameters as shown in~\cref{tab:compare_rbc_table_1}.
The selected DFTHC values lead to lookup-table sizes that are already substantially smaller than previously reported DF or THC methods, and an exhaustive search for optimal $(R,B,C)$ would lead to even further improvement.
For the typical DFTHC solutions that we found, the three contributions $\textsc{Rot},\textsc{Qroam},\textsc{Sel}$ capture $\sim 90\%$ of the Toffoli cost, with $\textsc{Sel}$ being the dominant cost accounting for $\sim 60\%$.
For example,~\cref{tab:femoco54_cost_breakdown} gives a full cost breakdown.
\begin{table}
    \centering
    \scalebox{0.7}{
    \begin{tabular}{|c|c|c|c|c|c|}
    \hline \hline
    & Subroutine & Cost & Persistent Ancilla & Temporary Ancilla & Calls/Walk \\
    \hline
    \multirow{4}{*}{\rotatebox[origin=c]{90}{Outer PREP}}
    & Uniform State prep  & $4\lceil \log(N+RC)\rceil=36$  & $\lceil \log(N + RC)\rceil + 2 = 11$ &  $\lceil \log(N + RC)\rceil - 2 = 7$ & 1\\
    & AliasSampling & $\left\lceil\frac{N+RC}{2^{k_1}}\right\rceil + 2^{k_1} b_1 = 177$ , $b_1 = 24$, $k_1 = 2$ & $b_{1}(\text{QROM-out}) + b_{k_{1}}(\text{uniform}) = 39$ & $2^{k_{1}}b_{1} + \lceil \log_{2}(\frac{N + RC}{2^{k_{1}}}) \rceil = 103$ & 1 \\
    & AliasSampling$^{\dagger}$ & $b_1 + \left\lceil \frac{N+RC}{2^{k_5}}\right\rceil + 2^{k_5} = 61$ &  & & 1 \\
    & Uniform State prep$^{\dagger}$ & $ 4\lceil \log(N+RC)\rceil = 36$ & & &  1 \\
    & Total & 310 & 50 & 110 &  \\
    \hline
    \multirow{4}{*}{\rotatebox[origin=c]{90}{Inner PREP}}
    & for $x_{o}$ do: QROAM$(b)$  & $RC \left\lceil \frac{B+1}{2^{k_2}}\right\rceil + 2^{k_2} b_2 = 860, b_2 = 20, k_2=4$ & $b_2 + b_{k2} + \lceil \log(R) \rceil + \lceil \log(C) \rceil + 2 = 46$ & $\left\lceil \log \left(RC \left\lceil \frac{B+1}{2^{k_2}}\right\rceil\right)\right\rceil  + (2^{k_2}-1)b_2 = 310 $ & 2 \\
    & ( for $x_{o}$ do: QROAM$(b)$)$^{\dagger}$ & $b_2 + N + \left\lceil \frac {RC}{2^{k_4}}\right\rceil + 2^{k_4}(B+1) = 254, k_4 = 2$ & & & 2 \\
    & uniform($b$) & $4\lceil \log(B + 1)\rceil = 20$ & $\lceil \log (B + 1)\rceil +2 = 7$ & $\lceil \log (B + 1)\rceil - 2 = 3$ & 2 \\
    & uniform($b$)$^{\dagger}$ & $4\lceil \log(B + 1)\rceil = 20$ & & &  2 \\
    & Total & 1154 & 53 & 313$^{*}$ & \\
    \hline 
    \multirow{2}{*}{\rotatebox[origin=c]{90}{RPREP}}
    & QROM($\vec{u}$) & $N+RB = 324$ & $(N-1)b_{\text{rot}} = 795$ &$\max\left( \lceil \log(N+RC) \rceil=9,  \lceil \log(RB) \rceil=9 \right)$& 2 \\
    & QROM($\vec{u}$)$^{\dagger}$ &  $R + B=37$ &  &  & 2 \\
    & Total & 361 & 795 & 9 & \\
    \hline
    \multirow{2}{*}{\rotatebox[origin=c]{90}{SEL}}
    & ROT & $4 b_{\text{rot}} (N - 1) = 3180$ & $b_{\text{rot}} = 15$ & 0 & 2 \\
    & CSWAP & $2N = 108$ & 0 & 0 & 2\\
    & Maj-control & 7 & 0 & 0 & 2 \\
    & Total & 3295 & 15 & 0 & \\ 
    \hline
     \multirow{2}{*}{\rotatebox[origin=c]{90}{REF}}
    & $T_{2}$ & $\lceil \log(B+1)\rceil + b_{k2} + 1 = 21$ & $0 $ & 0 & 1 \\
    & Walk & $\lceil \log(N+RC)\rceil + \lceil \log(B+1)\rceil + b_{k1}+b_{k2} + 2 = 46$ & $0 $ & 0 & 1 \\
        \hline 
    & Total & 9997 & 913 + 108 (system) = 1021 & max(110, 9) & \\
    \hline \hline
    \end{tabular}
    }
    \caption{FeMoco54 cost breakdown. $N = 54$, $(R, B, C) = (10, 27, 27)$, $b_{\text{rot}} = 15$,  $b_{\text{coeff}} = b_{k1} = b_{k2} = 15$. Total qubit costs is $1131$ because the 313 temporary qubits for inner state prep QROM can use the 795 persistent qubits for RPREP. We denote this reuse with an asterisks on the temporary ancilla for the inner PREP subroutine. }
    \label{tab:femoco54_cost_breakdown}
\end{table}

We use the following heuristic for choosing $(R,B,C)$.
\begin{itemize}
    \item Amdahl's law: $C=\Theta(N)$ so that the Toffoli counts of $\textsc{Rot}$ and $\textsc{Qroam}$ are roughly similarly with $\Theta(RB)$ scaling.
    \item Linear cost scaling: THC is notable for exhibiting an almost linear $\tilde{\Theta}(N)$ block-encoding Toffoli with $R_\text{THC}=\tilde{\Theta}(N)$.
    Hence we choose $B=\Theta(N)$ and $R=\tilde{\Theta}(1)$.
    \item Extensivity: Suppose a given $N_0$ orbital system is well-approximated by some $(R,B,C)=(R_0,\Theta(N_0),\Theta(N_0))$. 
    Then given $K$ non-interacting copies of the same system, one expects a good approximation with $(R,B,C)=(R_0K,\Theta(N_0),\Theta(N_0))$.
    In other words, $B,C\lesssim N$ suffices.
\end{itemize}
We also carefully study the bits-of-precision needed in~\cref{sec:bits_of_precision_truncation}. 
Roughly speaking, chemical precision is attained with $b_\text{Rot}\in[11,18]$, $b_\text{coeff}\in[6,14]$, depending on the system. 
This is consistent with previous work~\cite{PRXQuantum.2.030305,low2021halvingcostquantummultiplexed} and all resource estimates use bit counts reported in Appendix~\ref{sec:bits_of_precision_truncation}.

\subsection{Numerical optimization}\label{sec:DFTHC_optimization}
DFTHC allows us to simultaneously find electronic structure SOS with good gaps, block-encoding costs, and normalization factors, without directly solving the SDP in Eq.~\eqref{eq:sos_sdp_program}.
We find good DFTHC approximations $h^{(2)\prime}$ of the exact two-body integrals $h^{(2)}$ using a cost function that minimizes the Frobenius error $\epsilon_{\text{fro}} =\|h^{(2)}-h^{(2)\prime}\|_\text{fro}$ together with regularization on $\Lambda$ and $E_\text{gap}$.

It is only necessary to approximate the two-body terms in DFTHC as we have freedom to choose the one-body terms.
By expanding~\cref{eq:DFTHC_representation} and then comparing to the original exact Hamiltonian~\cref{eq:spinfree_Hamiltonian},
\begin{align}
H_\text{DFTHC}&=\sum_{pq}\left(h^{(1)\prime}_{pq}+\sum_{rc}W^{(rc)}\sum_{b\in[B]}w^{(rc)}_{b}u^{(r)}_{b,p}u^{(r)}_{b,q}\right)E_{pq}+\frac{1}{2}\sum_{rc}\left(\sum_{b\in[B]}w^{(rc)}_{b}\sum_{pq}u^{(r)}_{b,p}u^{(r)}_{b,q}E_{pq}\right)^2,
\\
H&=H_\text{DFTHC}+\frac{1}{2}\sum_{pqrs}(h^{(2)}_{pqrs}-h^{(2)\prime}_{pqrs})E_{pq}E_{rs},\label{eq:hamiltonian_error}
\end{align}
we see for any computed $w,u$ tensors, the choice $h^{(1)\prime}_{pq}=h^{(1)}_{pq}-\sum_{rc}W^{(rc)}\sum_{b\in[B]}w^{(rc)}_{b}u^{(r)}_{b,p}u^{(r)}_{b,q}$ leads to a one-body contribution equal to the exact Hamiltonian, with some error in the two-body DFTHC tensor $h^{(2)\prime}$.

We reduce the contribution of error terms by using the identity $W^{(rc)}=w^{(rc)}_B+\sum_{b\in[B]}w^{(rc)}_b$ to rewrite
\begin{align}
h^{(1)\prime}_{pq}&=h^{(1)}_{pq}+\sum_{r}h^{(2)\prime}_{pqrr}-\sum_{rc}w^{(rc)}_B\sum_{b\in[B]}w^{(rc)}_{b}u^{(r)}_{b,p}u^{(r)}_{b,q},\label{eq:h1_corrected}
\\
2\sum_{j}w^{(j)}_-
&=\|h^{(1)\prime}\|_1-\mathrm{Tr}[h^{(1)\prime}]
=\|h^{(1)\prime}\|_1-\mathrm{Tr}[h^{(1)}]-\sum_{pq}h^{(2)\prime}_{ppqq}+\sum_{rc}w^{(rc)}_B\sum_{b\in[B]}w^{(rc)}_b,
\\
\frac{1}{2}\sum_{rc}\left|W^{(rc)}\right|^2
&=\frac{1}{2}\sum_{pq}h^{(2)\prime}_{ppqq}+\frac{1}{2}\sum_{rc}|w^{(rc)}_{B}|^2-\sum_{rc}w^{(rc)}_{B}\sum_{b\in[B]}w^{(rc)}_b,
\end{align}
where $\|\cdot\|_1$ is the Schatten one-norm. 
Hence,
\begin{align}
E_\text{gap}=E_\text{gs}+\|h^{(1)\prime}\|_1-\mathrm{Tr}[h^{(1)}]-\frac{1}{2}\sum_{pq}h^{(2)\prime}_{ppqq}
+\frac{1}{2}\sum_{rc}|w^{(rc)}_{B}|^2.\label{eq:Egap_corrected}
\end{align}
Then using prior knowledge of the limit where DFTHC is exact, we make the replacements $\sum_{r}h^{(2)\prime}_{pqrr}\leftarrow\sum_{r}h^{(2)}_{pqrr}$.
A similar replacement was made previous work~\cite{berry2019qubitization} when going from the normal-ordered representation to~\cref{eq:spinfree_Hamiltonian} or to the Majorana representation~\cite{PhysRevResearch.3.033055}.

We expand the space of good solutions using symmetry shifting (BLISS)~\cite{patel2024guaranteed,loaiza2023block}, which is based on the observation that the spectrum of the symmetry sector with fixed particle number $\eta$ is invariant under any operator that multiplies the annihilator $(\hat{N}-\eta)$.
In this work, we consider first and second order annihilators constructed from particle number conservation, which correspond to a continuous set of spectrally equivalent Hamiltonian where original electronic structure tensor $(h^{(2)},h^{(1)}, h^{(0)})$ may be replaced by any $(h^{(2)}_\text{S},h^{(1)}_\text{S},h^{(0)}_\text{S})$ in
\begin{align}
\mathcal{H}_\eta=\left\{\left(h^{(2)}+\frac{h^{(\text{sym})}\otimes I+I\otimes h^{(\text{sym})}}{2},h^{(1)}-\frac{\eta}{2}h^{(\text{sym})}+\beta_1I,h^{(0)}-\frac{\eta}{2}\beta_1\right)\bigg|\forall \beta_1\in\mathbb{R},\;h^{(\text{sym})}=h^{(\text{sym})\dagger}\in\mathbb{R}^{N\times N}\right\}.
\end{align}

We optimize the DFTHC ansatz for good block-encoding parameters by defining the non-linear program
\begin{align}\label{eq:DFTHC_program}
\argmin_{\substack{u\in\mathbb{R}^{R\times B\times N},\\w\in\mathbb{R}^{R\times (B+1) \times C},\\(h^{(2)}_\text{S},h^{(1)}_\text{S},h^{(0)}_\text{S})\in\mathcal{H}_\eta}}
\left[\frac{\|h^{(2)}_\text{S}-h^{(2)'}\|_{\text{fro}}}{\epsilon_{\text{reg}}}+\frac{\Lambda}{\lambda_{\text{reg}}}
+\mathrm{Relu}\left(\frac{E_{\text{gap}}-E_\text{reg}}{E_\text{reg}}\right)\right],
\quad
\mathrm{Relu}(x)\coloneqq \begin{cases}
x,&x\ge0,\\
0,&x<0.
\end{cases}
\end{align}
where $\cdot_\text{reg}$ are regularization hyperparameters.
Equation \eqref{eq:DFTHC_program} minimizes the error between the DFTHC approximation $h^{(2)'}$ which is a function of $u$ and $w$, and any exact two-body tensor $h^{(2)}_\text{S}$ in the $\eta$-particle symmetry sector.
At the same time, it places some priority on minimizing the block-encoding normalization factor $\Lambda$, which is implicitly a function of $u,w,h^{(2)}_\text{S},h^{(1)}_\text{S}$ as seen in~\cref{eq:h1_corrected}.
This program also attempts to get solutions close to the optimal gap through $E_\text{gap}$, which implicitly depends on all parameters.
Note the $\mathrm{Relu}$ function, which avoids optimizing for unrealistic gaps less than the optimal.
Whereas prior approach to solving similar nonlinear linear programs for DF or THC required chemically motivated initial conditions $(u_0,w_0)$ for $(u,w)$ from which gradient descent begins, we find that it suffices to make the much simpler choice of initial parameter values $\mathcal{P}=(u_0,w_0,\beta_{1,0},h^{(\text{sym)}}_0)$, where $u^{(r)}_b$ is a random unit vector, $w_0, h^{(\text{sym)}}_0\sim\mathcal{N}(0.01)$ and  $\beta_1\sim\mathcal{N}(0.05)$ to be element-wise sampled from a distribution where $\mathcal{N}(\sigma)$ is the normal distribution with mean zero and standard deviation $\sigma$.

\begin{table}
\centering
    \begin{tabularx}{\textwidth}
    {|Y|Y|c|c|c|c|c|c|c|c|c|c|c|c|c|}
    \hline\hline
    Molecule & $(N,R,B,C)$ & $\lambda_\text{reg}$ [mHa] & $\epsilon_\text{reg}$ [mHa] & $E_\text{reg}$ [Ha] & Time per step [ms] & Steps to solution\\
    \hline
      Fe$_2$S$_2$ [30e, 20o] & $(20,14,15,5)$ & 20 & 10 & $[1,4]$& 0.8 & $10^{[5,6]}$\\
      Fe$_4$S$_4$ [54e, 36o]& $(36,9,18,18)$ & 20 & 10 & $1.87\times[1,2]$ & 1.2 & $10^{[5,6]}$\\
      FeMoco [54e, 54o]& $(54,10,27,27)$ & 50 & 10 & $3.46\times[1,1.6]$ & 1.3 & $10^{[5,6]}$\\  
      FeMoco [113e, 76o]& $(76,15,57,19)$ & 100 & 10 & $[4,8]$ & 4.4 & $10^{[5,6]}$\\
      CO$_2$[XVIII] [64e, 56o]& $(56,5,28,28)$ & 50 & 10 & $[1,4]$ & 1.8 & $10^{[5,6]}$\\
      CO$_2$[XVIII] [100e, 100o]& $(100,9,75,25)$ & 50 & 10 & $[2,8]$ & 8.2 & $10^{[5,6]}$\\
    CO$_2$[XVIII] [150e, 150o]& $(150,13,112,37)$ & 200 & 10 & $[4,16]$ & 49.0  & $10^{[5,6]}$\\
      CPD1-P450X [63e, 58o]& $(58,9,29,14)$ & 50 & 10 & $4.70\times[1,1.5]$ & 1.3& $10^{[5,6]}$\\
       \hline \hline
\end{tabularx}
\caption{\label{tab:implementation_details}Runtime for evaluating a single gradient descent $\textsc{Adam}$ step of~\cref{eq:DFTHC_program} for typical examples in a $\textsc{Jax}$ implementation running on a single $\textsc{Nvidia L4}$ GPU, and typical hyperparameters. In all cases, the learning rate is geometrically decreasing from $10^{-1}$ to a final value that is typically $10^{-4,-5,-6}$.}
\end{table}

We approximately solve~\cref{eq:DFTHC_program} by first-order gradient descent using the $\textsc{Adam}$ optimizer implemented in the $\textsc{Optax}$ and $\textsc{Jax}$ libraries for the $\textsc{Python}$ programming language.
The most expensive operation in each step is the tensor contraction in~\cref{eq:DFTHC_h2_tensor}, which costs $\mathcal{O}(RCN^4)$ multiplications.
A major benefit is this approach is auto-differentiation, which allows for easily computing gradients even through complicated functions, such as the Schatten norm in~\cref{eq:Egap_corrected}.
Our approach is also GPU-accelerated, which allows us to collect robust statistics and optimize over a large number of $(R,B,C)$ parameters for systems up to $N=150$ orbitals, as shown in~\cref{tab:implementation_details} together with typical hyperparameters.

\subsection{Selection of good solutions}\label{sec:selection_good}
Although finding the optimal solution to~\cref{eq:DFTHC_program} could take exponential time in $N,R,B,C$, we emphasize that it suffices to find approximate solutions more rapidly, so long as they are chemically accurate.
Chemical accuracy for each DFTHC solution is validated by calculating the difference in CCSD(T) correlation energy $\epsilon_{\text{corr}}$ compared to the original representation $H$, as commonly recommended~\cite{berry2019qubitization,PRXQuantum.2.030305}, as a proxy to the difference in exact correlation energy $\epsilon^{*}_\text{corr}$. 
Prior work~\cite{Rocca2024SymmetryCompressedDF,PRXQuantum.4.040303} computed good DF or THC solutions by scanning over ranks, searching for good hyperparameters, and finally picking the lowest quantum cost solution with $|\epsilon_\text{corr}|$ below some threshold $\epsilon_\text{th}$.
In other words, using $\cdot^\prime$ to refer to variables computed from a DFTHC solution,
\begin{align}
E_\text{gs}=E_\text{HF}+E_\text{corr},
\quad
E_\text{gs}^\prime=E_\text{HF}^\prime+E_\text{corr}^\prime,
\quad
\epsilon_{\text{corr}}=E_\text{corr}-E_\text{corr}^\prime,
\end{align}
where $E_{\text{HF}}$ is the Hartree-Fock energy, which can be computed in $\mathcal{O}(N^3)$ time.
Hence, using a quantum computer to estimate $\hat{E}_\text{gs}^\prime$ with standard deviation $\sigma_\text{PEA}$ from the PEA, we may obtain the true ground-state energy $\hat{E}_\text{gs}=\hat{E}_\text{gs}^\prime+(E_\text{HF}-E_\text{HF}^\prime)+ \epsilon_\text{corr}^*$ with standard deviation
\begin{align}\label{eq:linear_error}
\sigma_\text{E}\le\sigma_\text{PEA}+ |\epsilon_\text{corr}^*|
\approx
\sigma_\text{PEA}+ |\epsilon_\text{corr}|
\le \sigma_\text{PEA}+\epsilon_\text{th}.
\end{align}

\begin{figure}
    \centering
    \includegraphics[]{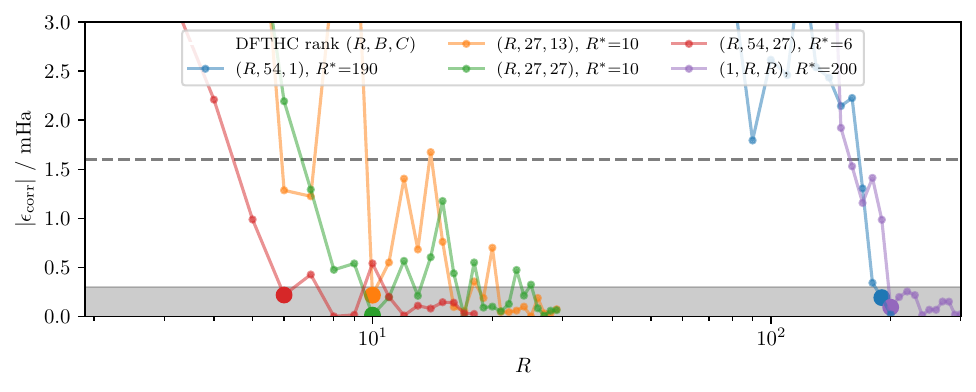}
    \vspace{-.5cm}
    \caption{\label{fig:FeMoCo_DFTHC_best_solution_scan} FeMoCo-54 CCSD(T) correlation energy error vs.\ rank of DFTHC truncation (small dots). The dashed line at $1.6$mHa is the chemical accuracy threshold and the shaded region is the threshold $0.3$mHa used to select the rank $R^*$ (large dots) of good solutions.}
\end{figure}

We use the same approach to generate candidate DFTHC solutions below the threshold $\epsilon_\text{th}\le 0.3$mHa as illustrated for FeMoCo-54 in~\cref{fig:FeMoCo_DFTHC_best_solution_scan}.
Following the heuristics in~\cref{sec:DFTHC}, we choose $(R,B,C)$ values with $B\in\{N,\lfloor 3N/4\rfloor,\lfloor N/4\rfloor\}$ and $C\in\{\lfloor N/2\rfloor,\lfloor N/4\rfloor\}$, and scan over $R$.
To make a comparison with prior art, we also study DF-like $(R,N,1)$ and THC-like $(1,R_\text{THC},R_\text{THC})$ factorizations.
The best Toffoli gate costs for points below the $\epsilon_\text{th}$ threshold is reported in~\cref{tab:dfthcblisssossa_costs} together with relevant parameter like $E_\text{gap}$ -- see~\cref{tab:compare_rbc_table_1} and~\cref{tab:compare_rbc_table_2} for best solutions for each choice of $(B,C)$.
As the absolute energy of each DFTHC factorization may shift by a large amount $E_\text{HF}^\prime-E_\text{HF}$, we also recompute $E_\text{gap}$ using the CCSD(T) energy to approximate the ground-state energy. 
CCSD(T) is not a upper bound on $E_\text{gs}^{\prime}$, but we expect the correction to be small compared $E_\text{gap}$, which is on the order of Hartrees.
CCSD(T) in the spin-$0$ sector is run on the systems FeMoCo-54 and the XVIII complexes. For other molecules, convergence requires targeting the high-spin sector: as described in prior art~\cite{PRXQuantum.2.030305,berry2024rapid}, we choose $n_\uparrow-n_\downarrow$ for Fe$_2$S$_2$, Fe$_4$S$_4$, FeMoCo-76, and CPD1-P450X to be $8$, $16$, $35$, and $5$ respectively.

Drawing general conclusions about the best hyperparameters is not straightforward, as $\epsilon_\text{corr}$ fluctuates significantly in~\cref{fig:FeMoCo_DFTHC_best_solution_scan} and does not improve monotonically with rank.
Similar to DF and THC, these fluctuations arise as the DFTHC program of~\cref{eq:DFTHC_program} is nonlinear and the final solution is sensitive to initial conditions.
This is easily resolved by collecting statistics on $\epsilon_\text{corr}$, by restarting the DFTHC program for multiple different initial conditions $\mathcal{P}$, in a randomized fashion as described in~\cref{sec:DFTHC_optimization}.
This allows us to make statistically robust conclusions in~\cref{sec:robust_selection} about the scaling of various parameters of interest, such as the standard deviation  $\sigma_\text{corr}$, $\Lambda$, and $E_\text{gap}$ with DFTHC rank $R$, and enables the use of a different bound on error contributions like
\begin{align}\label{eq:squared_error_main_text}
\sigma_E\le\sqrt{\sigma_\text{PEA}^2+\sigma_\text{corr}^2}+ |\bar{\epsilon}_\text{corr}| \le \epsilon_\text{chem}.
\end{align}
This bound also tolerates larger errors, say $\sigma_\text{PEA}=1.4$mHa and $\sigma_\text{corr}=0.7$mHa.
In~\cref{tab:dfthcblisssossa_robust_costs} of Appendix~\ref{sec:robust_selection}, we present an alternate set of resource estimates based on~\cref{eq:squared_error_main_text}, as they represent average-case solutions rather than best-case solutions and are of independent interest. 
\section{Resource estimates}\label{sec:results}
In~\cref{tab:dfthcblisssossa_costs}, we report the quantum resource estimates of the DFTHC+BLISS+SA protocol for estimating the ground-state energy, and compare these with prior art using the double-factorization, tensor hypercontraction, and BLISS where applicable.
We consider a variety of chemical systems: Iron-Sulfur complexes (Fe$_{2}$S$_{2}$~\cite{li2017spin}, Fe$_{4}$S$_{4}$~\cite{li2017spin}, 54 orbital FeMoco active space~\cite{Reiher2017Elucidating}, 76 orbital FeMoco active space~\cite{li2019electronic}), a representation of the cytochrome P450 enzyme active site (CPD1)~\cite{goings2022reliably}, and the CO$_{2}$-fixation Ruthenium catalyst~\cite{PhysRevResearch.3.033055}.  Beyond being chemically distinct and a robust test of the method, the varying size and degree of correlation in each of the systems allows us to extrapolate scalings in~\cref{tab:dfthcblisssossa_costs}, based on the eight system considered spanning an order of magnitude difference in number of orbitals and electrons.

The DFTHC resources were computed by scanning over $(R,B,C)$ values to find lowest cost solutions balancing gap, rotation, and coefficient oracle costs along with the difference in correlation energy with respect to the untruncated integrals $\epsilon_{\text{corr}}$. We report the LCU rescaling value for the SOS Hamiltonian $\Lambda$, differences between DFTHC two-electron integral tensors, the effective DFTHC+SA LCU rescaling value $\lambda_{\text{eff}}$, block-encoding space and time resources, and improvement factors over prior work.  The improvement factors are computed as the ratio of the products of block-encoding and $\lambda$. 
The bits for LCU coefficients and rotations $b_\text{coeff}$ and $b_\text{coeff}$ respectively were estimated according to~\cref{sec:bits_of_precision_truncation},
where we design a randomized rounding scheme that converts bit truncation errors to a variable $\epsilon_\text{trunc}$ contribution to $\epsilon_\text{corr}$.
We choose the maximum number of bits that satisfies the error bound
\begin{align}
\sigma_E&\le\sqrt{\sigma_\text{PEA}^2+\sigma_\text{trunc}^2}+ |\epsilon_\text{corr}| + |\bar{\epsilon}_\text{trunc}|\le \epsilon_\text{chem}.
\end{align}
We obtain a $\Lambda$ that is smaller than those computed by prior techniques, which combined along with a small $E_{\text{gap}}$, indicates that the DFTHC factorization of spin-free algebra is sufficient for finding spectrum amplifiable representations where speedups in phase estimation are possible.

Our results are at least a factor of 3 times cheaper than the best costs reported in the literature for this set of systems. For the FeMoco-76 orbital system from Li \textit{et al.}~\cite{li2019electronic} we find a factor of 4 improvement over recent work integrating tensor hypercontraction and BLISS~\cite{caesura2025faster} resulting in a cost of 9.99$\times 10^{8}$ Toffoli gates for ground-state energy estimation. For smaller Iron-Sulfur systems (Fe$_{2}$S$_{2}$, Fe$_{4}$S$_{4}$, FeMoco54) and the CO$_{2}$ systems where symmetry shifting is used to a lesser extent in prior art, DFTHC+BLISS+SA constitutes a substantial improvement by a factor of 10 to 195. The SA effective LCU normalization, $\lambda_{\text{eff}}$ determined by DFTHC are within a factor of 2 of the lower bounds to the optimal values provided in Table~\ref{tab:var_energies}. This does not indicate a limit of the SA technique -- only that for a given SOS algebra we have found a tight gap and a highly compressed form of the the two-electron integral tensor.

Utilizing the same quantum error correction model as Ref.~\cite{PRXQuantum.2.030305} we can estimate the physical resources required for ground-state energy estimation using DFTHC+BLISS+SA. Using four CCZ factories, a physical error rate of 0.001, a surface code cycle time of 1 microsecond and reaction time of 10 microseconds we estimate that the FeMoco76 example will require 4.5 million physical qubits running for 8.6 hours on a set of $d=27$ surface code patches. FeMoco54 is estimated to require 3.2 million physical qubits running for 2.7 hours on a set of $d=27$ surface code patches. For reference, prior THC based runtime estimates from Reference~\cite{PRXQuantum.2.030305}, not considering the additional space-time tradeoffs made in that work, result in a FeMoco54 runtime of 2 days and FeMoco76 runtime of 12 days~\cite{PRXQuantum.2.030305}.

\begin{table}
    \centering 
    \begin{tabularx}{\textwidth}{|Y|c|c|c|c|c|c|c|c|c|}
    \hline \hline
     &         \multirow{2}{*}{Molecule}                     &  Fe$_2$S$_2$-20             & Fe$_4$S$_4$-36                &  \multicolumn{2}{c|}{FeMoCo}                       &  CPD1-P450X  & \multicolumn{3}{c|}{CO$_2$ [XVIII]}     \\
     \cline{5-6}\cline{8-10}
&&[30e, 20o]&[54e, 36o]&[54e, 54o]&[113e, 76o]&[63e, 58o]& [64e, 56o]& [100e, 100o]& [150e, 150o]
     \\
     \hline \hline           
     \multirow{3}{*}{\rotatebox[origin=c]{0}{DF}}&$\lambda_{\text{DF}}$         &  41.13 ~\cite{berry2024rapid}&  154.83~\cite{berry2024rapid} &  78.00~\cite{Rocca2024SymmetryCompressedDF} & 584.54~\cite{berry2024rapid}     & 111.3~\cite{Rocca2024SymmetryCompressedDF} & 293.50~\cite{PhysRevResearch.3.033055} & 782.1~\cite{PhysRevResearch.3.033055} & 1923.8~\cite{PhysRevResearch.3.033055} \\
     &$C_{\text{B.E.}}$  &  7177                        & 16505                         &  19589                                      & 40442                            & 21736                                      & 54227                                  & 97679 & 182005 \\
     & Qubits                       & 989                          & 3115                          & 3722                                        & 6404                             & 2596                                       & 3700                                   & 18017 & 34218 \\
     \hline
     \multirow{3}{*}{\rotatebox[origin=c]{0}{THC}}&$\lambda_{\text{THC}}$        & 63.81~\cite{berry2024rapid}  & 168.34 ~\cite{berry2024rapid} & 306.30~\cite{PRXQuantum.2.030305}           & 198.90~\cite{caesura2025faster}  & 130.9~\cite{caesura2025faster} &         --                               &-- & --\\ 
     &$C_{\text{B.E.}}$ & 4626                         & 8573                          & 11016                                       & 13763                            & 8334                           &                      --                  & --&-- \\
     & Qubits                       & 579                          & 1081                          & 2142                                        & 1512                             & 1357                           &                     --                   &-- &-- \\
     \hline
     \multirow{11}{*}{\rotatebox[origin=c]{90}{DFTHC+BLISS+SA}}&
     $(R,B,C)$                   &$(14, 15, 5)$        &$(9, 18, 18)$        &$(10, 27, 27)$       &$(15, 57, 19)$       & $(9, 29, 14)$       &$(5, 28, 28)$        & $(8, 75, 25)$             & $(9,112,37)$ \\
     &$\epsilon_\text{corr}$/mHa & 0.0131              &0.0041               &0.1254               &0.1362               & 0.2766              &0.2697               & 0.042                     & 0.062 \\
     &$\epsilon_{\text{fro}}$/Ha&0.0221               &0.1669               &0.2383               &0.0825               &0.3083               &0.4182               & 0.2187                     & 0.3510 \\
     &$\Lambda$/Ha               &17.5299              &49.8149              &58.3440              &179.7296             &97.4395              &55.4773              & 155.5                     & 336.1 \\
     &$E_\text{gap}$/Ha          &1.2381               &2.3070               &4.0535               &5.3820               &5.6837               &2.6918               & 4.565                     & 6.454 \\
     &$\lambda_{\text{eff}}$/Ha  &6.4690               &14.9842              &21.3674              &43.6538              &32.7923              &17.0712              & 37.68                     & 65.87 \\
     &$C_{\text{B.E}}$          & 3906                & 7322                & 10169               & 14563               & 9535                & 7651                & 17975                     & 27237 \\ 
     & Qubits                    & 466                 & 873                 &  1137               & 1459                & 1150                & 924                 & 1960                      & 2870  \\
     & Total Toffoli             & 3.97$\times 10^{7}$ & 1.72$\times 10^{8}$ & 3.41$\times 10^{8}$ & 9.99$\times 10^{8}$ & 4.91$\times 10^{8}$ & 2.05$\times 10^{8}$ & 1.06$\times 10^{9}$       & 2.81$\times 10^{9}$ \\
     & Improvement (DF)   & 11.7 & 23.3 & 7.0 & 37.1 & 7.7 & 121.9  & 112.7 & 195.2 \\
     & Improvement (THC)  & 11.7 & 13.2 & 15.5 & 4.3 & 3.5 &    --    &   --    &  --\\    
     \hline \hline
    \end{tabularx}
    \scalebox{.8}{\includegraphics[]{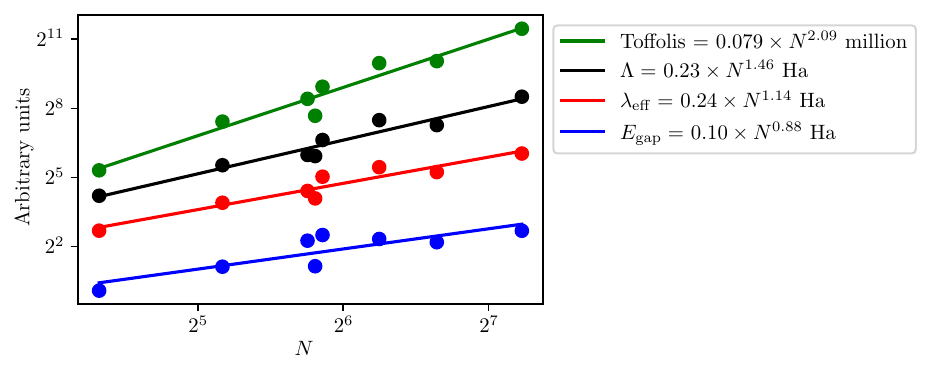}}
    \vspace{-0.5cm}
    \caption{
    (top) Improvements in simulating electronic structure to chemical accuracy with DFTHC and spectral amplification compared to best prior double-factorization methods and tensor hypercontraction methods. Improvement factors are computed as the ratio of the product the LCU 1-norm $\lambda_\cdot$ and the block-encoding cost $C_\text{B.E.}$.  The source for each cost is provided as a citation reference in the $\lambda$-row. The number of calls to the block-encoding is computed as $\lceil \frac{\pi \lambda}{2\sigma_\text{PEA}}\rceil $~\cite{BabbushPRX18}.  In line with previous work we take $\sigma_\text{PEA} = 1.0\text{mHa}$.
    (bottom) Scaling of parameters Toffoli cost (green), block-encoding normalization $\Lambda$, spectrum amplified normalization $\lambda_\text{eff}$, and SOS gap $E_\text{gap}$
    }
    \label{tab:dfthcblisssossa_costs}
\end{table}

The DFTHC protocol requires selecting a large enough $(R, B, C)$ such that a faithful representation of the original Hamiltonian is preserved while optimizing for the desired phase estimation cost. Prior works have used the difference in correlation energy, $\epsilon_{\text{corr}}$, with respect to untruncated representation of the two-electron integrals in a high-spin electronic configuration. We follow this trend and report factorized solutions with $\epsilon_{\text{corr}}$ of $\leq 0.3$ mHa in a similar fashion to Ref.~\cite{caesura2025faster}.  Recovering correlations from a quantum computer is a viable way to compute high accuracy energy differences and thus correlation energy difference is a reasonable target when using an approximate wavefunction with a faithful representation of the correlation energy and multireference nature of the target eigenstate. High-spin electronic configuration are selected such that CCSD(T) converges without issue. In this regime the wavefunction is primarily single reference and thus correlation energy based metrics can differ substantially from low-spin, potentially, multireference targets for phase estimation. In Table~\ref{tab:ccsd_t_dmrg_correlation_energies} we demonstrate a substantial difference in correlation energy differences computed from high-spin CCSD(T) energies and low-spin DMRG wavefunctions.  This suggests that higher truncation thresholds may be needed for high accuracy electronic structure calculations on quantum computers. We raise this question for further research.

\begin{table}[htb]
    \centering
    \begin{tabular}{|c|c|c|c|c|D{.}{.}{3.6}|c|D{.}{.}{2.6}|D{.}{.}{3.6}|D{.}{.}{3.6}|}
    \hline \hline
& \multirow{2}{*}{$(R,B,C)$} &\multicolumn{3}{c|}{In units of [Ha]} &\multicolumn{5}{c|}{In units of [mHa]}
\\
\cline{3-10}
&& $E_{\text{scf}}$ & $E_{\text{CCSD(T)}}(\text{corr})$ & $E_{\text{DMRG}}(\text{corr})$ & \multicolumn{1}{c|}{$\epsilon_{\text{scf}}$} & $\epsilon_{\text{CCSD(T)}}(\text{corr})$ & \multicolumn{1}{c|}{$\epsilon_{\text{DMRG}}(\text{corr})$} & \multicolumn{1}{c|}{$\epsilon_{\text{CCSD(T)}}$} & \multicolumn{1}{c|}{$\epsilon_{\text{DMRG}}$}
\\
\hline \hline
\multirow{4}{*}{\rotatebox{90}{Fe$_{2}$S$_{2}$}} 
 & (12,20,10) & -116.381358 & -0.159411 & -0.217612 &   6.516516 & 0.077150 & 0.028436 &   6.593666 &  6.544952 \\
 & (14,10,10) & -116.387588 & -0.159194 & -0.219803 &   0.285888 & 0.294471 & 2.162635 &   0.580359 &  1.876747 \\
 & (14,15,5)  & -116.344198 & -0.159501 & -0.218300 &  43.675643 & 0.013070 & 0.659010 &  43.662573 & 43.016632 \\
 & (11,15,10) & -116.372589 & -0.159387 & -0.215839 &  15.284690 & 0.101229 & 1.801982 &  15.385919 & 17.086672 \\
 \hline
\multirow{4}{*}{\rotatebox{90}{Fe$_{4}$S$_{4}$}} 
 & (16,27,9)  & -326.903132 & -0.178192 & -0.219675 & 101.275269 & 0.405329 &  2.645671 & 101.680599 & 103.920941 \\
 & (9,18,18)  & -326.772089 & -0.178601 & -0.213938 & 232.317519 & 0.004119 &  8.383212 & 232.313399 & 240.700731 \\
 & (12,18,18) & -326.865711 & -0.173126 & -0.210854 & 138.695868 & 0.242423 & 11.467001 & 138.938291 & 150.162870 \\
 & (12,27,18) & -326.944912 & -0.178400 & -0.218820 &  59.494843 & 0.197676 &  3.500753 &  59.692519 &  62.995596 \\
      \hline \hline
    \end{tabular}
    \caption{Various energies and energy difference between a DFTHC representation of the two-electron integrals and the true two-electron integrals. All simulations used the exact one-electron integrals. The DMRG calculations for Fe$_{2}$S$_{2}$ used a bond dimension of $M=2000$ while Fe$_{4}$S$_{4}$ used $M=1000$.  The key difference in correlation energies is shown in columns labeled $\epsilon_{\text{CCSD(T)}}(\text{corr})$ and  $\epsilon_{\text{DMRG}}(\text{corr})$. The variance and orders-of-magnitude difference suggest high spin CCSD(T) correlation energy may not be an optimal metric when selecting tensor factorization thresholds even with a randomization protocol proposed in Section~\ref{sec:selection_good}. The last two columns show absolute energy differences of substantial magnitude suggesting that low threshold truncations may be representing a Hamiltonian with fundamentally different physics.    }
    \label{tab:ccsd_t_dmrg_correlation_energies}
\end{table}

\section{Discussion}
In this work we have reduced the cost of ground-state energy estimation of second quantized electronic structure Hamiltonians by utilizing ideas from SA. For SA to provide a net gain, an SOS Hamiltonian with a sufficiently small gap and a sufficiently simple $H_{\text{sqrt}}$ is required. Our work demonstrates that an SOS algebra that is similar to the algebra needed to express the spin-free chemistry Hamiltonian is sufficient for tight energy gaps and low block-encoding costs yielding reduced quantum resources. To ensure the block-encoding of $H_{\text{sqrt}}$ is low-cost we developed a new tensor factorization that provides an optimal trade-off between the best prior art (DF and THC factorizations). The new factorization efficiently compresses the two-electron integral tensor, results in a tight SOS gap $\Delta_{\rm gap}\ll 1$, and balances the trade-off between the two dominant table lookup oracles. The efficiency of this new integral compression scheme is demonstrated by the fact that the effective LCU normalization is within a factor of two of the optimal value for all systems studied. These results establish that the major obstacles of combining SA and ground-state energy estimation can be surmounted. 

Application of DFTHC and SA to FeMoco continues the trend of cost reductions. In this case, a factor of 4 over prior art employing THC combined with block-invariant symmetry shifting~\cite{caesura2025faster}. With this reduction in costs the phase estimation component ($9.99 \times 10^{8}$ Toffoli gates) of ground-state energy estimation is now similar cost as state preparation (between 7.6$\times 10^{8}$ and 1.3$\times 10^{9}$) for MPS with estimated ground-state overlaps between $0.95$ and $0.99$~\cite{berry2024rapid}. 

Even more refinements to DFTHC and SA are possible. For DFTHC, improved optimal parameter search that minimizes space-time instead of simply the Toffoli complexity can lead to improved circuits for execution on real systems. Furthermore, it may be possible to convert the large non-linear optimization to a low-rank semidefinite program which also includes various hyperparameter optimizations~\cite{burer2003nonlinear}. Such a conversion would provide a provable polynomial-time pre-processing that optimizes quantum algorithm parameters and not just $E_{\text{gap}}$. SA can be further optimized by exploring minimal expansions to the SOS algebra to decrease the ground-state energy of $H_{\text{SA}}$. In terms of physical resources, the cost models of Ref.~\cite{PRXQuantum.2.030305} are now out of date with the invention of magic state cultivation in Ref.~\cite{gidney2024magic}. Considering the small space-time volume of magic state cultivation, it is possible that the predicted 8 hours of runtime is a substantial overestimate. At the logical level, with the reduction in space-time costs for T-gates a re-analysis of space-time trade-offs at the QROM level, which will then completely dominate the spacetime volume of $\textsc{Sel}$, will be required.

Applying SA beyond second quantized chemistry will require detailed examination of the trade-off between the growing SOS complexity needed to express a Hamiltonian close to a frustration free form and the commensurate increased block-encoding cost. Possible avenues forward are improving the pre-processing cost of using the full spin-ful level-2 algebra or, based on recent work on perturbative analysis of SOS representations~\cite{hastings2024improving, hastings2023field, hastings2022perturbation}, including a subset of level-3 algebra components. Application of SA to lattice models will require a careful trade-off between SOS algebra size and block-encoding cost due to the existing methods already providing linear scaling in system size~\cite{BabbushPRX18}. We expect SA to be integral to further reducing ground-state energy estimation costs.

\section*{Acknowledgements}

DWB worked on this project under a sponsored research agreement with Google Quantum AI. DWB is also supported by Australian Research Council Discovery Projects DP210101367 and DP220101602.

\bibliography{library}
\appendix

\section{Equivalent quantum walk representations of spectral amplifiable Hamiltonians} \label{sec:equivalent_SA_walks}
In this section, we present alternate methods to generate quantum walks specialized to `spectrum-amplifiable' Hamiltonians $H_{\rm SA}$ as in ~\cref{eq:gap_amplifiable_Hamiltonian}, given block-encodings of its constituent parts $\textsc{Be}[O_\alpha/\lambda_\alpha]$. 
\cref{sec:phase_esitimate_sos} presents a specific block-encoding
\begin{align}\label{eq_block_encoding_canonical}
\textsc{Be}\left[\frac{2H}{\sum_\alpha\lambda_\alpha^2}-\mathbb{I}\right],
\end{align}
which can be converted into a quantum walk with eigenphases $\pm\arccos\left[\frac{2H}{\sum_\alpha\lambda_\alpha^2}-1\right]$ though qubitization~\cite{low2019hamiltonian}. 
The quadratic amplification of gradient for phases near $\pi$ then provides a quadratic advantage in phase estimation for estimating the ground-state energy. 
One may then interpret the quantum walk as the source of advantage stemming from the nonlinear dependence of phase on the eigenvalue of the original Hamiltonian.
Similarly, it is possible to consider the source of advantage by defining an appropriate square-root of Hamiltonian.

\subsection{Quantum walk on Hermitian square-rooted Hamiltonians}\label{sec:quantum_walk_hermitian_square_root}
One possible definition for the square-root of~\cref{eq:gap_amplifiable_Hamiltonian} is the Hermitian operator
\begin{align}\label{eq:Hermitian_gap_ampliable_Hamiltonian}
H_{\mathrm{sqrt}} = \sum_{\alpha=0}^{L-1}\left(   \ket{\chi_\alpha}\bra 0 \otimes O^{}_\alpha +\ket 0 \bra{\chi_\alpha} \otimes  O_\alpha^{\dagger}  \right).
\end{align}
Note that we are free to replace $\ket{\chi_\alpha}$ with any set of mutually orthogonal quantum states  that are also orthogonal to the multi-qubit computational basis state $\ket{0}$.
This squares to 
\begin{align}
H_{\mathrm{sqrt}}^{2} = 
\begin{pmatrix}
H_{\rm SA} & {\bf 0} \\
{\bf 0} & \ddots
\end{pmatrix},
\end{align}
where the upper diagonal block is the original gap-amplifiable Hamiltonian and ${\bf 0}$ is an all-zero matrix of corresponding dimension. 
We can obtain $H_{\rm SA}$ by applying a projection over the the ancilla space specified by the ancillary state $\ket 0$.
As such, if $H_{\rm SA}$ has eigenvalues $E_j \ge 0$ and eigenstates $\ket{\psi_j}$, then $H_{\mathrm{sqrt}}$ has eigenvalues $\pm\sqrt{E_j}$ with closely-related eigenstates $\ket{\phi_j^{\pm}}$.
To prove this, first observe that the subspace spanned by $\{|0\rangle |\psi_{j}\rangle, H_{\mathrm{sqrt}}|0\rangle |\psi_{j}\rangle \}$ is invariant under $H_{\mathrm{sqrt}}$:
\begin{align}
H_{\mathrm{sqrt}}|0\rangle |\psi_{j}\rangle &= \sum_{\alpha}\ket{\chi_\alpha} \otimes \left(O_\alpha|\psi_{j}\rangle\right)\\
H_{\mathrm{sqrt}}^{2}|0\rangle |\psi_{j}\rangle &= H_{\mathrm{sqrt}}\sum_{\alpha}\ket{\chi_\alpha} \otimes \left(O_\alpha|\psi_{j}\rangle\right) = \sum_{\alpha}\left(\mathbb{I} \otimes O_\alpha^{\dagger}O^{}_\alpha\right) |0\rangle \otimes |\psi_{j}\rangle = E_{j}|0\rangle \otimes |\psi_{j}\rangle.
\end{align}
Second, we diagonalize $H_{\mathrm{sqrt}}$ on the subspace spanned by this set of states, to find that it has eigenvalues and eigenvectors
\begin{align}
|\phi_{j}^{\pm}\rangle &= \frac{1}{\sqrt{2}}\left(\vert \psi_{j}\rangle \pm \frac{1}{\sqrt{E_{j}}}H_{\mathrm{sqrt}}\vert \psi_{j}\rangle \right),
\\
H_{\mathrm{sqrt}}\vert \phi_{j}^{\pm}\rangle &= \pm \sqrt{E_{j}}\vert \phi_{j}^{\pm}\rangle. 
\end{align}

We now discuss a block-encoding for~\cref{eq:Hermitian_gap_ampliable_Hamiltonian}.
Let ${\cal B}$ also denote an ancillary system.
Given access to controlled block-encodings $\textsc{Be}[O_\alpha/\lambda_\alpha]$, where $\bra{0}_\mathcal{B}\textsc{Be}[O_\alpha/\lambda_\alpha]\ket{0}_\mathcal{B}=O_\alpha/\lambda_\alpha$ and $\lambda_\alpha \ge \|O_\alpha\|$, unitary iteration and coherent-alias-sampling state preparation~\cite{BabbushPRX18} allows us to synthesize the SELECT and PREPARE circuits:
\begin{align}
\textsc{Sel}&=\ketbra{0}\otimes {\mathbb I}+\sum_{\alpha}\ketbra{\alpha} \otimes\textsc{Be}[O_\alpha/\lambda_\alpha],\\
\textsc{Prep}:\ket{0}_\mathrm{a}&\mapsto\sum_{\alpha}\frac{\lambda_\alpha}{\lambda_\mathrm{sqrt}}\ket{\alpha}_\mathrm{a}\ket{\text{garb}_\alpha}_\mathrm{a}=\sum_{\alpha}\frac{\lambda_\alpha}{\lambda_\mathrm{sqrt}}\ket{\chi_\alpha}_\mathrm{a}=\ket{\Psi}_{\rm a},
\quad \lambda_\mathrm{sqrt} = \sqrt{\sum_{\alpha}\lambda_\alpha^2} \, .
\end{align}
Let $X_c=\ket{0}\bra{1}_c+\ket{1}\bra{0}_c$ be the `Pauli X' and let $\textsc{CPrep}=\ketbra{0}_c\otimes {\mathbb I}+\ketbra{1}_c\otimes\textsc{Prep}$, where $c$ is an ancillary qubit.
Observe that
\begin{align}\nonumber
\textsc{CPrep}\label{eq:cprep}(X_c\otimes\ketbra{0}_\mathrm{a})\textsc{CPrep}^\dagger&=\ket{0}\bra{1}_c\otimes \ketbra{0}_\mathrm{a} \textsc{Prep}^\dagger+\ket{1}\bra{0}_c\otimes\textsc{Prep}\ket{0}  \bra{0}_\mathrm{a}
\\\nonumber
&=\ket{0}\bra{1}_c\otimes \ket{0}\bra{\Psi}_\mathrm{a}+\ket{1}\bra{0}_c\otimes\ket{\Psi}\bra{0}_\mathrm{a}
\\
&=\ket{00}\bra{1\Psi}_{\mathrm{ac}}+ \ket{1\Psi}\bra{00}_\mathrm{ac} \;.
\end{align}
Let $\textsc{And}:={\mathbb I}_b\otimes\ketbra{0}_\mathrm{a}+X_b\otimes({\mathbb I}_\mathrm{a}-\ketbra{0}_\mathrm{a})$ be a two-qubit operator, where $b$ is another ancilla qubit, with $\mathbb I_{a,b}$ denoting the $2 \times 2$ identity. 
Hence
\begin{align}
\bra{b}_\mathrm{b}\textsc{CPrep}(X_c\otimes \textsc{And})\textsc{CPrep}^\dagger\ket{0}_\mathrm{b}
=
\textsc{CPrep}(X_c\otimes \ketbra{0}_\mathrm{a})\textsc{CPrep}^\dagger \;,
\end{align}
so $\textsc{CPrep}(X_c\otimes \textsc{And})\textsc{CPrep}^\dagger$ that block-encodes~\cref{eq:cprep}.

Hence, the circuit
\begin{align}
U=\textsc{Sel} \cdot \textsc{CPrep} \cdot (X_c\otimes\textsc{And})\cdot \textsc{CPrep}^\dagger \cdot \textsc{Sel}^\dagger,
\end{align}
is a block-encoding of $H_{\mathrm{sqrt}}/\lambda_\mathrm{sqrt}$:
\begin{align}\nonumber
\bra{0}_\mathrm{b}\bra{0}_\mathcal{B}U\ket{0}_\mathcal{B}\ket{0}_\mathrm{b}&=
\bra{0}_\mathcal{B}\textsc{Sel}
(\ket{1\Psi}\bra{00}_\mathrm{ac}+\ket{00}\bra{1\Psi}_{\mathrm{ac}})\textsc{Sel}^\dagger\ket{0}_\mathcal{B}
\\\nonumber
&=\bra{0}_\mathcal{B}\textsc{Sel}\left(\sum_{\alpha}\frac{\lambda_\alpha}{\lambda_\mathrm{sqrt}}\ket{\alpha}\ket{\text{garb}_\alpha}\ket{1}\bra{0}+\ket{0}\sum_{\alpha}\frac{\lambda_\alpha}{\lambda_\mathrm{sqrt}}\bra{\alpha}\bra{\text{garb}_\alpha}\bra{1}\right)\textsc{Sel}^\dagger\ket{0}_\mathcal{B}
\\\nonumber
&=\bra{0}_\mathcal{B}\left(\sum_{\alpha}\frac{\lambda_\alpha}{\lambda_\mathrm{sqrt}}\ket{\alpha}\ket{\text{garb}_\alpha1}\bra{0}\otimes \textsc{Be}\left[\frac{O_\alpha}{\lambda_\alpha}\right]+\ket{0}\sum_{\alpha}\frac{\lambda_\alpha}{\lambda_\mathrm{sqrt}}\bra{\alpha}\bra{\text{garb}_\alpha1}\otimes\textsc{Be}^\dagger\left[\frac{O_\alpha}{\lambda_\alpha}\right]\right)\ket{0}_\mathcal{B}
\\\nonumber
&=\frac{1}{\lambda_\mathrm{sqrt}}\sum_{\alpha}\left(\ket{\chi_\alpha}\bra{0}\otimes O_\alpha+\ket{0}\bra{\chi_\alpha}\otimes O^\dagger_\alpha\right)
\\
&=\frac{H_{\mathrm{sqrt}}}{\lambda_\mathrm{sqrt}}.
\end{align}

Observe that $U^2={\mathbb I}$ is self-inverse. Hence, following qubitization~\cite{low2019hamiltonian}, the quantum walk $W=\textsc{Ref}U$ of $H_\mathrm{sqrt}$ is obtained by applying a reflection $\textsc{Ref}=2\proj{0}_{\mathrm{b}\mathcal{B}}-I$.
This quantum walk has eigenphases $\pm\arccos(\sqrt{E_j})\approx\pm(\frac{\pi}{2}-\sqrt{E_j})$. 
Hence, applying the PEA with precision $\delta$ using $c/\delta$ queries to $W$ learns $\sqrt{E_j}/\lambda_\mathrm{sqrt}$, for some constant $c$, which also translates to a quadratic advantage in the error of $E_j$. 
In the limit of small $\delta$, the precision $\epsilon$ of learning $E_j$ is 
\begin{align}\label{eq:sensitivity}
\epsilon=\left|\frac{d\sqrt{E_j}/\lambda_\mathrm{sqrt}}{dE_j}\right|^{-1}\delta=2\sqrt{E_j\lambda_\mathrm{sqrt}^2}\delta.
\end{align}
Overall, each quantum walk step has cost dominated by two queries to $\textsc{Sel}$ and two queries to $\textsc{Prep}$.
Hence, the overall query complexity is $4c\sqrt{E_j\sum_{\alpha}\lambda_\alpha^2}/\epsilon$ to $\textsc{Sel}$ and to $\textsc{Prep}$.

\subsection{Quantum walk on rectangular matrices}\label{sec:quantum_walk_rectangular}
One may more directly define a non-Hermitian matrix square-root by choosing
\begin{align}\label{eq:hsga_rect}
H_{\mathrm{sqrt}}' &= \sum_{\alpha=0}^{L-1}|\chi_\alpha\rangle  \otimes O_\alpha,
\end{align}
which has a singular value decomposition
\begin{align}
H'_\mathrm{sqrt}=\sum_j\sqrt{E_j}\ket{\tilde{\psi}_j}\bra{\psi_j},\quad \ket{\tilde{\psi}_j}=\frac{H'_\mathrm{sqrt}\ket{\psi_j}}{|H'_\mathrm{sqrt}\ket{\psi_j}|}.
\end{align}
Just like~\cref{eq:Hermitian_gap_ampliable_Hamiltonian}, we are free to replace $\ket{\chi_\alpha}$ with any set of mutually orthogonal quantum states like basis states.
As expected, the singular values (SVs) of $H'_{\mathrm{sqrt}}$ are equivalent to those of $H_{\rm SA}$ following
\begin{align}
\mathrm{SV}(H_{\mathrm{sqrt}}' )&=\sqrt{\mathrm{SV}\left(H_{\mathrm{sqrt}}^{\prime\dagger} H_{\mathrm{sqrt}}^{\prime} \right)}=\sqrt{\mathrm{SV}\left(\sum_{\alpha} \bra{\chi_\alpha}   \otimes O_\alpha^\dagger \sum_{\alpha'}\ket{\chi_{\alpha'}}  \otimes O_{\alpha'}\right)}
=\sqrt{\mathrm{SV}\left(\sum_{\alpha} O_\alpha^\dagger  O_\alpha\right)} = 
\sqrt{\mathrm{SV}(H_{\rm SA})}.
\end{align}
By forming a quantum walk over rectangular matrices~\cite{GSLW2019QSVT}, we can again use the PEA to estimate the energies $\sqrt{E_{j}}$.
Compared to~\cref{sec:quantum_walk_hermitian_square_root}, the block-encoding of the rectangular matrix $H_{\mathrm{sqrt}}'$ is more straightforward.

To see this, given access to controlled block-encodings $\textsc{Be}[O_\alpha/\lambda_\alpha]$ where $\bra{0}_\mathcal{B}\textsc{Be}[O_\alpha/\lambda_\alpha]\ket{0}_\mathcal{B}=O_\alpha/\lambda_\alpha$, unitary iteration and coherent alias sampling state preparation~\cite{BabbushPRX18} allows us to synthesize the SELECT and PREPARE circuits
\begin{align}
\textsc{Sel}&=\sum_{\alpha}\ket{\alpha}\bra{\alpha}\otimes\textsc{Be}[O_\alpha/\lambda_\alpha],\\
\textsc{Prep}\ket{0}_\mathrm{a}&=\sum_{\alpha}\frac{\lambda_\alpha}{\lambda_\mathrm{sqrt}}\ket{\alpha}_\mathrm{a}\ket{\text{garb}_\alpha}_\mathrm{a}
=
\sum_{\alpha}\frac{\lambda_\alpha}{\lambda_\mathrm{sqrt}}\ket{\chi_\alpha}_\mathrm{a},
\quad
\lambda_\mathrm{sqrt} := \sqrt{\sum_{\alpha}\lambda_\alpha^2}.
\end{align}
Then we have the block-encoding 
\begin{align}
U\coloneqq\textsc{Sel}\cdot \textsc{Prep}^\dagger=\textsc{Be}\left[\frac{H_\mathrm{sqrt}'}{\lambda_\mathrm{sqrt}}\right],
\end{align}
following
\begin{align}
\bra{0}_\mathcal{B}\textsc{Sel}\cdot \textsc{Prep}^\dagger\ket{0}_\mathrm{a}\ket{0}_\mathcal{B}&=\frac{\sum_{\alpha}|\chi_\alpha\rangle \otimes O_\alpha}{\lambda_\mathrm{sqrt}}
=\frac{H_\mathrm{sqrt}'}{\lambda_\mathrm{sqrt}},
\end{align}
where we used the freedom to choose any set of mutually orthogonal states $\ket{\chi_\alpha}$.

Let us now define the quantum walks $W_1\coloneqq\textsc{Ref}_{\mathcal{B}}U$ and $W_2\coloneqq\textsc{Ref}_{\mathrm{a}\mathcal{B}}U^\dagger$ over non-Hermitian matrices following~\cite{GSLW2019QSVT}.
We find it useful to introduce the following states
\begin{align}
\ket{\perp_j}&\coloneqq\frac{({\mathbb I}-\proj{0}_{\mathrm{a}\mathcal{B}})U^\dagger\ket{0}_{\mathcal{B}}\ket{\tilde{\psi}_j}}{|({\mathbb I}-\proj{0}_{\mathrm{a}\mathcal{B}})U^\dagger\ket{0}_{\mathcal{B}}\ket{\tilde{\psi}_j}|}=\frac{({\mathbb I}-\proj{0}_{\mathrm{a}\mathcal{B}})U^\dagger\ket{0}_{\mathcal{B}}\ket{\tilde{\psi}_j}}{\sqrt{1-E_j/\lambda_\mathrm{sqrt}^2}}.
\\
\ket{\tilde{\perp}_j}&\coloneqq\frac{({\mathbb I}-\proj{0}_\mathcal{B})U\ket{0}_\mathrm{a}\ket{\psi_j}}{|({\mathbb I}-\proj{0}_\mathcal{B})U\ket{0}_\mathrm{a}\ket{\psi_j}|}=\frac{({\mathbb I}-\proj{0}_\mathcal{B})U\ket{0}_\mathrm{a}\ket{\psi_j}}{\sqrt{1-E_j/\lambda_\mathrm{sqrt}^2}}.
\end{align}
Consider the eigenstates $\ket{\psi_j}$ of $H_{\rm SA}$ of eigenvalue $E_j\ge 0$. 
Then $W_1$ maps the the subspace spanned by $\{\ket{0}_{\mathrm{a}\mathcal{B}}\ket{\psi_j}, \ket{\perp_j}\}$ to 
$\{\ket{0}_\mathcal{B}\ket{\tilde{\psi}_j},\ket{\tilde{\perp}_j}\}$. 
We can show this by direct computation:
\begin{align}\label{eq:walk_1}
W_1\ket{0}_{\mathrm{a}\mathcal{B}}\ket{\psi_j}
&=
\textsc{Ref}_{\mathcal{B}}\left(\frac{\sqrt{E_j}}{\lambda_\mathrm{sqrt}}\ket{0}_\mathcal{B}\ket{\tilde{\psi}_j}+\sqrt{1-\frac{E_j}{\lambda_\mathrm{sqrt}^2}}\ket{\tilde{\perp}_j}\right)
=
\frac{\sqrt{E_j}}{\lambda_\mathrm{sqrt}}\ket{0}_\mathcal{B}\ket{\tilde{\psi}_j}-\sqrt{1-\frac{E_j}{\lambda_\mathrm{sqrt}^2}}\ket{\tilde{\perp}_j},
\\\nonumber
W_1\ket{\perp_j}
&=
\textsc{Ref}_{\mathcal{B}}U\frac{(I-\proj{0}_{\mathrm{a}\mathcal{B}})U^\dagger\ket{0}_{\mathcal{B}}\ket{\tilde{\psi}_j}}{\sqrt{1-E_j/\lambda_\mathrm{sqrt}^2}}
\\\nonumber
&=
\frac{\ket{0}_{\mathcal{B}}\ket{\tilde{\psi}_j}}{\sqrt{1-E_j/\lambda_\mathrm{sqrt}^2}}
-
\textsc{Ref}_{\mathcal{B}}U\frac{(\proj{0}_{\mathrm{a}\mathcal{B}})U^\dagger\ket{0}_{\mathcal{B}}\ket{\tilde{\psi}_j}}{\sqrt{1-E_j/\lambda_\mathrm{sqrt}^2}}
\\\nonumber
&=\frac{\ket{0}_{\mathcal{B}}\ket{\tilde{\psi}_j}}{\sqrt{1-E_j/\lambda_\mathrm{sqrt}^2}}
-
(2\proj{0}_{\mathcal{B}}-I)U\ket{0}_{\mathrm{a}\mathcal{B}}\frac{\sqrt{E_j}/\lambda_\mathrm{sqrt}\ket{\psi_j}}{\sqrt{1-E_j/\lambda_\mathrm{sqrt}^2}}
\\\nonumber
&=\frac{\ket{0}_{\mathcal{B}}\ket{\tilde{\psi}_j}}{\sqrt{1-E_j/\lambda_\mathrm{sqrt}^2}}
-
2\proj{0}_{\mathcal{B}}U\ket{0}_{\mathrm{a}\mathcal{B}}\frac{\sqrt{E_j}/\lambda_\mathrm{sqrt}\ket{\psi_j}}{\sqrt{1-E_j/\lambda_\mathrm{sqrt}^2}}
+
U\ket{0}_{\mathrm{a}\mathcal{B}}\ket{\psi_j}\frac{\sqrt{E_j}/\lambda_\mathrm{sqrt}}{\sqrt{1-E_j/\lambda_\mathrm{sqrt}^2}}
\\
&=-\sqrt{1-\frac{E_j}{\lambda_\mathrm{sqrt}^2}}\ket{0}_{\mathcal{B}}\ket{\tilde{\psi}_j}
+
\frac{\sqrt{E_j}}{\lambda_\mathrm{sqrt}}\ket{\tilde{\perp}_j},
\end{align}
and similarly for $W_2$, which maps the the subspace spanned by 
$\{\ket{0}_\mathcal{B}\ket{\tilde{\psi}_j},\ket{\tilde{\perp}_j}\}$ to $\{\ket{0}_{\mathrm{a}\mathcal{B}}\ket{\psi_j}, \ket{\perp_j}\}$. 

We can diagonalize $W_1$ and $W_2$ on these subspaces by defining
\begin{align}
\ket{\phi^{\pm}_j}\coloneqq\frac{\ket{0}_{\mathrm{a}\mathcal{B}}\ket{\psi_j}\pm i\ket{\perp_j}}{\sqrt{2}},
\quad
\ket{\tilde{\phi}^{\pm}_j}\coloneqq\frac{\ket{0}_{\mathcal{B}}\ket{\tilde{\psi}_j}\pm i\ket{\tilde{\perp}_j}}{\sqrt{2}}.
\end{align}
Then by substituting from~\cref{eq:walk_1},
\begin{align}\label{eq:walk_eigenstates}
W_1\ket{\phi^{\pm}_j}
&=
\frac{\sqrt{E_j}}{\lambda_\mathrm{sqrt}}\frac{\ket{0}_\mathcal{B}\ket{\tilde{\psi}_j}\pm i\ket{\tilde{\perp}_j}}{\sqrt{2}}\pm i\sqrt{1-\frac{E_j}{\lambda_\mathrm{sqrt}^2}}\frac{\ket{0}_\mathcal{B}\ket{\tilde{\psi}_j}\pm i\ket{\tilde{\perp}_j}}{\sqrt{2}}
=e^{i\arccos(\pm\sqrt{E_j}/\lambda_\mathrm{sqrt})}\ket{\tilde{\phi}^{\pm}_j},
\\
W_2\ket{\tilde{\phi}^{\pm}_j}
&=e^{i\arccos(\pm\sqrt{E_j}/\lambda_\mathrm{sqrt})}\ket{\phi^{\pm}_j}.
\end{align}
Hence, the quantum walks $W_1$ and $W_2$ apply eigenphases $\arccos(\pm\sqrt{E_j}/\lambda_\mathrm{sqrt})$ on the alternating subspaces $\ket{\phi_j^{\pm}}$ and $\ket{\tilde{\phi}_j^{\pm}}$.

We may thus perform phase estimation on the quantum walks, so long as we alternate between $W_1$ and $W_2$.
For instance, if the PEA calls for the $k^{\text{th}}$ power of a unitary to be applied, we would replace this with the length $k$ sequence $\cdots W_2W_1W_2W_1$, ending with $W_1$ if $k$ is odd and with $W_2$ if $k$ is even.
Hence, we may estimate $E_j$ with the same sensitivity as~\cref{eq:sensitivity}, but with half the number of queries to $\textsc{Sel}$ and $\textsc{Prep}$ due to the simpler block-encoding of the rectangular $H'_{\mathrm{sqrt}}$ compared to the Hermitian $H_{\mathrm{sqrt}}$ in~\cref{eq:Hermitian_gap_ampliable_Hamiltonian}.

Moreover, we may also choose to perform phase estimation on the product $W_2W_1$.
From~\cref{eq:walk_eigenstates},
\begin{align}\nonumber
W_2W_1\ket{\phi^{\pm}_j}&=e^{i2\arccos(\pm\sqrt{E_j}/\lambda_\mathrm{sqrt})}\ket{\phi^{\pm}_j}
\\\nonumber
&=
(\cos(2\arccos(\pm\sqrt{E_j}/\lambda_\mathrm{sqrt}))+i\cdots)\ket{\phi^{\pm}_j}
\\\nonumber
&=
(T_2[\pm\sqrt{E_j}/\lambda_\mathrm{sqrt}]+i\cdots)\ket{\phi^{\pm}_j}
\\\nonumber
&=
(2E_j/\lambda_\mathrm{sqrt}^2-1)+i\cdots)\ket{\phi^{\pm}_j}
\\
&=
e^{i\arccos(\pm(2E_j/\lambda_\mathrm{sqrt}^2-1))}\ket{\phi^{\pm}_j},
\end{align}
where we have used the definition of Chebyshev polynomials $T_k[\cos(\theta)]=\cos(k\theta)$ with $T_2[x]=2x^2-1$.
If we expand the quantum circuit
\begin{align}
W_2W_1&=\textsc{Ref}_{\mathrm{a}\mathcal{B}}(U^\dagger\textsc{Ref}_{\mathcal{B}}U),
\end{align}
the attentive reader will notice that the term in round brackets is a block-encoding $\textsc{Be}[2H_{\text{SA}}/\lambda_\mathrm{sqrt}^2-{\mathbb I}]$ following
\begin{align}
\frac{2H_{\text{SA}}}{\lambda_\mathrm{sqrt}^2}-{\mathbb I}
=\bra{0}_{\mathrm{a}\mathcal{B}}W_2W_1\ket{0}_{\mathrm{a}\mathcal{B}}
=\bra{0}_{\mathrm{a}\mathcal{B}}\textsc{Ref}_{\mathrm{a}\mathcal{B}}U^\dagger\textsc{Ref}_{\mathcal{B}}U\ket{0}_{\mathrm{a}\mathcal{B}}
=\bra{0}_{\mathrm{a}\mathcal{B}}U^\dagger\textsc{Ref}_{\mathcal{B}}U\ket{0}_{\mathrm{a}\mathcal{B}}.
\end{align}

\section{DFTHC block-encoding cost}\label{sec:Block-encoding-costs}
For block-encoding we will leverage the representation of the spin-free operator in Majorana representation. There two high level components to the block-encoding: 1) the linear combination of $T_{2}(\textsc{Be}[O_{\alpha}^{\dagger}], \textsc{Be}[O_{\alpha}])$ and 2) implementation of each qubitization step on each $O_{\alpha}$. The reflection on the ancilla to construct the walk operator is controlled on all ancilla associated with the linear combination of $T_{2}$ and qubitization signal states. In the Majorana representation a DFTHC ansatz with $r\in[R],c\in[C],b\in[B]$ has SOS generators
\begin{align}
O_{\text{SF},(rc)}&=w^{(rc)}_{B}I+\frac{i}{2}\sum_{\sigma\in\{0,1\}}\sum_{b\in[B]}w^{(rc)}_b\tilde{\gamma}_{\vec{u}^{(r)}_{b}\sigma1}\tilde{\gamma}_{\vec{u}^{(r)}_{b}\sigma0},\qquad \tilde{\gamma}_{\vec{u}\sigma x}=\sum_{p}u_p\gamma_{p\sigma x}, \label{eq:O_sf}\\
O_{\text{D}_1^{\sigma},r}&
=
\frac{\sqrt{w^{(r)}_+}}{2} \left(\tilde{\gamma}_{\vec{u}^{(r)}_+\sigma0}+i\tilde{\gamma}_{\vec{u}^{(r)}_+\sigma1}\right), \label{eq:O_D1}\\
O_{\text{Q}_1^{\sigma},r}&
=
\frac{\sqrt{w^{(r)}_-}}{2} \left(\tilde{\gamma}_{\vec{u}^{(r)}_-\sigma0}-i\tilde{\gamma}_{\vec{u}^{(r)}_-\sigma1}\right). \label{eq:O_Q1}
\end{align}
As previously mentioned, the optimal $\text{D}_1$ and $\text{Q}_1$ generators are found by an eigendecomposition of the one-body $N\times N$ coefficient matrix. Hence, we can identify $w^{(r)}_+$  with the positive semidefinite $N_{\text{D}_1}$ eigenvalues of the coefficient matrix, and $w^{(r)}_-$ with the absolute value of the $N_{\text{Q}_1}=N-N_{\text{D}_1}$ remaining eigenvalues, which are negative. In total, there are $N+RC(B+1)$ different coefficients $w$ and $N+RB$ real unit vectors $\vec{u}$ across the different generators. 

In the following, we will use qubit registers  $\ket{r}_{\mathbf{r}}, \ket{G}_{\mathbf{G}},\ket{b}_{\mathbf{b}},\ket{c}_{\mathbf{c}}$ storing integer the integers $r,b,c$.  It is useful to index the different operators with a single integer $x_{\text{o}}\in[X_{\text{o}}]$ that is a function of $G,r,c$ as follows:
\begin{align}\label{eq:x_operator}
x_\text{o}(G,r,c)=
\begin{cases}
r,&G=G_{\text{D}_1},\\
N_{\text{D}_1}+r,&G=G_{\text{Q}_1},\\
N+rC+c,&G=G_\text{SF},\\
\end{cases}
\quad X_{\text{o}}=N+RC.
\end{align}
We adopt the convention that $b=B$ indexes the identity term. We note that the integers last line $\{N+RB+r\}_{r\in[R]}$ can be replaced by set of integers distinct from the earlier cases. 

The sum-of-square Hamiltonian is then 
\begin{align}
H_{\text{SOS}}&=\sum_g\sum_{rc}O^{\dagger}_{g,rc}O_{g,rc}=\sum_{x_\text{o}}O^{\dagger}_{x_\text{o}}O_{x_\text{o}}.
\end{align}

For each term, we associate a certain block-encoding normalization factor that will be later relevant.
\begin{align}\label{eq:DFTHC_normalization_factors}
\lambda_{\text{SF}}&=\sum_{rc}\lambda_{\text{SF},rc}^2, &\lambda_{\text{SF},rc}&=\frac{1}{\sqrt{2}}\left(|w^{(rc)}_B|+\sum_{b=0}^{B-1} |w_b^{(rc)}|\right),
\\
\lambda_{\text{D}_1}&=\sum_{r}\lambda_{\text{D}_1^0,r}^2+\sum_{r}\lambda_{\text{D}_1^1,r}^2,&\lambda_{\text{D}_1^\sigma,r}&=\sqrt{w_+^{(r)}},
\\
\lambda_{\text{Q}_1}&=\sum_{r}\lambda_{\text{Q}_1^0,r}^2+\sum_{r}\lambda_{\text{Q}_1^1,r}^2,&\lambda_{\text{Q}_1^\sigma,r}&=\sqrt{w_-^{(r)}}.
\end{align}
The overall block-encoding normalization of the sum-of-squares Hamiltonian will be
\begin{align}
\Lambda=\frac{1}{2}\left(\lambda_{\text{Q}_1}+\lambda_{\text{D}_1}+\lambda_{\text{SF}}\right).
\end{align}
The key idea is that using an efficiently computable indexing scheme to each distinct coefficient and unit vector, we may form table-lookups of the smallest possible size to this data, and hence block-encode with the smallest constant factors. 

The method for block-encoding the Hamiltonian uses similar principles to those used for double factorization in prior work \cite{PhysRevResearch.3.033055,PRXQuantum.2.030305}, with some key differences.
The main parts that are in common are as follows.
\begin{enumerate}
    \item A register is prepared with the appropriate amplitudes over $x_\text{o}$, which is used to control the block-encoding of $O_{x_\text{o}}$.
    \item The single step of oblivious amplitude amplification is used to provide the factor of 2 improvement, as explained above.
    This means that we block encode $O_{x_\text{o}}$, reflect on the ancilla qubits of the block-encoding, then apply the inverse of the block-encoding.
    \item A control register is prepared with an equal superposition to represent $\sigma$ for the spin, and used to control a swap between the system registers representing spin up and spin down.
    \item A superposition over $b$ needs to be prepared.
    \item Each $O_{x_\text{o}}$ involves application of Majorana operators in a transformed basis, with the choice of basis controlled by $x_\text{o}$.
    This is achieved by applying QROM on $x_\text{o}$ to output the rotation angles to change the basis, then applying a sequence of Givens rotations.
    These rotations are only applied on one half the qubits representing spin up (for example), because of the previous step.
    \item After the basis transformation, the Majorana operators need only be applied to a single qubit representing a single (transformed) spin-orbital.
\end{enumerate}

Next we detail the subtleties involved in each of these steps.

\subsection{Preparing the superposition over \texorpdfstring{$x_\text{o}$}{xo}}

The new subtlety is that the contiguous register containing $x_\text{o}$ is used to index multiple variables $G,r,c$.
Recall that $G$ is used to select between the operators $O_{\text{SF},(rc)}$, $O_{\text{D}_1,r}$, and $O_{\text{Q}_1,r}$.
The variable $r$ is used for further indexing over all three of these operators, and $c$ is used to index only $O_{\text{SF},(rc)}$.
At this stage we do not yet output these variables, and just prepare the state with the appropriate weightings
\begin{align}\label{eq:target_prep_state}
\textsc{Prep}\ket{0}
&=\sum_{x_{\text{o}}=0}^{X_{\text{o}}-1}\frac{\lambda^{(G_{x_\text{o}}r_{x_\text{o}}c_{x_\text{o}})}}{\sqrt{2\Lambda}}\ket{x_{\text{o}}}\ket{\text{garbage}_{x_{\text{o}}}}_\text{garb},
\end{align}
where the composite index $(G_{x_\text{o}}r_{x_\text{o}}c_{x_\text{o}})$ from ~\cref{eq:x_operator} selects one of the coefficients $\lambda_{\text{SF},rc},2\lambda_{\text{D}_1^\sigma,r},2\lambda_{\text{Q}_1^\sigma,r}$ from~\cref{eq:DFTHC_normalization_factors}.
The dominant cost of this step is the alias sampling step which invokes QROAM~\cite{Low2024tradingtgatesdirty,berry2019qubitization}.
This is the control state which we will use to index controlled block-encoding. 

\subsection{The step of oblivious-amplitude-amplification/qubitization}
The use of oblivious amplitude amplification is the same as in previous work~\cite{PhysRevResearch.3.033055}, with the only difference being the number of qubits reflected upon due to the different operators, and the non-Hermitian nature of the block encoded operators giving $T_{2}(O_{\alpha}^{\dagger},O_{\alpha})$.
The qubits we need to reflect on are as follows.
\begin{enumerate}[(i)]
    \item The $\lceil \log(B+1)\rceil$ qubits that the state for $b$ is prepared on.
    \item The $b_{k2}$ qubits used for the equal superposition state in the coherent alias sampling.
    \item One qubit that is rotated as part of the preparation of the equal superposition over $b$.
    \item One qubit for the sum over spins in $O_{\text{SF},rc}$ (see below).
\end{enumerate}
These registers are similar to those used in prior work.

The preparation of equal superposition states (as part of coherent alias sampling) has a small amplitude for failure, and operations should be controlled on success of these preparations.
That results in a very small (less than about $0.1\%$) increase in $\lambda$, and has a smaller impact than contamination from unwanted operations if the control was not used.
The reflection in this part needs to be controlled on the success of preparation of the register for $x_\text{o}$, so has a Toffoli cost of
\begin{equation}
    \lceil \log(B+1)\rceil + b_{k2} + 1 \, .
\end{equation}
In prior work there was one extra control which yielded an additional Toffoli of complexity (on $\ell=0$ in Ref.~\cite{PRXQuantum.2.030305}).

\subsection{Selecting between the spins}
A new subtlety in the selection of the spins is that for $O_{\text{SF},(rc)}$ the summation over the spin $\sigma$ is required for the block-encoding of that operator, whereas $O_{\text{D}_1^{\sigma},r}$ and $O_{\text{Q}_1^{\sigma},r}$ are dependent on $\sigma$, and the summation of spin is part of the outer sum (with $x_\text{o}$).
This means that we need separate control variables for the spin for SF versus $\text{D}_1$ and $\text{Q}_1$.

That is, we have one control qubit in a $\ket{+}$ state, that is used to control the swap of the spin up and spin down parts of the system register for $\text{D}_1$ or $\text{Q}_1$.
A second control qubit is used to control the swap for the case of $\text{SF}$.
These controlled swaps are performed in the block-encoding of $O_{x_\text{o}}$.
The second qubit is part of the block-encoding as well, so is reflected upon in the oblivious amplitude amplification.
That ensures that the summation over $\sigma$ for $\text{SF}$ is obtained.
The first qubit is not reflected upon in the step of oblivious amplitude amplification, ensuring that it is part of the outer summation.
When we perform the block-encoding of $O_{x_\text{o}}$, we extract the variables $G,r,c$, so we have the variable $G$ selecting between these operators available to control upon.

To be more specific, the Majorana operators of different spin can be related by
\begin{equation}
    \gamma_{p1 x}=\left(\prod_{j=0}^{N-1}\textsc{Swap}_{j,j+N}\right)\gamma_{p0 x}\left(\prod_{j=0}^{N-1}\textsc{Swap}_{j,j+N}\right)\cdot\vec{Z}_{N-1} \, ,
\end{equation}
where $\vec{Z}_{N-1}$ accounts for the extra string of operators needed for the Jordan-Wigner representation.
In the Jordan-Wigner representation, $\gamma_{p\sigma 0}=X_{\sigma N + p}\vec{Z}_{\sigma N + p-1}$ and $\gamma_{p\sigma 1}=Y_{\sigma N + p}\vec{Z}_{\sigma N + p-1}$, where $\vec{Z}_{\sigma N + p-1} = Z_{\sigma N + p-1}\cdots Z_0$.

Because it can be performed with Clifford gates, it does not add to the Toffoli complexity considered in our costings.
Let us define the controlled operations
\begin{align}
U_\text{Swap}&\coloneqq\sum_{\sigma\in\{0,1\}}\proj{\sigma}\otimes \left(\prod_{j=0}^{N-1}\textsc{Swap}_{j,j+N}\right)^\sigma,
\\
U_1&\coloneqq\sum_{\sigma\in\{0,1\}}\proj{\sigma}\otimes\vec{Z}^{\sigma}_{N-1}.
\end{align}
Then we can apply the appropriate operators using $U_\text{Swap}$ and $U_1$, for example
\begin{align}\label{eq:sel_prime_sigmax}
\sum_{\sigma}\proj{\sigma}\otimes \gamma_{p\sigma x}&=U_\text{Swap}\gamma_{p0x}U_\text{Swap} U_1 \, .
\end{align}

To select the appropriate $\sigma$ register to use, we will encode $G$ using two qubits, with one selecting between $\text{SF}$ and the pair $\text{D}_1$ and $\text{Q}_1$.
Calling that $G_0$, we will have registers with $\ket{G_0}\ket{\sigma_0}\ket{\sigma_1}$.
By applying two Toffolis (and $X$ gates), we obtain an output qubit $\ket{\sigma}$, where $\sigma=1$ if $\sigma_{G_0}=1$.
After the controlled swaps, $\ket{\sigma}$ may be erased using an $X$ measurement and Clifford phase fixups in the usual way, so there is no extra Toffoli cost for erasure.

Therefore, producing the qubit with $\ket{\sigma}$ adds an extra cost of two Toffolis to the block-encoding of each $O_{x_\text{o}}$, for a total of 4 Toffolis for the overall block-encoding of the Hamiltonian.
The four controlled swaps add a cost of $4N$ Toffolis.

\subsection{Preparing the superposition over \texorpdfstring{$b$}{b}}
This preparation of the inner superposition is similar to prior work with two key differences.
The first is that we use this step to also output $G,r,c$, which are used in later steps.
Prior to this we have just prepared a superposition over the contiguous register $x_\text{o}$, and the unary iteration over this variable in this step is used to output $G,r,c$ as well as the information needed for the coherent alias sampling for preparing the appropriate superposition over $b$.

The second difference is that no superposition over $b$ is needed for the operators $O_{\text{D}_1^{\sigma},r},O_{\text{Q}_1^{\sigma},r}$.
When iterating over the values of $x_\text{o}$ corresponding to these operators, we can simply not output values used for coherent alias sampling for $b$.
However, it is convenient to output the set of rotations needed in the next step when iterating over these values of $x_\text{o}$.

\subsection{Rotating the basis}
Before rotating the basis, we need to first output the set of rotations needed in an ancilla register to be used as control.
The distinction from prior work is what registers are used to control the output of these rotations.
For the operator $O_{\text{SF},(rc)}$, the rotations are selected using $r$ and $b$, but the are \emph{not} selected using $c$.
This reduces the cost of the iteration needed to output the set of rotations.

Another difference is that for $\text{D}_1,\text{Q}_1$, the rotations are selected only using $r$ and $b$ is not used.
A further subtlety is that we are using $b=B$ to select the identity term.
The way this is accounted for is that we prepare the superposition over $b$ including $b=B$, but then in the iteration to output the rotations, for $b=B$ we instead flip an ancilla qubit to flag that we have the identity.
The qubit is used as part of the control in the application of the Majorana operators (see below).
The signs needed in the superposition over $b$ (i.e., the signs of $w^{(rc)}_b$) can be obtained simply by performing the sign flips directly in the iteration over $b$, and does not add to the Toffoli cost.
(There is no need to output signs in an ancilla.)

Otherwise the rotation of the basis is applied in a similar way as in prior work.
Note that for any unit vector $\vec{u}$, there exists a unitary rotation $U(\vec{u})$ comprised of $N-1$  different arbitrary single-qubit rotations with rotation angle $\theta_j$ and $\mathcal{O}(N)$ two-qubit Clifford gates such that $U^\dagger(\vec{u})\gamma_{00x}U(\vec{u})=\tilde{\gamma}_{\vec{u}0x}$. In particular any sequence of rotation angles $\vec{\theta}=(\theta_0,\cdots,\theta_{N-2})$ corresponds to some $u=f(\vec{\theta})$. By having each of the $j\in[N-1]$ rotation angles $\theta_j$ be specified by a $b_{\text{rot}}$-bit binary numbers, one may further define the controlled rotation~\cite{PhysRevResearch.3.033055}
\begin{align}
U_{\text{Rot}}\ket{\vec{\theta}}\ket{\psi}&=\ket{\vec{\theta}}U(u(\vec{\theta}))\ket{\psi} ,\\
U^\dagger_{\text{Rot}}\gamma_{00x}U_{\text{Rot}}&=\sum_{\vec{\theta}}\proj{\vec{\theta}}\otimes \tilde{\gamma}_{f(\vec{\theta})0x},
\end{align}
and each $U_{\text{Rot}}$ can be implemented with $2(N-1)b_{\text{rot}}$ Toffoli gates. To simplify notation, we use $\vec{u}$ and its representation through $\vec{\theta}$ interchangeably. For instance, $\ket{\vec{u}}=\ket{\vec{\theta}=f^{-1}(\vec{u})}$. As there are multiple solutions $\vec{\theta}$  representing the same $u$,  we arbitrarily select any one solution from $f^{-1}(u)$.

Note also that the transformation can be applied in common between the Majorana operators $\tilde{\gamma}_{\vec{u}0x}$ with $x=0$ and $1$.
we can transform between the two operators using the phase gate $S$, because $X=S^\dagger Y S$.
One could apply $S$ and $S^\dagger$ before and after (respectively) on all system qubits, but because the tensor product of $S$ commutes through the Givens rotations, they need only be applied to the single qubit representing the transformed spin-orbital.

\subsection{Applying the Majorana operators on a single qubit}

This last step is significantly changed here as compared to prior work.
In Ref.~\cite{PhysRevResearch.3.033055} the number operator was written as a sum of the identity and $Z$, and the identity was removed from the sum so that only the $Z$ gate needed to be applied.
Due to the transformation of basis, it need only be applied to a single qubit.

In our operators, we have a similar term in $O_{\text{SF},(rc)}$, with the product of Majorana operators $\tilde{\gamma}_{\vec{u}^{(r)}_{b}\sigma1}\tilde{\gamma}_{\vec{u}^{(r)}_{b}\sigma0}$. 
The basis transformation means that these Majorana operators are only applied to a single qubit, and the sequence of $Z$ gates on the other qubits is not required.
This means that $i \tilde{\gamma}_{\vec{u}^{(r)}_{b}\sigma1}\tilde{\gamma}_{\vec{u}^{(r)}_{b}\sigma0}$ simplifies to just a $Z$ gate on a single qubit, similar to prior work.

The difference is that we also have $\text{D}_1$ and $\text{Q}_1$, where the operators $\tilde{\gamma}_{\vec{u}^{(r)}_{b}\sigma1}$ and $\tilde{\gamma}_{\vec{u}^{(r)}_{b}\sigma0}$ need to be applied individually.
However, these are simplified to $Y$ and $X$ operators on a single qubit, so are still simple to apply.
A single control qubit is used to provide the sum $\tilde{\gamma}_{\vec{u}^{(r)}_-\sigma0}\pm i\tilde{\gamma}_{\vec{u}^{(r)}_-\sigma1} = X \pm iY$.
A qubit flagging the difference between $\text{D}_1$ and $\text{Q}_1$ can be used to control the sign in this implementation.
A second qubit flagging whether $\text{SF}$ is performed can be used to control whether the $Z$ gate is performed, or whether it is $X$ or $Y$.

To be more specific, we have the following control qubits.
\begin{enumerate}[(i)]
    \item The first qubit for $G$, which distinguishes between $\text{SF}$ and the pair
    $\text{D}_1,\text{Q}_1$.
    \item The second qubit for $G$, which distinguishes between $\text{D}_1$ and $\text{Q}_1$.
    \item A qubit which flags whether we have the identity component of $\text{SF}$.
    \item A qubit in superposition used to distinguish between $\tilde{\gamma}_{\vec{u}^{(r)}_{b}\sigma1}$ and $\tilde{\gamma}_{\vec{u}^{(r)}_{b}\sigma0}$ for $\text{D}_1,\text{Q}_1$.
\end{enumerate}

\begin{figure}\label{fig:Majorana}
\includegraphics[width=0.3\textwidth]{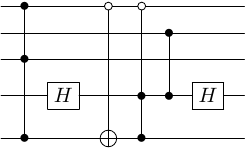}
\caption{The quantum circuit used to apply the Majorana operators on the single qubit, with the target qubit at the bottom.}
\end{figure}

The quantum circuit used to apply the Majorana operators with these control qubits is then shown in Fig.~\ref{fig:Majorana}.
The first operation is to apply the $Z$ gate for $\text{SF}$.
The second controlled operation is giving the required $X$ operation for $\tilde{\gamma}_{\vec{u}^{(r)}_-\sigma0}$.
The third controlled operation applies a $Z$ gate, provided we have $\text{D}_1,\text{Q}_1$, and the fourth qubit is selecting $\tilde{\gamma}_{\vec{u}^{(r)}_-\sigma1}$ (so $iY=ZX$ is needed).
The fourth controlled operation applies the correct sign factor in $X \pm iY$ needed to select between $\text{D}_1$ and $\text{Q}_1$.
This application of the Majorana operators uses only two Toffolis.
We also need controls by successful preparation of the equal superposition states for $x_\text{o}$ and $\ket{b}$.
A single Toffoli is used to flag success for both, and control by this flag will add another 4 Toffolis for a total of 7.

\subsection{Total costs}
\subsubsection{Summary of costs}

Now we enumerate the total costs for the block-encoding, with the complete circuit illustrated in Fig.~\ref{fig:fullcircuit}.
A more detailed explanation of the block-encoding costs is given in Appendix \ref{app:costs}.
The total costs for constructing the walk operator are as follows.
\begin{enumerate}
    \item Preparation of the control register for the outer sum over $x_\text{o}$ and erasure, corresponding to everything in Table~\ref{tab:outer_prepare_costs}.
    \item The reflection in the block-encoding of the Hamiltonian for $T_{2}(O_{\alpha}^{\dagger}, O_{\alpha})$, with cost
    \begin{equation}
    \lceil \log(B+1)\rceil + b_{k2} + 1 \, .
    \end{equation}
    \item The four controlled swaps for the implementation of the spin, with cost $4(N+1)$.
    \item Preparation of the control register for $b$ in the inner sum, given by the costs in Table~\ref{tab:b_weight_prep_via_unary_qrom}, with a factor of 2 for the two calls.
    \item The basis rotation, including the output and erasure of the rotation angles, as well as the rotations themselves, as given in Table~\ref{tab:rotation_qrom} (again with a factor of 2).
    \item The Majorana operators on the single qubit with cost of 7 Toffolis. This is called twice per $T_{2}$.
    \item A reflection on all the ancilla for the block-encodings to construct the walk operator of Toffoli cost
    \begin{equation}
    \lceil \log(N+RC)\rceil + \lceil \log(B+1)\rceil + b_{k1}+b_{k2} + 2 \, .
    \end{equation}
\end{enumerate}

\begin{table}[H]
    \centering
    \begin{tabular}{|c|c|c|c|c|c|}
    \hline \hline
    Operation & Cost & Persistent Ancilla & Temporary Ancilla & Calls & Step \\
    \hline
    Uniform($x_\text{o}$)  & $4\lceil \log(N+RC)\rceil$  & $\lceil \log(N + RC)\rceil + 2$ &  $\lceil \log(N + RC)\rceil - 2$ & 1 & \ref{step1} \\
    AliasSampling($x_\text{o}$) & $\left\lceil\frac{N+RC}{2^{k_1}}\right\rceil + 2^{k_1} b_1$  & $b_{1} + b_{k_{1}}$ & $2^{k_{1}}b_{1} + \lceil \log_{2}(\frac{N + RC}{2^{k_{1}}}) \rceil$ & 1 & \ref{step2} \\
    AliasSampling($x_\text{o}$)$^{\dagger}$ & $b_1 + \left\lceil \frac{N+RC}{2^{k_5}}\right\rceil + 2^{k_5} $ & -- & -- & 1 & \ref{step10} \\
    Uniform($x_\text{o}$)$^{\dagger}$ & $ 4\lceil \log(N+RC)\rceil$ & -- & -- &  1 & \ref{step11} \\
    \hline \hline
    \end{tabular}
    \caption{Preparation costs for the state preparation over unitaries $T_{2}(O_{\alpha}^{\dagger},O_{\alpha})$.
    The right column gives the corresponding step number from Appendix \ref{app:costs},
    and the `Calls' column gives the number of calls for each step of the walk (so the multiplying factor for the Toffoli cost).
    The selection register is $x_\text{o}$, which takes on $N + RC$ values.
    The quantity $b_1 = \lceil \log(N+RC)\rceil + b_{k1}$, where
    $b_{k1}$ is the number of bits used for the `keep' register in coherent alias sampling. The AliasSampling involves an inequality test at the cost of $b_{k1}$ and controlled swap costing $\lceil \log(N + RC)\rceil$.
    The $b_{1}$ term in the fourth line is for inverting the inequality tests and controlled swaps.
    The integers $k_1$ and $k_5$ may be chosen to minimise costs.
    The ancillas are not listed for the inverse operations because they are considerably smaller than used for the forward operations, so do not affect the maximum qubit usage of the algorithm.}
    \label{tab:outer_prepare_costs}
\end{table}

\begin{table}[H]
    \centering
    \scalebox{1.0}{
    \begin{tabular}{|c|c|c|c|c|c|}
    \hline \hline
    Operation & Toffoli & Persistent Ancilla & Temporary Ancilla & Calls & Step \\
    \hline
    Uniform($b$) & $4\lceil \log(B+1)\rceil$ & $\lceil \log (B+1)\rceil +2$ & $\lceil \log (B+1)\rceil - 2$ & 2 & \ref{step3} \\
    AliasSampling($b$)  & $RC \left\lceil \frac{B+1}{2^{k_2}}\right\rceil + 2^{k_2} b_2$ & $b_2 + b_{k2} + \lceil \log R \rceil + \lceil \log C \rceil + 2 $ & $\left\lceil \log \left(RC \left\lceil \frac{B+1}{2^{k_2}}\right\rceil\right)\right\rceil  + (2^{k_2}-1)b_2  $ & 2 & \ref{step4} \\
    AliasSampling($b$)$^{\dagger}$ & $b_2 + N + \left\lceil \frac {RC}{2^{k_4}}\right\rceil + 2^{k_4}(B+1)$ & -- & -- & 2 & \ref{step8} \\
    Uniform($b$)$^{\dagger}$ & $4\lceil \log(B+1)\rceil$ & -- & -- &  2 & \ref{step9} \\
    \hline \hline
    \end{tabular}
    }
    \caption{Preparation costs for the $b$ register.
    Unary iteration is performed over the first $RC$ values of $x_{o}$, combined with QROAM over $b$, to output data needed for the coherent alias sampling state preparation on $b$.
    The quantity $b_{2} = \lceil \log(B+1)\rceil + b_{k_{2}}$, where $b_{k_{2}}$ is the bit size of the `keep' register. While performing the unary iteration over $x_{o}$, we also output $G,r,c$ with no further Toffoli cost.
    The right column gives the corresponding step from Appendix \ref{app:costs}.
    }
    \label{tab:b_weight_prep_via_unary_qrom}
\end{table}

\begin{table}[H]
    \centering
    \begin{tabular}{|c|c|c|c|c|c|}
    \hline \hline
    Operation & Toffoli & Persistent Ancilla & Temporary Ancilla & Calls & Step \\
    \hline
    QROM($\vec{u}$) & $N+RB$ & $(N-1)b_{\text{rot}}$ &$\max\left( \lceil \log(N+RC) \rceil,  \lceil \log(RB) \rceil \right)$& 2 & \ref{step5} \\
    basis rotations & $2(N-1)b_{\text{rot}}$ & -- & $b_{\text{rot}}$ & 2 & \ref{step6} \\
    QROM($\vec{u}$)$^{\dagger}$ &  $R+B$ & -- & -- & 2 & \ref{step7} \\
    \hline \hline
    \end{tabular}
    \caption{QROM to output rotations, implement them, and erase the data. The integer $b_{\text{rot}}$ is the bits of precision used for each rotation angle.}
    \label{tab:rotation_qrom}
\end{table}

\subsubsection{More details of block-encoding costs}\label{app:costs}

Here we give further details of the block-encoding costs, excluding the controlled swaps for the spin, reflection for the step of oblivious amplitude amplification, and implementation of the Majorana operator on a single qubit, as these are detailed above.

\begin{enumerate}
\item \label{step1} First we prepare an equal superposition for $x_\text{o}$ with cost
\begin{equation}\label{eq:eqprepcost}
  4\lceil \log(N+RC)\rceil  \, .
\end{equation}
This is omitting small $O(1)$ costs, and also ignoring savings when $N+RC$ has factors of 2.
For this costing see Section III E 2 of Ref.~\cite{SandersPRQ20} (pages 42 and 43), where the cost is $4k+2s-13$ Toffolis for preparation on $k$ qubits and using $s$ bits of accuracy for a rotation.
That rotation can be used with $s=7$ to give very accurate results, for a cost of $4k+1$.
This is why we give the cost above as 4 times the number of qubits, ignoring $O(1)$ costs.

\textbf{Qubit cost:} The number of qubits used in this step is $\lceil \log(N+RC)\rceil$ to store $x_o$, the extra qubit that is rotated in the equal superposition, and a qubit which is used to flag success of the preparation.
Extra temporary qubits are used in the coherent arithmetic (inequality test), which may be ignored as they are fewer than those used in later steps.
We also use a phase gradient register for rotations, which is used in the entire algorithm so the qubits for this register are quantified separately.
The total qubit cost is therefore
\begin{equation}
    \lceil \log(N+RC)\rceil +2
\end{equation}
persistent qubits, and
\begin{equation}
    \lceil \log(N+RC)\rceil -2
\end{equation}
temporary qubits for the arithmetic.
It is also possible that the number of temporary qubits used for the rotation on the ancilla qubit can be larger for small values of $N+RC$.

\item \label{step2} Perform QROAM on $x_\text{o}$ with cost
\begin{equation}
    \left\lceil\frac{N+RC}{2^{k_1}}\right\rceil + (2^{k_1}-1)b_1 \, ,
\end{equation}
for some choice of integer $k_1$, where $b_1$ is the output size
\begin{equation}
b_1 = \lceil \log(N+RC)\rceil + b_{k1} \, ,
\end{equation}
with $b_{k1}$ the number of bits for the `keep' register in coherent alias sampling.  Then the inequality test is cost $b_{k1}$ and the controlled swap is cost $\lceil \log(N+RC)\rceil$, for total cost in step 2 of 
\begin{equation}
   \left\lceil\frac{N+RC}{2^{k_1}}\right\rceil + 2^{k_1} b_1  \, .
\end{equation}

\textbf{Qubit cost:} The persistent qubits used in this step are
\begin{equation}
    b_1 + b_{k1}
\end{equation}
with $b_1$ for the output of the QROM, and $b_{k1}$ qubits used for the equal superposition register.
(Those must be retained because they are not reset.)
There are also
\begin{equation}
   \left\lceil \log \left(\frac{N+RC}{2^{k_1}}\right)\right\rceil  + (2^{k_1}-1)b_1
\end{equation}
temporary qubits for the output of the QROM as well as the temporary qubits used in the unary iteration.
Recall that there are $2^{k_1}b_1$ qubits output, then we swap the $b_1$ output qubits into the correct register, and $(2^{k_1}-1)b_1$ qubits are erased (via an $X$ measurement and a later phase fixup).
There are also temporary qubits used in the inequality test for coherent alias sampling, but these are much less than those used in the QROM so may be ignored.

\item \label{step3} For the block-encoding of $O_{x_\text{o}}$, we first prepare an equal superposition over $b$ with cost
\begin{equation}
4\lceil \log (B+1)\rceil \, .
\end{equation}
This preparation does not need to be controlled.

\textbf{Qubit cost:} Similar to step 1, there is a persistent ancilla cost of
\begin{equation}
\lceil \log (B+1) \rceil +2
\end{equation}
qubits for $b$ and the rotated qubit, and
\begin{equation}
\lceil \log (B+1) \rceil -2
\end{equation}
temporary qubits in the arithmetic.

\item \label{step4} Next we can perform unary iteration over $x_\text{o}$, combined with QROAM on $b$.  This has Toffoli cost
\begin{equation}
RC \left\lceil \frac{B+1}{2^{k_2}}\right\rceil + (2^{k_2}-1)b_2 \, .
\end{equation}
Here $k_2$ is a new choice of integer, and $b_2$ is the output size for coherent alias sampling on $b$ of
\begin{equation}
b_2 = \lceil \log (B+1)\rceil + b_{k2}\, ,
\end{equation}
with $b_{k2}$ being another choice of the number of bits for the keep register.  Again there will be a $b_2$ cost for the inequality test as well as a controlled swap, to give total cost
\begin{equation}
RC \left\lceil \frac{B+1}{2^{k_2}}\right\rceil + 2^{k_2} b_2 \, .
\end{equation}
The idea of performing the unary iteration in this step is furthermore that we can output other useful information.  We can output $G,r,c$ with no further Toffoli cost, since we are already iterating over $x_\text{o}$.
We separate out the QROM for the final $N$ values of $x_\text{o}$ into a separate step.

Depending on the values of the parameters, it may also be suitable to output data for all values of $b$ at once, then use the controlled swaps to swap the desired data into the correct location.
That is, use the entire $b$ register like the less significant bits in QROAM.
Then the Toffoli cost is instead
\begin{equation}
RC + (B+1) b_2 \, .
\end{equation}

\textbf{Qubit cost:} The persistent ancilla cost is
\begin{equation}
    b_2 + b_{k2} + \lceil \log R \rceil + \lceil \log C \rceil + 2 \, .
\end{equation}
This includes $b_2$ for the QROM output, $b_{k2}$ for the equal superposition state, and the qubits for storing the output values of $G,r,c$.
The temporary qubits needed are
\begin{equation}
   \left\lceil \log \left(RC \left\lceil \frac{B+1}{2^{k_2}}\right\rceil\right)\right\rceil  + (2^{k_2}-1)b_2 \, .
\end{equation}
Here $(2^{k_2}-1)b_2$ qubits are used as temporary qubits in the QROM output, and the first term is temporary qubits in the unary iteration.

\item \label{step5} 
In this step we consider the output of the set of rotations.
First, we have unary iteration through the final $N$ values of $x_\text{o}$ and to output the set of rotations for $G_{\text{D}_1}$ and $G_{\text{Q}_1}$.
As part of this we also output the $G,r$ values.

Next, perform QROM on the $r,b$ registers to output the set of rotations.  In previous work there was no advantage to QROAM since the size of the output set of rotations is quite large, so here we cost this as regular QROM (unary iteration) with cost $RB$.
This gives a total Toffoli cost for this step of
\begin{equation}
    N+RB \, .
\end{equation}

We also will flip an ancilla qubit for the case that the identity is needed, corresponding to $b=B$.
This only needs one additional Toffoli, because we may iterate over $b$ first then iterate over $r$.
That is, for each value of $b$ in the unary iteration from $0$ to $B-1$, we use it to control an iteration over $r$.
The cost of the controlled iteration over $r$ is $R-1$, which is multiplied by $B$ for the number of value of $b$ where we iterate over $r$.
The total number of value of $b$ that we iterate over is $B+1$, with no iteration over $r$ needed for $b=B$.
The cost of the (uncontrolled) iteration over $B+1$ values of $b$ is $B-1$, for a total of
\begin{equation}
    B(R-1)+(B-1)=RB-1.
\end{equation}
Together with the $N$ cost for $\text{D}_1$ and $\text{Q}_1$, the cost is $N+RB$ as above, omitting the $-1$ for simplicity.

\textbf{Qubit cost:} The persistent ancilla cost is, for the output rotations
\begin{equation}
    (N-1)b_{\text{rot}} \, .
\end{equation}
We do not include another ancilla cost for $G,r$ output here as that was accounted for the preceding step.
There is a temporary ancilla cost of
\begin{equation}
    \lceil \log(N+RC) \rceil \, ,
\end{equation}
for the first unary iteration, and
\begin{equation}
    \lceil \log R \rceil  +\lceil \log (B+1) \rceil   \, ,
\end{equation}
for the second unary iteration.
The total temporary ancilla cost here is the maximum of these.
(Part of the temporary ancillas are used from the preceding step to save Toffolis.)

\item \label{step6} Perform the rotations for the select part of the block-encoding.
For $N$ orbitals with $b_{\text{rot}}$ bits of accuracy for the rotations, the Toffoli cost of the forward and reverse Givens rotations is
\begin{equation}
2(N-1)b_{\text{rot}} \, .
\end{equation}

\textbf{Qubit cost:} There is just a temporary ancilla cost of $b_{\text{rot}}$ for the additions into the phase gradient register.

\item \label{step7} The second QROM from \ref{step5} can be erased with phase fixup cost
\begin{equation}
R+B \, .
\end{equation}
This can be achieved by performing binary-to-unary conversion on one register, then unary iteration on the other register with controlled phases on the unary qubits.
There is a temporary ancilla cost of $\min(R,B)$, with the minimum corresponding to which register we perform the binary to unary conversion on.
The steps from here on have negligible temporary ancilla costs as compared to the earlier steps, so will be omitted from this discussion.

\item \label{step8} Invert the coherent alias sampling from \ref{step4}. Inverting the controlled swaps and inequality tests has cost $b_2$.
For the phase fixup, we note that typically $RC$ is larger than $B$, and also that the method of phase fixup from Ref.~\cite{berry2019qubitization} is to perform binary-to-unary conversion on one subset of qubits, and unary iteration on the other set of qubits (see Figure 6 of Ref.~\cite{berry2019qubitization}).
Here, we can perform binary-to-unary conversion on the register containing $B$, and unary iteration on the $x_\text{o}$ register.
That gives the phase fixup cost as
\begin{equation}
RC + B + 1 \, .
\end{equation}
We can also combine some of the qubits from $x_\text{o}$ with $B$ in this procedure to give a cost
\begin{equation}
\left\lceil \frac {RC}{2^{k_4}}\right\rceil + 2^{k_4}(B+1) \, .
\end{equation}
Similar to the procedure for the QROM, we can perform iteration over the final $N$ values, but here to perform the phase fixup.
This gives a total cost (with the inversion of swaps and inequality test)
\begin{equation}
b_2 + N + \left\lceil \frac {RC}{2^{k_4}}\right\rceil + 2^{k_4}(B+1) \, .
\end{equation}
In the case that $RC$ is divisible by some power of 2, we can use a similar phase fixup approach for $N$ to give cost
\begin{equation}
b_2 + \left\lceil \frac {N}{2^{k'_4}}\right\rceil + 2^{k'_4} + \left\lceil \frac {RC}{2^{k_4}}\right\rceil + 2^{k_4}(B+1) 
\end{equation}
where $RC$ is divisible by $2^{k'_4}$.

\item \label{step9} Invert the equal superposition from \ref{step3} with cost
\begin{equation}
4\lceil \log (B+1)\rceil
\end{equation}
Note that all these costs from \ref{step3} to \ref{step9} are multiplied by two for the squared operator.

\item \label{step10} Invert the coherent alias sampling from \ref{step2} with cost
\begin{equation}
b_1 + \left\lceil \frac{N+RC}{2^{k_5}}\right\rceil + 2^{k_5} \, ,
\end{equation}
where $b_1$ is for inverting the controlled swaps and inequality test.

\item \label{step11} Invert the equal superposition preparation from \ref{step1} with cost again
\begin{equation}
  4\lceil \log(N+RC)\rceil \, .
\end{equation}
\end{enumerate}
Further minor savings can be made by retaining ancillas to eliminate the non-Clifford cost of inverting the inequality test and controlled swaps for the coherent alias sampling.
That would remove the $b_2$ cost from step \ref{step8} and $b_1$ cost from step \ref{step10}, at the expense of retaining the same number of ancillas.

Note that we order the values so that the alternatives for $G_{SF}$ start at $x_\text{o}=0$.
This ensures that we do not use the temporary qubits for QROM at the same time as the output angles.
In important cases we can also obtain $r,c$ from the appropriate subset of bits.
If $C$ is a power of 2 (like 2 or 4), then using $x_\text{o}=rC+c$ this works, and
similarly, if $r$ is a power of 2 and we use $x_\text{o}=cR+r$.

In addition to the block-encoding costs above, a reflection is used to provide the complete qubitised operator.
The registers that need be reflected upon are as follows.
\begin{enumerate}[(i)]
    \item Registers of size $\lceil \log(N+RC)\rceil$ and $\lceil \log(B+1)\rceil$ used for $x_\text{o}$ and $b$, respectively.
    \item The $b_{k1}$ and $b_{k2}$ qubits used for the equal superposition states (that we preform the inequality test with) in the two steps of coherent alias sampling.
    \item Two qubits that are rotated in the preparation in the equal superposition states for $x_\text{o}$ and $b$.
    \item Two qubits for the spin (with one for $\text{SF}$ in the inner sum, and one for $\text{D}_1$ and $\text{Q}_1$ in the outer sum).
\end{enumerate}
The reflection on these qubits needs to be controlled by a single qubit for the phase estimation, and each step there is another Toffoli used for the unary iteration on the control register for phase estimation.
That gives a Toffoli cost in each step of
\begin{equation}
    \lceil \log(N+RC)\rceil + \lceil \log(B+1)\rceil + b_{k1}+b_{k2} + 2 \, .
\end{equation}

\section{Robust selection of DFTHC solutions}\label{sec:robust_selection}
Finding the rank $R^*$ of the best DFTHC approximation is not straightforward as $\epsilon_\text{corr}$ fluctuates significantly and does not improve monotonically with rank.
For instance, best solutions selected by thresholding on the lowest cost with $|\epsilon_\text{corr}|\le0.3$mHa in~\cref{tab:compare_rbc_table_1} and~\cref{tab:compare_rbc_table_2} show no clear pattern of what choice of hyperparameters $B,C$ are best.
Similar to DF and THC, these fluctuations arise as the DFTHC program~\cref{eq:DFTHC_program} is nonlinear and the final solution is sensitive to hyperparameters and the initial values of, say, $u,w$.
However, we can obtain robust estimates of rank and mitigate the effect of fluctuations by averaging $\epsilon_\text{corr}$ over many samples for each $(R,B,C)$, rather than taking the best value.
This is accomplished by converting $\epsilon_\text{corr}$ to a random variable by initializing the DFTHC solver with randomly selected parameters values $\mathcal{P}$ for each sample as described in~\cref{sec:DFTHC_optimization}.
Hence, each $\epsilon_\text{corr}$ is drawn from a distribution with mean $\bar{\epsilon}_\text{corr}$ and variance $\sigma^2_\text{corr}$ and in~\cref{fig:robust_selection1}, we select $R_{\le\epsilon_\text{th}}$ according to the criteria $\sigma_\text{corr}\le\epsilon_\text{th}$ and $\bar{\epsilon}_\text{corr}\ll \epsilon_\text{th}$.
Although this leads to a significantly larger rank than $R^{*}$, another  major advantage is that we may instead bound error contributions to $\sigma_\text{E}$ differently like
\begin{align}\label{eq:squared_error}
\sigma_E\le\sqrt{\sigma_\text{PEA}^2+\sigma_\text{corr}^2}+ |\bar{\epsilon}_\text{corr}| \le \epsilon_\text{chem}.
\end{align}
We find empirically that $|\bar{\epsilon}_\text{corr}|$ concentrates to zero, up to statistical accuracy.
Hence, this enables significantly larger choice of $\sigma_\text{PEA}=1.4$mHa, and $\sigma_\text{corr}\le 0.7$mHa.
The costs of simulation in~\cref{tab:dfthcblisssossa_robust_costs} under this criteria is not directly comparable to prior work, but its statistical robustness might prevent an unintentional bias of solutions towards good $\epsilon_\text{corr}$ but bad $\epsilon_\text{corr}^{*}$.

\begin{table}[H]
    \centering
    \begin{tabularx}{\textwidth}
    {|c|c|Y|Y|Y|Y|Y|Y|Y|Y|Y|Y|Y|Y|Y|}
    \hline
    &\multirow{2}{*}{$(R,B,C)$}&DF-like &\multicolumn{3}{c|}{$C=\lfloor N/4\rfloor$}&\multicolumn{3}{c|}{$C=\lfloor N/2\rfloor$}& THC-like
    \\
&&$(M,N,1)$ 
& $(M,\lfloor\frac{N}{2}\rfloor,\cdot)$ & $(M,\lfloor\frac{3N}{4}\rfloor,\cdot)$ & $(M,N,\cdot)$ 
& $(M,\lfloor\frac{N}{2}\rfloor,\cdot)$ & $(M,\lfloor\frac{3N}{4}\rfloor,\cdot)$ & $(M,N,\cdot)$ 
& $(1,M,M)$ 
\\
\hline\hline
\multirow{13}{*}{\rotatebox{90}{Fe$_2$S$_2$-20o30e}} & $M_{|\epsilon_{\text{corr}}|\le 0.3\text{mHa}}$ & 85 & \cellcolor{red}15 & 14 & 13 & \cellcolor{gray}14 & 11 & 12 & 80 \\
 & $\epsilon_{\text{corr}}$/mHa & 0.016 & \cellcolor{red}0.938 & -0.019 & 0.253 & \cellcolor{gray}0.288 & 0.094 & 0.187 & 0.073 \\
 & $\epsilon_{\text{in}}$/mHa & 3.0 & \cellcolor{red}43.9 & 22.1 & 18.7 & \cellcolor{gray}37.4 & 20.7 & 8.5 & 15.6 \\
 & $\lambda$/Ha & 15.9 & \cellcolor{red}19.4 & 17.5 & 17.5 & \cellcolor{gray}19.8 & 19.5 & 16.3 & 16.1 \\
 & $E_{\text{gap}}$/Ha & 0.988 & \cellcolor{red}1.03 & 1.02 & 1.028 & \cellcolor{gray}0.977 & 0.992 & 0.984 & 1.002 \\
 & $\lambda_{\text{eff}}$/Ha & 5.52 & \cellcolor{red}6.24 & 5.89 & 5.9 & \cellcolor{gray}6.15 & 6.14 & 5.58 & 5.58 \\
\cline{2-10}
 & $(b_\text{coeff},b_\text{rot})$ & $(11,15)$ & \cellcolor{red}$(11,15)$ & $(11,15)$ & $(11,15)$ & \cellcolor{gray}$(11,15)$ & $(11,15)$ & $(11,15)$ & $(11,15)$ \\
 & ${\text{Rot}}$ Toffolis & 6800 & \cellcolor{red}600 & 840 & 1040 & \cellcolor{gray}560 & 660 & 960 & 320 \\
 & ${\text{Coeff}}$ Toffolis & 814 & \cellcolor{red}530 & 622 & 706 & \cellcolor{gray}734 & 786 & 972 & 1670 \\
 & $\textsc{Select}$ Toffolis & 2280 & \cellcolor{red}2280 & 2280 & 2280 & \cellcolor{gray}2280 & 2280 & 2280 & 2280 \\
\cline{2-10}
 & Total Toffolis & 9894 & \cellcolor{red}3410 & 3742 & 4026 & \cellcolor{gray}3574 & 3726 & 4212 & 4270 \\
 & $\lambda\times$ total Toffolis & $1.57 \times 10^{5}$ & \cellcolor{red}$6.62 \times 10^{4}$ & $6.55 \times 10^{4}$ & $7.05 \times 10^{4}$ & \cellcolor{gray}$7.08 \times 10^{4}$ & $7.27 \times 10^{4}$ & $6.87 \times 10^{4}$ & $6.87 \times 10^{4}$ \\
 & $\lambda_\text{eff}\times$ total Toffolis & $5.46 \times 10^{4}$ & \cellcolor{red}$2.13 \times 10^{4}$ & $2.20 \times 10^{4}$ & $2.38 \times 10^{4}$ & \cellcolor{gray}$2.20 \times 10^{4}$ & $2.29 \times 10^{4}$ & $2.35 \times 10^{4}$ & $2.39 \times 10^{4}$ \\
\hline\hline

\multirow{10}{*}{\rotatebox{90}{Fe$_4$S$_4$-36o54e}} & $M_{|\epsilon_{\text{corr}}|\le 0.3\text{mHa}}$ & 105 & 23 & 17 & 15 & \cellcolor{gray}9 & 12 & 12 & 175 \\
 & $\epsilon_{\text{corr}}$/mHa & -0.289 & 0.104 & -0.297 & -0.276 & \cellcolor{gray}-0.004 & 0.198 & -0.092 & -0.188 \\
 & $\epsilon_{\text{in}}$/mHa & 38.8 & 38.3 & 33.1 & 29.4 & \cellcolor{gray}166.9 & 43.7 & 23.9 & 20.5 \\
 & $\lambda$/Ha & 39.6 & 45.1 & 49.6 & 47.2 & \cellcolor{gray}49.8 & 46.2 & 44.3 & 40.4 \\
 & $E_{\text{gap}}$/Ha & 2.119 & 2.082 & 1.967 & 1.977 & \cellcolor{gray}2.099 & 1.926 & 1.902 & 1.946 \\
 & $\lambda_{\text{eff}}$/Ha & 12.78 & 13.55 & 13.84 & 13.52 & \cellcolor{gray}14.31 & 13.2 & 12.85 & 12.39 \\
\cline{2-10}
 & $(b_\text{coeff},b_\text{rot})$ & $(13,17)$ & $(13,17)$ & $(13,17)$ & $(13,17)$ & \cellcolor{gray}$(13,17)$ & $(13,17)$ & $(13,17)$ & $(13,17)$ \\
 & ${\text{Rot}}$ Toffolis & 15120 & 1656 & 1836 & 2160 & \cellcolor{gray}648 & 1296 & 1728 & 700 \\
 & ${\text{Coeff}}$ Toffolis & 1298 & 1280 & 1336 & 1478 & \cellcolor{gray}1126 & 1594 & 1878 & 4002 \\
 & $\textsc{Select}$ Toffolis & 4760 & 4760 & 4760 & 4760 & \cellcolor{gray}4760 & 4760 & 4760 & 4760 \\
\cline{2-10}
 & Total Toffolis & 21178 & 7696 & 7932 & 8398 & \cellcolor{gray}6534 & 7650 & 8366 & 9462 \\
 & $\lambda\times$ total Toffolis & $8.39 \times 10^{5}$ & $3.47 \times 10^{5}$ & $3.93 \times 10^{5}$ & $3.96 \times 10^{5}$ & \cellcolor{gray}$3.25 \times 10^{5}$ & $3.53 \times 10^{5}$ & $3.71 \times 10^{5}$ & $3.82 \times 10^{5}$ \\
 & $\lambda_\text{eff}\times$ total Toffolis & $2.71 \times 10^{5}$ & $1.04 \times 10^{5}$ & $1.10 \times 10^{5}$ & $1.14 \times 10^{5}$ & \cellcolor{gray}$9.35 \times 10^{4}$ & $1.01 \times 10^{5}$ & $1.07 \times 10^{5}$ & $1.17 \times 10^{5}$ \\
\hline\hline

\multirow{10}{*}{\rotatebox{90}{FeMoCo-54o54e}} & $M_{|\epsilon_{\text{corr}}|\le 0.3\text{mHa}}$ & 190 & \cellcolor{gray}10 & 7 & 13 & 10 & 8 & 6 & 200 \\
 & $\epsilon_{\text{corr}}$/mHa & 0.192 & \cellcolor{gray}0.217 & 0.11 & -0.214 & -0.014 & -0.166 & 0.219 & -0.094 \\
 & $\epsilon_{\text{in}}$/mHa & 36.2 & \cellcolor{gray}279.4 & 243.7 & 80.1 & 238.3 & 142.3 & 136.1 & 121.5 \\
 & $\lambda$/Ha & 57.9 & \cellcolor{gray}60.7 & 61.9 & 65.2 & 58.3 & 61.2 & 65.3 & 63.6 \\
 & $E_{\text{gap}}$/Ha & 3.567 & \cellcolor{gray}3.772 & 4.789 & 3.609 & 3.691 & 3.617 & 3.626 & 3.607 \\
 & $\lambda_{\text{eff}}$/Ha & 20.01 & \cellcolor{gray}21.06 & 23.88 & 21.39 & 20.42 & 20.73 & 21.46 & 21.11 \\
\cline{2-10}
 & $(b_\text{coeff},b_\text{rot})$ & $(9,16)$ & \cellcolor{gray}$(9,16)$ & $(9,16)$ & $(9,16)$ & $(9,16)$ & $(9,16)$ & $(9,16)$ & $(9,16)$ \\
 & ${\text{Rot}}$ Toffolis & 41040 & \cellcolor{gray}1080 & 1120 & 2808 & 1080 & 1280 & 1296 & 800 \\
 & ${\text{Coeff}}$ Toffolis & 1962 & \cellcolor{gray}1116 & 1162 & 1848 & 1620 & 1804 & 1810 & 4074 \\
 & $\textsc{Select}$ Toffolis & 6784 & \cellcolor{gray}6784 & 6784 & 6784 & 6784 & 6784 & 6784 & 6784 \\
\cline{2-10}
 & Total Toffolis & 49786 & \cellcolor{gray}8980 & 9066 & 11440 & 9484 & 9868 & 9890 & 11658 \\
 & $\lambda\times$ total Toffolis & $2.88 \times 10^{6}$ & \cellcolor{gray}$5.45 \times 10^{5}$ & $5.61 \times 10^{5}$ & $7.46 \times 10^{5}$ & $5.53 \times 10^{5}$ & $6.04 \times 10^{5}$ & $6.46 \times 10^{5}$ & $7.41 \times 10^{5}$ \\
 & $\lambda_\text{eff}\times$ total Toffolis & $9.96 \times 10^{5}$ & \cellcolor{gray}$1.89 \times 10^{5}$ & $2.16 \times 10^{5}$ & $2.45 \times 10^{5}$ & $1.94 \times 10^{5}$ & $2.05 \times 10^{5}$ & $2.12 \times 10^{5}$ & $2.46 \times 10^{5}$ \\
\hline\hline

\multirow{10}{*}{\rotatebox{90}{FeMoCo-76o113e}} & $M_{|\epsilon_{\text{corr}}|\le 0.3\text{mHa}}$ & 130 & 27 & 15 & \cellcolor{red}18 & 16 & 18 & 7 & \cellcolor{gray}220 \\
 & $\epsilon_{\text{corr}}$/mHa & 0.257 & 0.063 & 0.134 & \cellcolor{red}0.359 & 0.152 & 0.059 & -0.132 & \cellcolor{gray}-0.038 \\
 & $\epsilon_{\text{in}}$/mHa & 287.2 & 65.7 & 82.5 & \cellcolor{red}43.5 & 131.0 & 45.3 & 120.0 & \cellcolor{gray}234.1 \\
 & $\lambda$/Ha & 140.6 & 185.6 & 179.7 & \cellcolor{red}181.7 & 206.0 & 162.0 & 197.2 & \cellcolor{gray}144.7 \\
 & $E_{\text{gap}}$/Ha & 6.149 & 4.529 & 5.037 & \cellcolor{red}4.53 & 4.571 & 4.526 & 4.982 & \cellcolor{gray}5.335 \\
 & $\lambda_{\text{eff}}$/Ha & 41.13 & 40.75 & 42.25 & \cellcolor{red}40.32 & 43.15 & 38.03 & 44.05 & \cellcolor{gray}38.93 \\
\cline{2-10}
 & $(b_\text{coeff},b_\text{rot})$ & $(9,15)$ & $(9,15)$ & $(9,15)$ & \cellcolor{red}$(9,15)$ & $(9,15)$ & $(9,15)$ & $(9,15)$ & \cellcolor{gray}$(9,15)$ \\
 & ${\text{Rot}}$ Toffolis & 39520 & 4104 & 3420 & \cellcolor{red}5472 & 2432 & 4104 & 2128 & \cellcolor{gray}880 \\
 & ${\text{Coeff}}$ Toffolis & 1970 & 2726 & 2476 & \cellcolor{red}3214 & 2970 & 3852 & 2830 & \cellcolor{gray}4484 \\
 & $\textsc{Select}$ Toffolis & 9000 & 9000 & 9000 & \cellcolor{red}9000 & 9000 & 9000 & 9000 & \cellcolor{gray}9000 \\
\cline{2-10}
 & Total Toffolis & 50490 & 15830 & 14896 & \cellcolor{red}17686 & 14402 & 16956 & 13958 & \cellcolor{gray}14364 \\
 & $\lambda\times$ total Toffolis & $7.10 \times 10^{6}$ & $2.94 \times 10^{6}$ & $2.68 \times 10^{6}$ & \cellcolor{red}$3.21 \times 10^{6}$ & $2.97 \times 10^{6}$ & $2.75 \times 10^{6}$ & $2.75 \times 10^{6}$ & \cellcolor{gray}$2.08 \times 10^{6}$ \\
 & $\lambda_\text{eff}\times$ total Toffolis & $2.08 \times 10^{6}$ & $6.45 \times 10^{5}$ & $6.29 \times 10^{5}$ & \cellcolor{red}$7.13 \times 10^{5}$ & $6.22 \times 10^{5}$ & $6.45 \times 10^{5}$ & $6.15 \times 10^{5}$ & \cellcolor{gray}$5.59 \times 10^{5}$ \\
\hline\hline

\end{tabularx}
\caption{\label{tab:compare_rbc_table_1}Best solutions found satisfying $|\epsilon_\text{corr}|\le 0.3$mHa. Highlighted columns (gray) minimize total Toffolis in the last row and (red) describes the best solution found if the target $\epsilon_\text{corr}$ was not attained.
$O_\text{Rot}$ and $O_\text{Coeff}$ Toffolis count the cost of computing and uncomputing twice.
Total Toffolis is the dominant costs in block-encoding $\textsc{Be}[H_\text{SA}/\Lambda-1]$, corresponding to the sum of $\textsc{Rot}$ Toffolis, $\textsc{Coeff}$ Toffolis and $8(N-1)b_\text{rot}$ from the dominant cost of two select unitaries.}
\end{table}

\begin{table}
    \centering\ContinuedFloat
    \begin{tabularx}{\textwidth}
    {|c|c|Y|Y|Y|Y|Y|Y|Y|Y|Y|Y|Y|Y|Y|}
    \hline
    &\multirow{2}{*}{$(R,B,C)$}&DF-like &\multicolumn{3}{c|}{$C=\lfloor N/4\rfloor$}&\multicolumn{3}{c|}{$C=\lfloor N/2\rfloor$}& THC-like
    \\
&&$(M,N,1)$ 
& $(M,\lfloor\frac{N}{2}\rfloor,\cdot)$ & $(M,\lfloor\frac{3N}{4}\rfloor,\cdot)$ & $(M,N,\cdot)$ 
& $(M,\lfloor\frac{N}{2}\rfloor,\cdot)$ & $(M,\lfloor\frac{3N}{4}\rfloor,\cdot)$ & $(M,N,\cdot)$ 
& $(1,M,M)$ 
\\\hline\hline

\multirow{10}{*}{\rotatebox{90}{XVIII-56o64e}} & $M_{|\epsilon_{\text{corr}}|\le 0.3\text{mHa}}$ & 110 & 8 & 6 & 5 & \cellcolor{gray}5 & 5 & 4 & 120 \\
 & $\epsilon_{\text{corr}}$/mHa & -0.253 & 0.013 & 0.121 & 0.222 & \cellcolor{gray}0.27 & -0.083 & -0.109 & -0.181 \\
 & $\epsilon_{\text{in}}$/mHa & 62.0 & 284.2 & 221.0 & 191.7 & \cellcolor{gray}418.2 & 243.3 & 184.7 & 237.3 \\
 & $\lambda$/Ha & 58.1 & 59.2 & 61.0 & 65.1 & \cellcolor{gray}55.5 & 59.9 & 62.8 & 59.7 \\
 & $E_{\text{gap}}$/Ha & 2.34 & 2.425 & 2.58 & 2.681 & \cellcolor{gray}2.569 & 2.299 & 2.294 & 2.554 \\
 & $\lambda_{\text{eff}}$/Ha & 16.33 & 16.77 & 17.55 & 18.48 & \cellcolor{gray}16.69 & 16.44 & 16.82 & 17.28 \\
\cline{2-10}
 & $(b_\text{coeff},b_\text{rot})$ & $(7,12)$ & $(7,12)$ & $(7,12)$ & $(7,12)$ & \cellcolor{gray}$(7,12)$ & $(7,12)$ & $(7,12)$ & $(7,12)$ \\
 & ${\text{Rot}}$ Toffolis & 24640 & 896 & 1008 & 1120 & \cellcolor{gray}560 & 840 & 896 & 480 \\
 & ${\text{Coeff}}$ Toffolis & 1432 & 992 & 1082 & 1138 & \cellcolor{gray}1114 & 1404 & 1446 & 2256 \\
 & $\textsc{Select}$ Toffolis & 5280 & 5280 & 5280 & 5280 & \cellcolor{gray}5280 & 5280 & 5280 & 5280 \\
\cline{2-10}
 & Total Toffolis & 31352 & 7168 & 7370 & 7538 & \cellcolor{gray}6954 & 7524 & 7622 & 8016 \\
 & $\lambda\times$ total Toffolis & $1.82 \times 10^{6}$ & $4.24 \times 10^{5}$ & $4.50 \times 10^{5}$ & $4.91 \times 10^{5}$ & \cellcolor{gray}$3.86 \times 10^{5}$ & $4.51 \times 10^{5}$ & $4.79 \times 10^{5}$ & $4.79 \times 10^{5}$ \\
 & $\lambda_\text{eff}\times$ total Toffolis & $5.12 \times 10^{5}$ & $1.20 \times 10^{5}$ & $1.29 \times 10^{5}$ & $1.39 \times 10^{5}$ & \cellcolor{gray}$1.16 \times 10^{5}$ & $1.24 \times 10^{5}$ & $1.28 \times 10^{5}$ & $1.38 \times 10^{5}$ \\
\hline\hline

\multirow{10}{*}{\rotatebox{90}{Cpd1X-58o63e}} & $M_{|\epsilon_{\text{corr}}|\le 0.3\text{mHa}}$ & 200 & \cellcolor{gray}9 & 6 & 5 & 8 & 6 & 4 & 120 \\
 & $\epsilon_{\text{corr}}$/mHa & -0.214 & \cellcolor{gray}0.277 & -0.19 & -0.019 & 0.098 & -0.206 & -0.223 & -0.183 \\
 & $\epsilon_{\text{in}}$/mHa & 16.0 & \cellcolor{gray}308.3 & 359.6 & 412.0 & 341.6 & 195.0 & 218.9 & 288.1 \\
 & $\lambda$/Ha & 96.2 & \cellcolor{gray}97.4 & 101.4 & 99.5 & 90.6 & 99.4 & 106.0 & 93.1 \\
 & $E_{\text{gap}}$/Ha & 4.815 & \cellcolor{gray}5.243 & 5.677 & 6.324 & 5.467 & 5.039 & 5.207 & 5.559 \\
 & $\lambda_{\text{eff}}$/Ha & 30.05 & \cellcolor{gray}31.53 & 33.44 & 34.9 & 30.99 & 31.25 & 32.82 & 31.69 \\
\cline{2-10}
 & $(b_\text{coeff},b_\text{rot})$ & $(10,15)$ & \cellcolor{gray}$(10,15)$ & $(10,15)$ & $(10,15)$ & $(10,15)$ & $(10,15)$ & $(10,15)$ & $(10,15)$ \\
 & ${\text{Rot}}$ Toffolis & 46400 & \cellcolor{gray}1044 & 1032 & 1160 & 928 & 1032 & 928 & 480 \\
 & ${\text{Coeff}}$ Toffolis & 2142 & \cellcolor{gray}1168 & 1184 & 1254 & 1596 & 1716 & 1622 & 2434 \\
 & $\textsc{Select}$ Toffolis & 6840 & \cellcolor{gray}6840 & 6840 & 6840 & 6840 & 6840 & 6840 & 6840 \\
\cline{2-10}
 & Total Toffolis & 55382 & \cellcolor{gray}9052 & 9056 & 9254 & 9364 & 9588 & 9390 & 9754 \\
 & $\lambda\times$ total Toffolis & $5.33 \times 10^{6}$ & \cellcolor{gray}$8.82 \times 10^{5}$ & $9.18 \times 10^{5}$ & $9.21 \times 10^{5}$ & $8.48 \times 10^{5}$ & $9.53 \times 10^{5}$ & $9.95 \times 10^{5}$ & $9.08 \times 10^{5}$ \\
 & $\lambda_\text{eff}\times$ total Toffolis & $1.66 \times 10^{6}$ & \cellcolor{gray}$2.85 \times 10^{5}$ & $3.03 \times 10^{5}$ & $3.23 \times 10^{5}$ & $2.90 \times 10^{5}$ & $3.00 \times 10^{5}$ & $3.08 \times 10^{5}$ & $3.09 \times 10^{5}$ \\
\hline\hline
\end{tabularx}
\caption{\label{tab:compare_rbc_table_2}(Continued).}
\end{table}

\begin{figure}
\centering
\begin{tabularx}{\textwidth}{lY}
    a) & \vspace{-0.5cm}\\
    &
\includegraphics[width=0.9\textwidth]{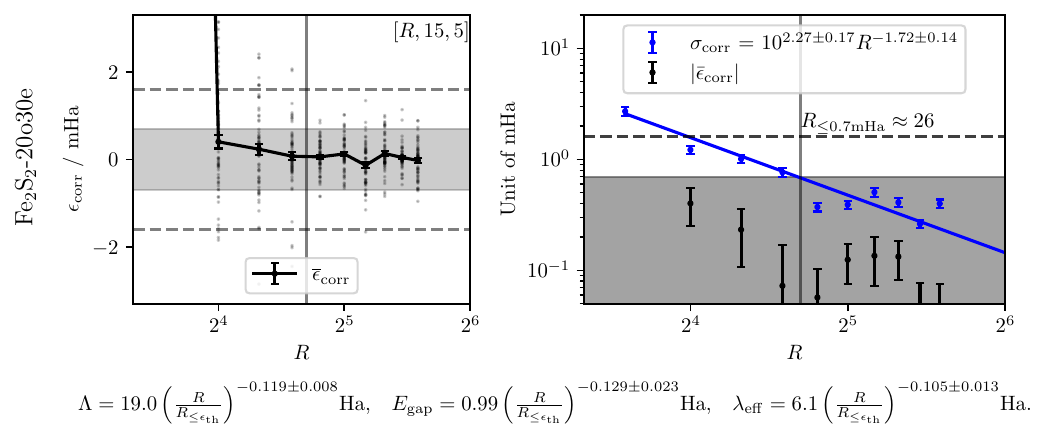}\\
b) & \vspace{-0.5cm}\\
&
\includegraphics[width=0.9\textwidth]{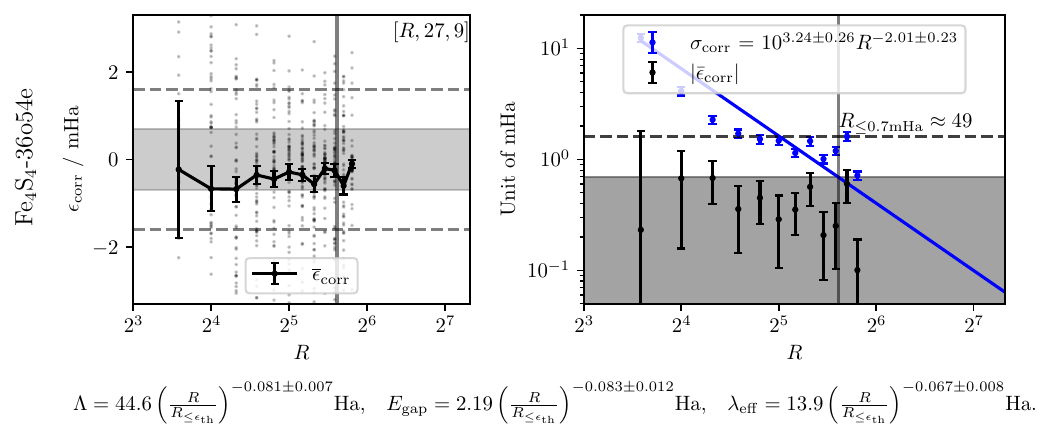}.\\
c) &\vspace{-0.5cm}\\
&
\includegraphics[width=0.9\textwidth]{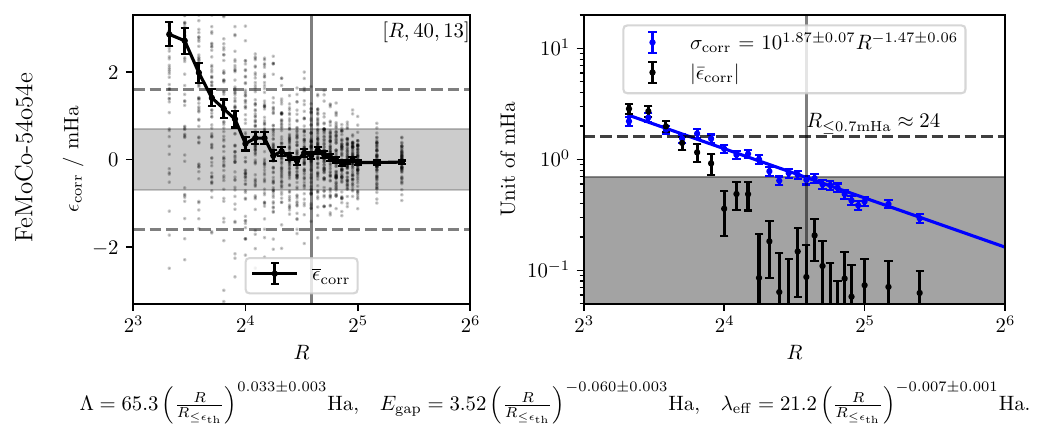}
\end{tabularx}
\caption{For molecule a) Fe$_2$S$_2$, b) Fe$_4$S$_4$, c) FeMoCo-54, d) FeMoCo-76, e) Cpd1X-58, and f) XVIII-56: (left) CCSD(T) correlation energy error vs.\ rank of DFTHC truncation (small dots). Error bars are the standard deviation of the mean $\bar{\epsilon}_\text{corr}$ estimated with $64$ samples. Dashed lines at $\pm 1.6$mHa are at the chemical accuracy threshold. (right) $\bar{\epsilon}_\text{corr}$ and estimated standard deviation $\sigma_\text{corr}$ of $\epsilon_\text{corr}$. The shaded region is the threshold $\epsilon_\text{th}=0.7$mHa used to select the rank of good solution. (below) Additional fits of other parameters relevant to estimation quantum simulation costs.\label{fig:robust_selection1}}
\end{figure}

\begin{figure}
\centering\ContinuedFloat
\begin{tabularx}{\textwidth}{lY}
    d) & \vspace{-0.5cm}\\
    &
\includegraphics[width=0.9\textwidth]{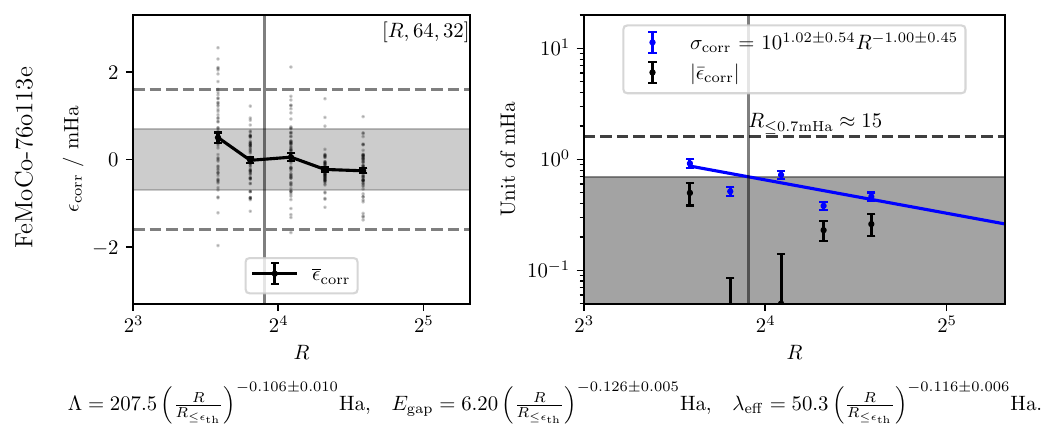}\\
e) & \\ &
\includegraphics[width=0.9\textwidth]{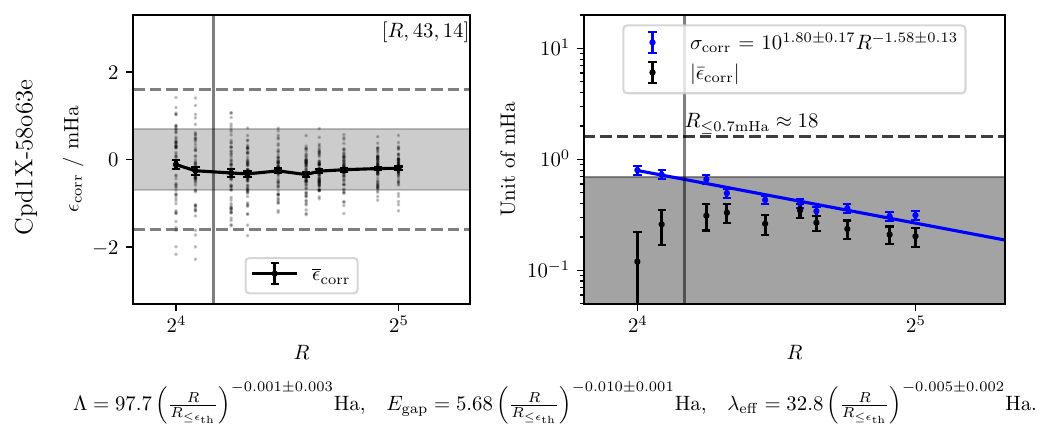} \\
f) & \\ &
\includegraphics[width=0.9\textwidth]{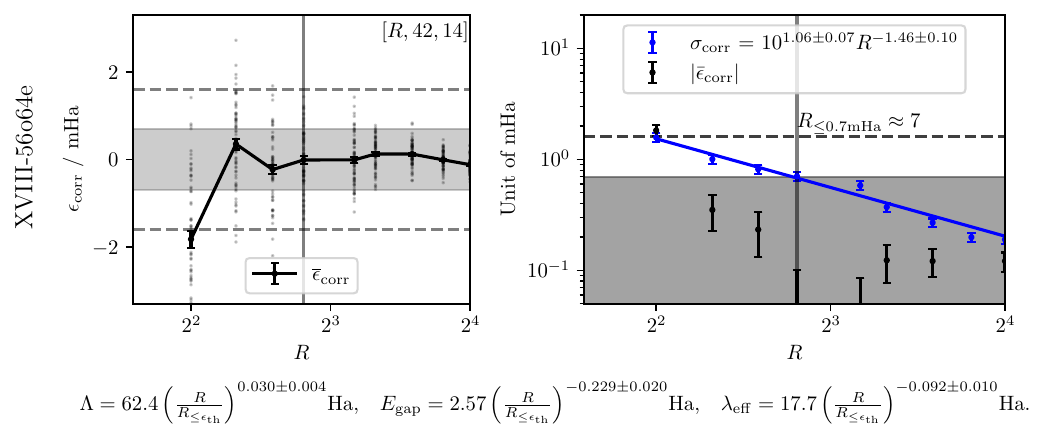}
\end{tabularx}
\caption{(continued).}
\end{figure}

\begin{table}
    \centering 
    \begin{tabularx}{\textwidth}{|c|Y|Y|Y|Y|Y|Y|}
    \hline \hline
                                   &  Fe$_2$S$_2$-20              & Fe$_4$S$_4$-36                &  FeMoco-54                                  &  FeMoco-76                       &  p450CPD1X-58  & CO$_2$[XVIII]-56   \\
     \hline \hline           
     $(R,B,C)$                   &$(26,15,5)$       &$(49,27,9)$       &$(24,40,13)$       &$(15,64,32)$      & $(18,43,14)$      &$(7,42,14)$        \\ 
     $\epsilon_\text{corr}$/mHa &0.6809               &0.6893               & 0.6865               & 0.6966              & 0.6597               & 0.6783                \\
     $\epsilon_{\text{fro}}$/mHa&5.7744               &4.3616               &42.8963               &48.9071              &49.6692              &181.6920              \\
     $(b_{\text{coeff}}, b_{\text{rot}})$ & (12, 16) & (14, 18) & (10, 17) & (11, 16) & (11, 17) & (8, 13) \\
     $E_\text{gap}$/Ha          &0.9924               & 2.1867              & 3.5156             & 6.1981             & 5.6782              & 2.5696              \\
     $\lambda_{\text{eff}}$/Ha  &6.1604               &14.0722              &21.4486             &50.7167             &33.3112              &17.9276            \\
     $C_{\text{walk}}$          & 4814                & 11493                & 12930               & 17588               & 12897                 & 8554                \\
      Qubits                    & 499                 & 926                 &  1201              & 1560                & 1285                & 989                 \\
      Total Toffoli             & 3.33$\times 10^{7}$ & 1.81$\times 10^{8}$ & 3.11$\times 10^{8}$ & 1.00$\times 10^{9}$ & 4.82$\times 10^{8}$ & 1.72$\times 10^{8}$  \\
     \hline \hline
    \end{tabularx}
    \caption{
    Improvements in simulating electronic structure to chemical accuracy with DFTHC and spectral amplification compared to best prior double-factorization methods and tensor hypercontraction methods. Improvement factors are computed as the ratio of the product the LCU 1-norm and the block-encoding cost rounded to the nearest integer.  The source for each cost is provided as a citation reference in the $\lambda$-row. The number of calls to the block-encoding is computed as $\lceil \frac{\pi \lambda_\text{eff}}{2\sigma_\text{PEA}}\rceil $~\cite{BabbushPRX18}.  In line with previous work we take $\sigma_\text{PEA} = 1.4\text{mHa}$ and $\sigma_{\text{trunc}}=0.331$mHa.
    }
    \label{tab:dfthcblisssossa_robust_costs}
\end{table}

\section{Truncation to finite bits-of-precision}\label{sec:bits_of_precision_truncation}
\begin{figure}
    \centering
    \includegraphics[]{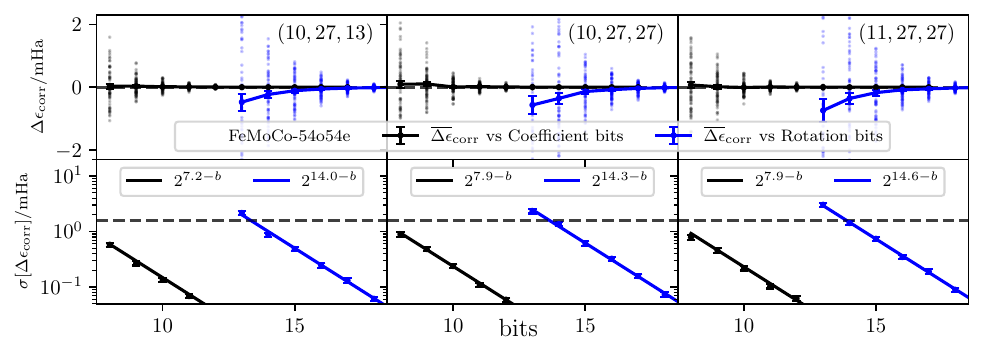}
    \vspace{-.5cm}
    \caption{\label{fig:FeMoCo_DFTHC_bits_scan}(top) Change in correlation energy due to randomized rounding of rotation and coefficients to $b_\text{rot}$ and $b_\text{coeff}$ bits respectively for selected DFTHC solutions with different $(R,B,C)$. The estimated mean change $\overline{\Delta\epsilon}_\text{corr}$ and the standard deviation of the mean are computed with $64$ samples for each bit setting. (bottom) Estimated standard deviation of $\Delta\epsilon_\text{corr}$ and best fit with respect to $b\in\{b_\text{rot},b_\text{coeff}\}$. Chemical accuracy at $1.6$ mHa is plotted for reference (dashed line).}
    \label{fig:enter-label}
\end{figure}

A final source of error $\sigma_\mathrm{trunc}$ arises from choosing a finite number of bits-of-precision $b_\text{rot}$ and $b_\text{coeff}$ to represent the $u$ and $w$ tensors in the block-encoding.
Although previous approaches used a deterministic rounding strategy~\cite{PRXQuantum.2.030305}, exploration over yet more hyperparameters $(b_\text{rot},b_\text{coeff},R,B,C)$ for each sample would introduce prohibitive cost.
We instead use an unbiased randomized rounding strategy which allows us to explore hyperparameters independently.
This randomization induces a distribution on $\epsilon_\mathrm{trunc}$ which allows us to use either of the bound
\begin{align}\label{eq:trunc_error}
\sigma_E&\le\sqrt{\sigma_\text{PEA}^2+\sigma_\text{trunc}^2}+ |\epsilon_\text{corr}| + |\bar{\epsilon}_\text{trunc}|\le \epsilon_\text{chem},
\\
\sigma_E&\le\sqrt{\sigma_\text{PEA}^2+\sigma_\text{corr}^2+\sigma_\text{trunc}^2}+ |\bar{\epsilon}_\text{corr}| + |\bar{\epsilon}_\text{trunc}|\le \epsilon_\text{chem}.
\end{align}
The change in correlation energy between the truncated and original DFTHC solutions by our randomized rounding strategy is presented in~\cref{fig:FeMoCo_DFTHC_bits_scan} for some FeMoCo-54 solution.
See~\cref{fig:bits_of_precision_1} for data on the other molecules.
We see that $\sigma_\text{trunc}$ scales like $O(2^{-b_\text{rot},-b_\text{coeff}})$ with bits-of-precision as presented, and $|\bar{\epsilon}_\text{trunc}|$ concentrates to zero rapidly, and so we assume it is exactly zero in the following for simplicity.
We assume that these solutions are typical for FeMoCo-54 to determine bits of precision required.
The linear error bound~\cref{eq:linear_error} leaves a budget of $\sigma_\mathrm{trunc}\le 0.830$mHa, which is achieved with $(b_\text{coeff},b_\text{rot})=(9,15)$, 
and the statistical bound~\cref{eq:squared_error} leave a budget of $\sigma_\mathrm{trunc}\le 0.331$mHa which is achieved with $(b_\text{coeff},b_\text{rot})=(10,17)$.

As discussed in the DFTHC block-encoding~\cref{fig:fullcircuit} each $\vec{u}^{(r)}_{b}$ is generated by a sequence of $N-1$ rotation angles $\vec{\theta}^{(r)}_b\in[0,2\pi)^{N-1}$ that realize the Majorana rotation from $\gamma_{000}$ to $\gamma_{\vec{u}00}$.
Discretizing these rotations angles to $b_\text{rot}$ bits of precision leads to a Toffoli cost of $(N-1)b_\text{bits}$.
For every angle $\theta\in[0,2\pi)$, we use the randomized rounding strategy in
~\cite{low2021halvingcostquantummultiplexed} by choosing the first $b_\text{rot}-1$ bits to be the integer $x=\lfloor(2^{b_\text{rot}-1})\theta/(2\pi)\rfloor$. 
The last bit is sampled randomly from the Bernouli distribution $\mathbb{B}(p)$ with probability $p=(2^{b_\text{rot}-1})\theta/(2\pi)\mod 1 $.
Then the rounded $\theta^{(b_\text{rot})}$ is sampled from the distribution $2\pi(x+\mathbb{B}(p))/2^{b_\text{rot}-1}$.
By a triangle inequality, each sampled vector generated by these truncated rotations deviates by at most $|\vec{u}^{(r)b_\text{rot}}_{b}-\vec{u}^{(r)}_{b}|=\mathcal{O}(N2^{-b_\text{rot}})$, but the expectation $\mathbb{E}[\|\vec{u}^{(r,b_\text{rot})}_{b}-\vec{u}^{(r)}_{b}\|]=\mathcal{O}(N4^{-b_\text{rot}})$ has effectively double the bits of precision.
A key difference from~\cite{low2021halvingcostquantummultiplexed} is that we sample each DFTHC sample by rounding once in the entire phase estimation algorithm instead of for every application of the block-encoding circuit.

The coefficients $w^{(rc)}_b$ are implemented by alias sampling~\cite{BabbushPRX18} to $b_\text{coeff}$ bits of precision.
As seen in the block-encoding of $O_{\text{SF},rc}$ in~\cref{eq:SOS_generator_SF}, for each component $(r,c)$, alias sampling prepares an approximation to the quantum state $\frac{1}{|w^{(rc)}|_1}\sum_{b}\sqrt{|w^{(rc)}_b|}\ket{b}\ket{\text{garb}_b}$.
To simplify notation, let us drop the $(r,c)$ superscript.
Then in essence, alias sampling on a dimension $B+1$ state discretizes the distribution $w$ into $M=(B+1)2^{b_\text{coeff}-1}$ bins, where the subtracted bit stores the sign of $w$. 
We now describe our randomized rounding strategy that assigns indices $b$ to each bins so that the fraction of bins marked with $b$ is $\frac{|w_b|}{|w|_1}$ in expectation.
First, for each $b$, assign $M_b=\lfloor M|w_b|/|w|_1\rfloor$ bins.
This can leave up to $K\le B$ unassigned bins.
Second, assign the remaining bins by sampling a Hamming weight $K$ bit-string according to the probability distribution $p_b\propto M|w_b|/|w|_1\mod{1}$ on $(B+1)$ bins.

We note that the block-encoding of one-body terms $H_{1}^\prime=\sum_{pq}h^{(1)'}_{pq}E_{pq}$ can essentially be treated as exact with minimal additional cost, which is not included in our later resources estimates.
First, observe that the weights $\sqrt{w_\pm^{(r)}}$ of one-body terms $O_{\mathrm{D}_1^\sigma,r},O_{\mathrm{Q}_1^\sigma,r}$ in~\cref{eq:SOS_generator_D1Q1} are prepared by outer $\text{Prep}$ alias sampling on only $N+RC$ elements, compared to inner $\text{Prep}$ alias sampling for $w^{(rc)}_b$ on $RC(B+1)$ elements.
Second, the bits of precision in outer $\text{Prep}$ can be chosen separately from that of inner $\text{Prep}$, and even doubling the bits of precision comes at a very small additive cost.
Hence, we may treat the coefficients $w_\pm^{(r)}$ and also $|w^{(rc)}|_1$ to be prepared with machine precision.
Third, the finite bits of precision $b_\text{rot}$ means that we block-encode an approximate one-body term
\begin{align}
H_{1}^{\prime(b_\text{rot})}&=\frac{1}{2}\sum_{s\in\{+,1\},r}sw^{(r)}_s\sum_{\sigma}(i\gamma_{\vec{u}^{(r)b_\text{rot}}_{\pm}\sigma0}\gamma_{\vec{u}^{(r)b_\text{rot}}_{\pm}\sigma1}+\mathbb{I})=\frac{1}{2}\sum_{pq}\sum_\sigma h^{(1)\prime(b_\text{rot})}_{pq}(i\gamma_{p\sigma0}\gamma_{q\sigma1}+\mathbb{I}),
\end{align}
with error $H_{1}^{\prime\prime} = H_1^{\prime}-H_{1}^{\prime(b_\text{rot})}=\frac{1}{2}\sum_{pq}\sum_\sigma h^{(1)\prime\prime}(i\gamma_{p\sigma0}\gamma_{q\sigma1}-1)$.
Fourth, the key idea is that we get a higher accuracy block-encoding of $H^{(1)\prime}$ by taking a linear combination of $H_{1}^{\prime(b_\text{rot})}$ and $H_{1}^{\prime\prime(b_\text{rot})}$.
This comes at a small additive cost of $N$ additional addresses to inner $\textsc{prep}$ and $\textsc{RPrep}$.
Using a triangle inequality and the fact $\|\gamma_{\vec{u}\sigma0}\|=1$, we see that this residual term is an exponentially small correction in the bits of rotation like
\begin{align}\nonumber
\|H_{1}^{\prime\prime}\|&=\frac{1}{2}\left\|\sum_{s\in\{+,1\},r}sw^{(r)}_s\sum_{\sigma}\left[
\left(\gamma_{\vec{u}^{(r)}_{\pm}\sigma0}
-\gamma_{\vec{u}^{(r)b_\text{rot}}_{\pm}\sigma0}\right)\gamma_{\vec{u}^{(r)}_{\pm}\sigma1}
+\gamma_{\vec{u}^{(r)b_\text{rot}}_{\pm}\sigma0}\left(\gamma_{\vec{u}^{(r)}_{\pm}\sigma1}-\gamma_{\vec{u}^{(r)b_\text{rot}}_{\pm}\sigma1}\right)
\right]\right\|
\\
&\le
2\sum_{s\in\{+,1\},r}|w^{(r)}_s|\|\gamma_{\vec{u}^{(r)}_{\pm}00}-\gamma_{\vec{u}^{(r)b_\text{rot}}_{\pm}00}\|
\le
2\sum_{s\in\{+,1\},r}|w^{(r)}_s||\vec{u}^{(r)}_{\pm}
-\vec{u}^{(r)b_\text{rot}}_{\pm}|
=\mathcal{O}(\|h^{(1)\prime}\|_1N2^{-b_\text{rot}}).
\end{align}
Hence, the change $\|h^{(1)\prime\prime}\|_1$ to block-encoding normalization is also small.
This process may be nested multiple times if the desired accuracy is still not achieved.

\begin{figure}
\centering
\begin{tabularx}{\textwidth}{lY}
a) & \\ &
\includegraphics[width=0.9\textwidth]{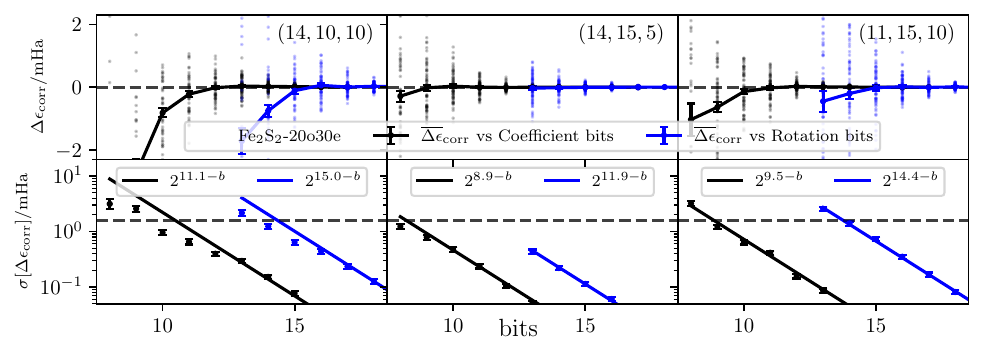}\\
b) & \\ & 
\includegraphics[width=0.9\textwidth]{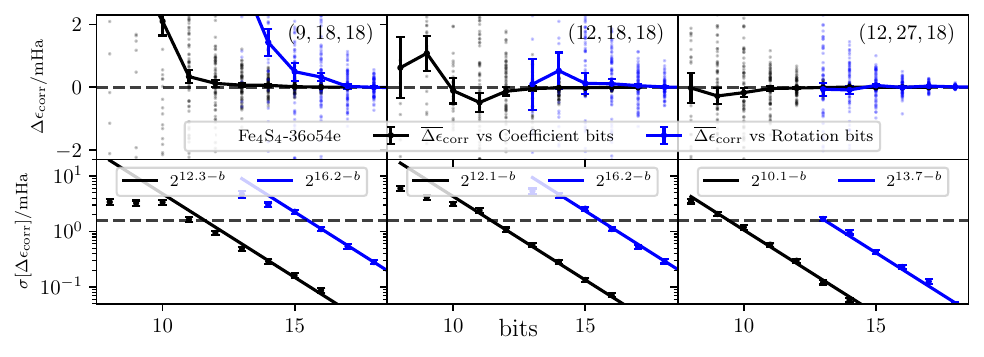}\\
c) & \\ &
\includegraphics[width=0.9\textwidth]{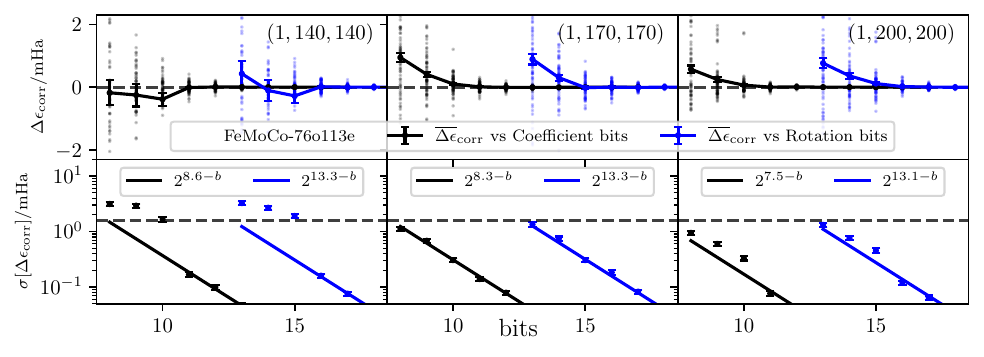}
\end{tabularx}
\caption{For molecule a) Fe$_2$S$_2$, b) Fe$_4$S$_4$, c) FeMoCo-76, d) Cpd1X-58, e) XVIII-56, f) XVIII-100 and XVIII-150: (top) change in correlation energy due to randomized rounding of rotation and coefficients to $b_\text{rot}$ and $b_\text{coeff}$ bits respectively for selected DFTHC solutions with different $(R,B,C)$. The estimated mean change $\overline{\Delta\epsilon}_\text{corr}$ and the standard deviation of the mean are computed with $64$ samples for each bit setting. (bottom) Estimated standard deviation of $\Delta\epsilon_\text{corr}$ and best fit with respect to $b\in\{b_\text{rot},b_\text{coeff}\}$. Chemical accuracy at $1.6$ mHa is plotted for reference (dashed line).\label{fig:bits_of_precision_1}}
\end{figure}

\begin{figure}
\centering\ContinuedFloat
\begin{tabularx}{\textwidth}{lY}
d) & \\ &
\includegraphics[width=0.9\textwidth]{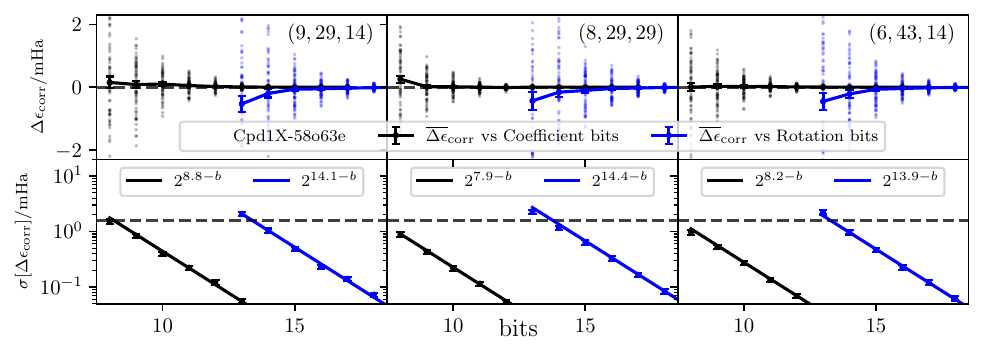}\\
e) & \\ &
\includegraphics[width=0.9\textwidth]{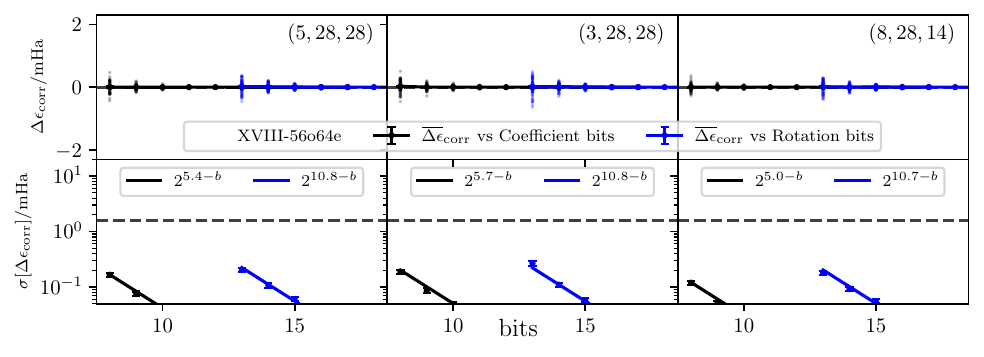}\\
f) & \\ &
\includegraphics[width=0.9\textwidth]{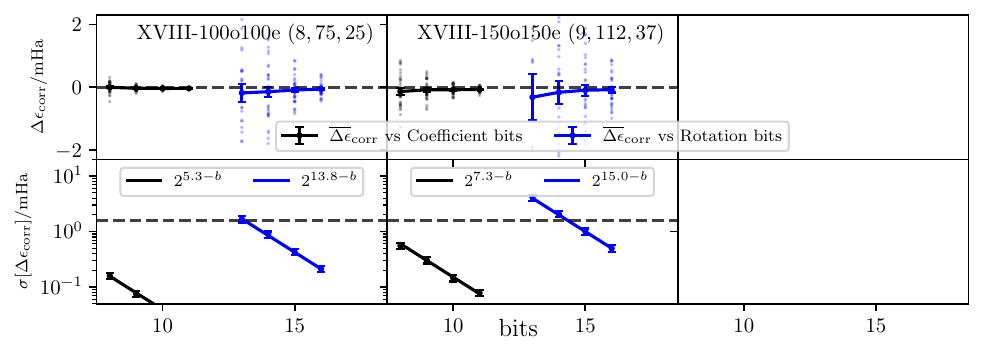}
\end{tabularx}
\caption{(continued).}
\end{figure}

\section{Estimating \texorpdfstring{$E_\text{gap}^*$}{Egap*} with DFTHC}\label{sec:E_gap_estimation}
For the large XVIII complexes on $100$ and $150$ orbitals, we obtain an estimate on the upper bound for the SOS gap $E_\text{gap}^{*}$ by using the DFTHC ansatz.
Instead of choosing the regularization parameter $E_\text{reg}$ in the DFTHC cost function~\cref{eq:DFTHC_program} to be $E_\text{gap}^{*}$, which is unavailable {\textit{a priori}} for these systems, we scan over the range of $E_\text{reg}$ presented in~\cref{tab:implementation_details}.
We compute the CCSD(T) energies $E_\text{gs}$ for DFTHC solutions obtained in this manner to estimate the actual gaps $E_\text{gap}$ as described in~\cref{sec:selection_good}, and find that all solutions have $E_\text{gap}>E_\text{gap}^*$, including those where $E_\text{reg}$ was chosen to be smaller than $E_\text{gap}^*$, such as for XVIII-56 and FeMoCo-76.
As $E_\text{gap}$ appears to be robust under DFTHC optimization and does not attain unphysical values, we estimate $E_\text{gap}^{*}$ as the smallest among all the found $E_\text{gap}$.
For this dataset, we chose the DFTHC bases and copies hyperparameters to be $B=\lfloor\frac{3N}{4}\rfloor$ and $C=\lfloor\frac{N}{4}\rfloor$ respectively, and tabulate in~\cref{tab:var_energies} the solutions found to have the best quantum simulation costs.
\begin{table}[H]
\small
    \centering
    \begin{tabularx}{\textwidth}
    {|c|Y|Y|Y|Y|Y|Y|Y|Y|}
    \hline\hline
    \multirow{3}{*}{$(R,B,C)=(\cdot,\lfloor\frac{3N}{4}\rfloor,\lfloor\frac{N}{4}\rfloor)$}
&XVIII & XVIII & XVIII & Fe$_2$S$_2$ & Fe$_4$S$_4$ & FeMoCo & FeMoCo & Cpd1X \\
&56o64e & 100o100e & 150o150e & 20o30e & 36o54e & 54o54e & 76o113e & 58o63e \\
&$(\cdot,42,14)$ & $(\cdot,75,25)$ & $(\cdot,112,37)$ & $(\cdot,15,5)$ & $(\cdot,27,9)$ & $(\cdot,40,13)$ & $(\cdot,57,19)$ & $(\cdot,43,14)$
\\\hline
$R_{|\epsilon_{\text{corr}}|\le 0.3\text{mHa}}$ & 9 & 8 & 9 & 17 & 16 & 7 & 15 & 6 \\
$\epsilon_{\text{corr}}$/mHa & -0.259 & -0.042 & -0.062 & -0.236 & 0.291 & 0.11 & 0.134 & -0.19 \\
$\epsilon_{\text{in}}$/mHa & 137.1 & 218.7 & 351.0 & 12.4 & 35.8 & 243.7 & 82.5 & 359.6 \\
$\Lambda$/Ha & 61.0 & 155.5 & 336.1 & 18.2 & 46.7 & 61.9 & 179.7 & 101.4 \\
$E_{\text{gap}}$/Ha & 2.222 & 4.565 & 6.454 & 1.066 & 2.19 & 4.789 & 5.037 & 5.677 \\
$\lambda_{\text{eff}}$/Ha & 16.47 & 37.68 & 65.87 & 6.23 & 14.3 & 24.36 & 42.55 & 33.92 \\
\hline
$(b_\text{coeff},b_\text{rot})$ & $(7,12)$ & $(8,16)$ & $(9,16)$ & $(11,15)$ & $(13,17)$ & $(9,16)$ & $(9,15)$ & $(10,15)$ \\
${\text{Rot}}$ Toffolis & 1512 & 3300 & 4032 & 1020 & 1728 & 1120 & 3420 & 1032 \\
${\text{Coeff}}$ Toffolis & 1330 & 2788 & 3846 & 688 & 1294 & 1162 & 2476 & 1184 \\
$\textsc{Select}$ Toffolis & 5280 & 12672 & 19072 & 2280 & 4760 & 6784 & 9000 & 6840 \\
\hline
Total Toffolis & 8122 & 18760 & 26950 & 3988 & 7782 & 9066 & 14896 & 9056 \\
$\lambda\times$ total Toffolis & $4.95 \times 10^{5}$ & $2.93 \times 10^{6}$ & $9.06 \times 10^{6}$ & $7.26 \times 10^{4}$ & $3.63 \times 10^{5}$ & $5.61 \times 10^{5}$ & $2.68 \times 10^{6}$ & $9.18 \times 10^{5}$ \\
$\lambda_\text{eff}\times$ total Toffolis & $9.46 \times 10^{4}$ & $4.63 \times 10^{5}$ & $1.26 \times 10^{6}$ & $1.76 \times 10^{4}$ & $7.87 \times 10^{4}$ & $1.56 \times 10^{5}$ & $4.48 \times 10^{5}$ & $2.17 \times 10^{5}$ \\
\hline\hline
\end{tabularx}
\includegraphics[width = \textwidth]
{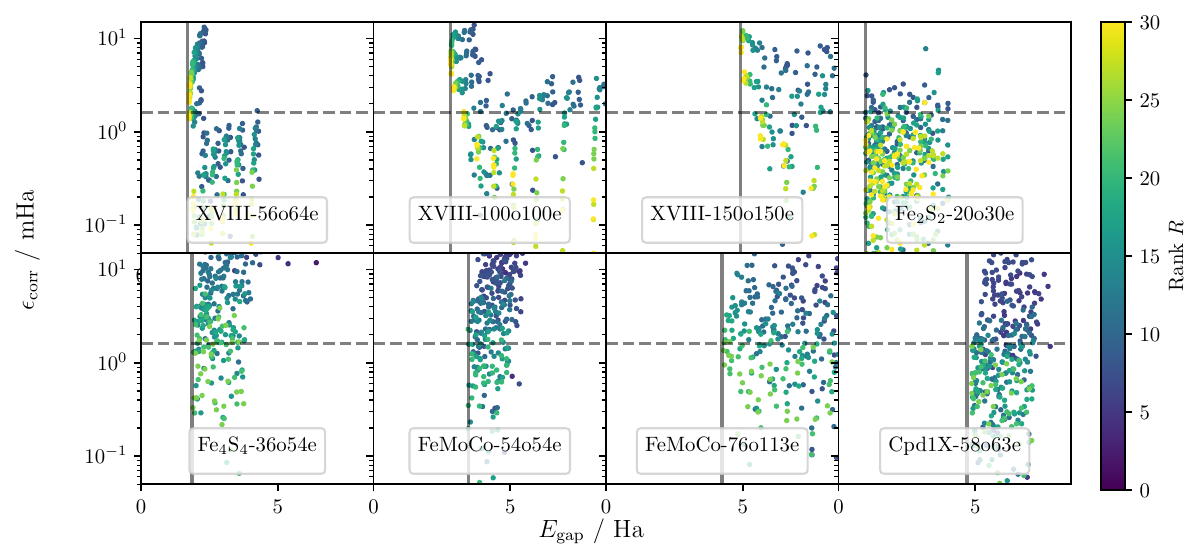}
\vspace{-0.5cm}
\caption{(top) Best DFTHC solutions found by scanning over $E_\text{reg}$ and $R$ and selecting those satisfying $|\epsilon_\text{corr}|\le 0.3$mHa. Three samples are generated for each choice ($E_\text{reg}$, $R$). 
(bottom) Plot of correlation energy error and computed $E_\text{gap}$ for all data in this scan. The dashed line is chemical accuracy $1.6$mHa and the solid line is $E_\text{gap}^*$ from~\cref{tab:var_energies} of the main text. For XVIII-100o and XVIII-150o, $E_\text{gap}^*$ is taken from the solution with the smallest $E_\text{gap}$.\label{tab:Gap_scan}}
\end{table}

\section{\texorpdfstring{SOS}{sos} Representation and moment method conversion}\label{sec:sos-dual-moment-conversion}
Herein we describe the protocol for constructing the SOS Hamiltonian from the solution of the typical polynomial optimization SDP. Given a semidefinite program of the form
\begin{align}\label{eq:sdp_primal}
\min \langle \boldsymbol{H}, \boldsymbol{X} \rangle \\
\mathrm{s.t.} \;\; \boldsymbol{X} \in \left(1, \boldsymbol{w} \right) \nonumber \\
\;\; \langle A_{0}, \boldsymbol{X} \rangle = 1  \nonumber \\
\;\; \langle A_{i}, \boldsymbol{X} \rangle = 0 \;\; i=1,...,m \nonumber\\
\;\; \boldsymbol{X} \in \mathcal{C} \nonumber
\end{align}
where $\boldsymbol{H}$ is a matrix representing an operator on the same spaces as $\boldsymbol{X}$, $\boldsymbol{w}$ is the primal variables, $A_{i}$ are any linear constraints, and $\boldsymbol{X} \in \mathcal{C}$ indicates that the primal variable is an element in the cone $\mathcal{C}$. In standard form all linear constraints equal zero except for the scalar augmenting the primal solutions.  The scalar in the primal allows all linear constraints to be expressed in standard form.  To derive the dual form we first write the Lagrangian
\begin{align}
L(\boldsymbol{X}, y, \boldsymbol{B}) = \langle H, \boldsymbol{X}\rangle  - \langle \boldsymbol{B}, \boldsymbol{X} \rangle + y_{0}\left(\langle A_{0}, \boldsymbol{X} \rangle - 1\right) + \sum_{i=1}^{m}y_{i}\langle A_{i}, \boldsymbol{X}\rangle
\end{align}
which for any given primal feasible $\boldsymbol{X}$ provides a lower bound to Eq.~\eqref{eq:sdp_primal}. Defining the dual function as the infimum over the primal variable and inverting the minimax the dual form of the primal problem is
\begin{align}\label{eq:sdp_dual}
\max -y_{0} \\
\mathrm{s.t.}\;\; \boldsymbol{H} = \boldsymbol{B} - y_{0}A_{0}  - \sum_{i=1}^{m}y_{i}A_{i} \nonumber \\
\;\; \boldsymbol{B} \in \mathcal{C}^{*} \nonumber
\end{align}
where $\mathcal{C}^{*}$ is the dual cone of $\mathcal{C}$. As a concrete example of both programs consider a $k$-local Hamiltonian $H$, the set of all $k$-local marginals $\mathcal{C}$ which implies $\boldsymbol{w}$ is a vector of matrices in $\mathbb{C}^{2^{k} \times 2^{k}}$ of length $\binom{n}{k}$.  The dual cone of the set of $k$-local marginals is defined as the set of all orthogonal operators
\begin{align}
\mathcal{C}^{*} = \left\{ \boldsymbol{B} \bigg| \langle \boldsymbol{B}_{i}, \boldsymbol{\rho}_{i} \rangle \geq 0  \;\; \forall i=1,...,\binom{n}{k}\;\;,\;\; \boldsymbol{\rho} \in \mathcal{C} \right\}.
\end{align}
Note that $B - \sum_{i=1}^{m}y_{i}A_{i} = H - y_{0}A_{0}$ and that $H - y_{0}A_{0} \mapsto \hat{H} - y_{0}\mathbb{I} \succeq 0$ by the fact that $y_{0}$ is a lower bound to the SDP. Here we use the hat ($\hat{.}$) to denote an operator. The linear constraints under the operator map $\sum_{i=1}^{m}y_{i}A_{i} \mapsto \sum_{i=1}^{m}y_{i}\hat{A_{i}}$ evaluate to zero except for linear constraints involving  the constant in the primal solution $\boldsymbol{w}_{1}$ or contraction maps. Thus defining $\tilde{B} = B - \sum_{\tilde{i}=1}^{\tilde{m}}y_{\tilde{i}}A_{\tilde{i}}$, where the tilde on the $i$ corresponds to the set of linear constraints involving $\boldsymbol{w}_{1}$ and contractions, defines an operator $\hat{\tilde{B}} \succeq 0$ and thus the dual SOS Hamiltonian. Care must be taken to construct $y_{\tilde{i}}\hat{A}_{\tilde{i}}$ as generically $A_{\tilde{i}} \mapsto \hat{A}_{\tilde{i}}$ may not be an SOS operator. The $H_{\text{sqrt}}$ is obtained by Cholesky decomposition on $\tilde{B}$.

\section{Spin-free formalism}\label{app:spin-free-sdp-def}
Here we derive the equality constraints of the semidefinite program to determine the optimal gap and $H_{\text{SOS}}$ representation using a spin-free algebra for the sum-of-squares generators. Consider a rank-2 spin-free (SF) particle-hole and hole-particle generators of the form
\begin{align}
\label{EQN:SPIN_FREE_G2_BASIS}
O_{\text{SF}} &=  \sum_{ij} \left ( g_{ij} \sum_{\sigma} a_{j\sigma}^\dagger a_{i\sigma}  + \bar{g}_{ij} \sum_{\sigma} a_{j\sigma} a^\dagger_{i\sigma} \right ) .
\end{align}
The resulting SOS is not flexible enough to describe the electronic Hamiltonian well, so we also introduce the following rank-1 generators
\begin{align}
\label{EQN:D1_Q1}
O_{D_\sigma} &=  \sum_{i}  d_{i\sigma} a_{i\sigma},  \\
O_{Q_\sigma} &=  \sum_{i}  q_{i\sigma} a^\dagger_{i\sigma},  \\
\end{align}
and the SOS Hamiltonian is
\begin{align}
\label{EQN:HSOS}
    \hat{H}_\text{SOS} & = O_{\text{SF}}^\dagger O_{\text{SF}} + \sum_\sigma \left ( O_{D_{\sigma}}^\dagger O_{D_{\sigma}} + O_{Q_{\sigma}}^\dagger O_{Q_{\sigma}}\right ),
\end{align}
with
\begin{align}
\label{EQN:SPIN_FREE_OPERATORS}
    O_{\text{SF}}^\dagger O_{\text{SF}}  &=  \sum_{i,j,k,l} \left ( G^{ik}_{lj}  \sum_{\sigma,\tau} \right . a^\dagger_{i\sigma}a_{k\sigma}a^\dagger_{j\tau}a_{l\tau} + {G^{\prime}}^{ik}_{jl}  \sum_{\sigma,\tau} a^\dagger_{i\sigma}a_{k\sigma}a_{l\tau} a^\dagger_{j\tau} \nonumber \\
    &\quad + {G^{\prime\prime}}^{ki}_{lj}  \sum_{\sigma,\tau} a_{k\sigma}a^\dagger_{i\sigma} a^\dagger_{j\tau}a_{l\tau} + {G^{\prime\prime\prime}}^{ki}_{jl}  \left . \sum_{\sigma,\tau} a_{k\sigma}a^\dagger_{i\sigma}a_{l\tau} a^\dagger_{j\tau} \right ) , \\
    O_{D_\sigma}^\dagger O_{D_\sigma} &= \sum_{i,j} D^{i\sigma}_{j\sigma} a^{\dagger}_{i\sigma} a_{j\sigma}\, , \\
    O_{Q_\sigma}^\dagger O_{Q_\sigma} &= \sum_{i,j} Q^{j\sigma}_{i\sigma} a_{j\sigma} a^\dagger_{i\sigma}\, .
\end{align}
The matrix representation of $O_{\text{SF}}^\dagger O_{\text{SF}}$ has four blocks, arranged as
\begin{align}
{\bf G_\text{SF}} = 
    \begin{pmatrix}
{\bf G} & {\bf G^\prime} \\
{\bf G^{\prime\prime}} & {\bf G^{\prime\prime\prime}} \\
    \end{pmatrix}.
\end{align}
After bringing the creation and annihilation operators in Eqs.~\ref{EQN:SPIN_FREE_OPERATORS} to a common order, we can see that 
\begin{align}
\hat{H}^{(2)}_\text{SOS} &= \sum_{i,j,k,l} (G^{ik}_{lj} - {G^{\prime}}^{ik}_{jl} - {G^{\prime\prime}}^{ki}_{lj} + {G^{\prime\prime\prime}}^{ki}_{jl} ) \sum_{\sigma,\tau} a^\dagger_{i\sigma} a_{k\sigma} a^\dagger_{j\tau} a_{l\tau}, \\
\hat{H}^{(1)}_\text{SOS} &= 2 \sum_{i,j} \sum_{\sigma} \left ( \sum_p \left ( {G^{\prime}}^{ij}_{pp} + {G^{\prime\prime}}^{pp}_{ji} - {G^{\prime\prime\prime}}^{pp}_{ij} - {G^{\prime\prime\prime}}^{ji}_{pp} \right ) + \frac{1}{2} \left ( D^{i\sigma}_{j\sigma} - Q^{j\sigma}_{i\sigma} \right )   \right )  a^\dagger_{i\sigma} a_{j\sigma}, \\
E_{\text{SOS}} &= 4 \sum_{pq} {G^{\prime\prime\prime}}^{pp}_{qq} + \sum_p \sum_\sigma Q^{p_\sigma}_{p_\sigma}\, .
\end{align}
Note that two-electron terms for a given $i$, $j$, $k$, and $l$ have the same coefficient regardless of the spins associated with the labels. The coefficients of $\hat{H}_{SOS}^{(2)}$ and $\hat{H}_{SOS}^{(1)}$ are then equated to the one- and two-electron integrals of the spin-free chemistry Hamiltonian.

\section{Correction for divergent derivative}\label{sec:spectrum_amplification_divergent_derivative}

In this appendix, we address the issue that the error propagation formula would predict an unreasonably large improvement arising from the divergent derivative of a function like $\sqrt{x}$ near zero.
It is possible to give a more accurate estimate of the uncertainty by deriving corrections to the error propagation formula.
To illustrate this method, we initially consider the estimation of $y=f(x)=\sqrt{x}$, with a normal distribution for the error in measuring $y$.
Then the estimate of $x$ can be taken to be the square of the estimated value of $y$.
If the estimation of $y$ has mean $y_0$ and standard error $\sigma_y$, then the mean value of $y^2$ will be $y_0+\sigma_y^2$, assuming the normal distribution.
The standard error in $x$ will then be
\begin{equation}
\sigma_x = 2 \sigma_y \sqrt{y_0^2+\sigma_y^2/2} \, .
\end{equation}
In contrast, if we were to assume the usual formula for propagation of uncertainty, we would obtain
\begin{equation}
\sigma_x = \sigma_y \left| \left(\left. \frac{df}{dx} \right|_{y=y_0}\right) \right|^{-1} \quad\implies\quad \sigma_x = 2\sigma_y y_0 \, .
\end{equation}
That is, the correction replaces $y_0$ with $\sqrt{y_0^2+\sigma_y^2/2}$, which means that the uncertainty in $x$ does not approach zero for small $y_0$.

More generally, an inverse function $f^{-1}(y)$ can be expanded in a quadratic about $y_0$, so that
\begin{equation}
x \approx f^{-1}(y_0) + (y-y_0) \left( \left.\frac{d f^{-1}}{dy} \right|_{y=y_0}\right)
+ \frac 12 (y-y_0)^2 \left( \left.\frac{d^2 f^{-1}}{dy^2} \right|_{y=y_0}\right).
\end{equation}
Then we have the mean value of $x$
\begin{equation}
\overline x \approx f^{-1}(y_0) 
+ \frac 12 \sigma_y^2 \left( \left.\frac{d^2 f^{-1}}{dy^2} \right|_{y=y_0}\right) ,
\end{equation}
and the variance is
\begin{align}
\langle (x-\overline x)^2 \rangle  &\approx
\left\langle \left[ (y-y_0) \left( \left.\frac{d f^{-1}}{dy} \right|_{y=y_0}\right)
+ \frac 12 [(y-y_0)^2-\sigma_y^2] \left( \left.\frac{d^2 f^{-1}}{dy^2} \right|_{y=y_0}\right) \right]^2 \right\rangle \nonumber \\
&\approx  \sigma_y^2 \left( \left.\frac{d f^{-1}}{dy} \right|_{y=y_0}\right)^2
+ \frac 14 (\kappa-1)\sigma_y^4 \left( \left.\frac{d^2 f^{-1}}{dy^2} \right|_{y=y_0}\right)^2  
\end{align}
where $\kappa$ is the kurtosis of the distribution (the fourth standardized moment of the distribution).

In the case where $f(x)=\sqrt{x}$ this gives $4\sigma_y^2 y_0^2 + 2\sigma_y^2$ as above, with the 2 coming from $\kappa-1$, so will be larger for distributions with excess kurtosis.
In Section \ref{sec:phase_esitimate_sos} the function is $\pm\arccos(h/\Lambda-1)$, which yields
\begin{equation}
\left.\frac{d f^{-1}}{dy} \right|_{y=y_0} = -\frac 12 \lambda\sin(y_0) = -\sqrt{h(2\Lambda-h)}, \qquad
\left.\frac{d^2 f^{-1}}{dy^2} \right|_{y=y_0} = -\frac 12 \lambda\cos(y_0) =\Lambda -h.
\end{equation}
That then gives
\begin{equation}
\sigma_{E_k}^2 = \sigma_\text{QPE}^2 E_{\min}(2\Lambda-E_{\min}) + \frac 14 (\kappa-1)\sigma_\text{QPE}^4 \left(2\Lambda -E_{\min}\right)^2 .
\end{equation}
Using $\kappa=3$ and approximating small $E_{\min}$ in the last term gives
\begin{equation}
\sigma_{E_k} = \sigma_\text{QPE} \sqrt{E_{\min}(2\Lambda-E_{\min}) + \Lambda^2\sigma_\text{QPE}^2 /2} .
\end{equation}

To give an estimate of the parameter range where this correction will be significant, let us solve for $\sigma_\text{QPE}$, and expand for small $\sigma_{E_k}$, to give
\begin{equation}
    \sigma_\text{QPE} \approx \frac{\sigma_{E_k}}{\sqrt{E_{\min}(2\Lambda-E_{\min})}} \left( 1 - \frac{\sigma_{E_k}^2 \Lambda^2}{4 E_{\min}^2(2\Lambda-E_{\min})^2} \right)
\end{equation}
In the cases we are interested in, $\Lambda\gg E_{\min}$, and the correction will be negligible provided
\begin{equation}
    E_{\min}\gg \sigma_{E_k} .
\end{equation}
If we substitute an initial choice for $\sigma_\text{QPE}=\sigma_{E_k}'/\lambda_\text{eff}$, where $\lambda_\text{eff}=\sqrt{E_\text{min}(2\Lambda-E_\text{min})}$, then the correction
\begin{align}
\sigma_{E_k}
=
\sigma_{E_k}^{\prime}\sqrt{1 + \frac{\Lambda^2\sigma_{E_k}^{\prime2}}{2\lambda_\text{eff}^4}}
\approx
\sigma_{E_k}^{\prime}\left(1 + \frac{\Lambda^2\sigma_{E_k}^{\prime2}}{4\lambda_\text{eff}^4}\right).
\end{align}
For typical parameters we consider like $\Lambda\sim 100$Ha, $\lambda_\text{eff}\sim 10$Ha, $\sigma_{E_k}'\sim 1$mHa, the fractional correction to $\sigma_{E_k}$ is negligible at $\sim 2.5\times 10^{-7}$.

\end{document}